\documentclass{article}

\usepackage{PRIMEarxiv}

\usepackage[utf8]{inputenc} % allow utf-8 input
\usepackage[T1]{fontenc}    % use 8-bit T1 fonts
\usepackage{hyperref}       % hyperlinks
\usepackage{url}            % simple URL typesetting
\usepackage{booktabs}       % professional-quality tables
\usepackage{amsfonts}       % blackboard math symbols
\usepackage{nicefrac}       % compact symbols for 1/2, etc.
\usepackage{microtype}      % microtypography
\usepackage{lipsum}
\usepackage{fancyhdr}       % header
\usepackage{graphicx}       % graphics
\graphicspath{{media/}}     % organize your images and other figures under media/ folder
\usepackage{subcaption}

\usepackage{bm}
\usepackage[short]{optidef}

\DeclareMathOperator*{\Poisson}{Poisson}

\DeclareMathOperator*{\ndash}{-}

\title{Ring Artifact Correction in Photon-Counting Spectral CT Using a Convolutional Neural Network With Spectral Loss
}

\author{
  Dennis Hein, Konstantinos Liappis, Alma Eguizabal, Mats Persson\\
  KTH Royal Institute of Technology \\
  \texttt{\{dhein,liappis,persson6\}@kth.se, alma.egui@gmail.com} \\
   \And
  Fredrik Grönberg \\
  GE Healthcare\\
  \texttt{fredrik.gronberg@ge.com} \\
}

\begin{document}
\maketitle

\begin{abstract}
Photon-counting spectral computed tomography is now clinically available. These new detectors come with the promise of higher contrast-to-noise ratio and spatial resolution and improved low-dose imaging. However, one important design consideration is to build detector elements that are sufficiently homogeneous.  In practice, there will always be a degree of inhomogeneity in the detector elements, which will materialize as variations in the energy bin thresholds. Without proper detector calibration, this will lead to streak artifacts in the sinograms and corresponding ring artifacts in the reconstructed images, which limit their clinical usefulness. Since detector calibration is a time-consuming process, having access to a fast ring correction technique may greatly improve workflow. In this paper, we propose a deep learning-based post-processing technique for ring artifact correction in photon-counting spectral CT. We train a UNet with a custom loss to correct for ring artifacts in the material basis images. Our proposed loss is made ``task-aware'' by explicitly incorporating the fact that we are working with spectral CT by combining a L1-loss operating on the material basis images with a perceptual loss, using VGG16 as feature extractor, operating on 70 keV virtual monoenergetic images. Our results indicate that using this novel loss greatly improves performance. We demonstrate that our proposed method can successfully produce ring corrected 40, 70, and 100 keV virtual monoenergetic images. 
\end{abstract}

% keywords can be removed
%\keywords{}

\section{Introduction}
X-ray computed tomography (CT) imaging is a widely used medical imaging modality providing healthcare with an important tool of diagnosis and treatment planning for a wide range of diseases such as stroke, cancer, and cardiovascular disease. The next generation X-ray CT scanners, based on photon-counting detectors, are now clinically available \cite{liu2022, higashigaito2022}. Advantages of photon-counting detectors, compared to standard energy-integrating detectors, include higher contrast-to-noise ratio and spatial resolution, improved low-dose imaging, and better characterization of tissue composition \cite{roessl2007,willemink2018,danielsson2021}. One important challenge in developing photon-counting detectors is to build detector elements that are sufficiently homogeneous \cite{persson2012}. Inhomogeneity in the detector elements will result in variations in the energy bin thresholds. Due to these variations, photons of a given energy may be registered in different bins in different detector elements. Hence, the detector elements miscount the number of photons. This leads to streak artifacts in the sinogram domain and, after tomographic reconstruction, corresponding ring artifacts in the image domain. These rings are very conspicuous and limit the clinical usefulness of the images. The generation of ring artifacts can be avoided through careful detector calibration. However, thorough calibration is a time consuming process. Thus, having a quick-to-apply method for ring correction may significantly improve workflow. 

Ring artifact reduction methods for X-ray CT can broadly be categorized into pre- and post-processing methods. Pre-processing methods operate in the sinogram domain and post-processing methods in the image domain. Some conventional ring artifact correction pre-processing methods included flat-field correction \cite{seibert1998} and various filtering methods \cite{raven1998,munch2009,anas2010,rashid2012}. Examples of post-processing methods include filtering-based methods such as \cite{sijbers2004, prell2009} and variation-based methods such as \cite{yan2016}. Shifting from a Cartesian to a polar coordinate system will transform the rings in the reconstructed images into streaks. Since streaks might be easier to detect and correct for, many post-processing methods first shift to polar coordinates.

Machine learning, and in particular deep learning, is increasingly playing an important role in medical imaging \cite{wang2016,wang2017}. A lot of this work focus on the problem of image denoising. In \cite{wolterink2017}, the authors use a generative adversarial network (GAN) \cite{goodfellow2014} to map lose-dose CT images to their normal dose counterpart. The GAN setup will encourage the distributions of the low-dose and normal-dose images to be similar. The authors argue that adding the adversarial loss prevents aggressive denosing and helps preserve fine details in the processed image. \cite{yang2018} is very similar to \cite{wolterink2017} but they substitute the L2-loss with a perceptual loss \cite{johnson2016}, and the GAN with the Wasserstein GAN (WGAN) \cite{arjovsky2017} with gradient penalty \cite{gulrajani2017}. Using a perceptual loss, which compares features extracted from the output and target image using a pretrained convolutional neural network, can help prevent over-smoothing and produce processed images of higher perceptual quality \cite{ledig2016,johnson2016}. \cite{kim2019} investigate the effect of different loss configurations for the low-dose CT image denoising problem. Their results indicate that using a perceptual loss is superior to pixel-wise loss functions and that adding an additional adversarial loss may further improve the noise and signal transfer properties of the network. 

Deep learning has also successfully been used for artifact correction and joint artifact and noise correction. Examples of deep learning-based ring artifact correction in the sinogram domain include \cite{nauwynck2020} and \cite{hein2022}. However, great care must be taken when processing data in the sinogram domain to ensure that no artifacts are induced in the reconstructed images as tiny, seemingly unimportant, mistakes in the sinogram can result in large reconstruction errors. For post-processing methods, as with conventional techniques, it is common to first transform the image from a Cartesian to a polar coordinate system. This is done in \cite{wang2019} where they, inspired by the super-resolution generative adversarial network (SRGAN) \cite{ledig2016}, train a deep residual neural network using a perceptual, an adversarial, and an unidirectional relative total variation loss. In \cite{trapp2022} the authors work in the image domain directly, without transforming to polar coordinates, and show that it is beneficial to train on random patches extracted from the images rather than the images themselves. Instead of focusing on one single domain, one may include information from both the sinogram and image domain in the ring correction pipeline. This is done in, for instance, \cite{fang2020} and \cite{chang2021}. In \cite{lv2020} the authors combine ring correction with the problem of image denoising in photon-counting spectral CT. In particular, they train a convolution residual network using a L2-loss to correct for noise and ring artifacts in reconstructed energy bin images.

In this paper, we add to this literature by developing a post-processing technique using deep learning for ring artifact correction in photon-counting \textit{spectral} CT. In particular, we train a 2D UNet using a custom spectral loss to correct for rings in the material basis images. The spectral loss combines a L1-loss operating on the material basis images and a perceptual loss operating on $70 \mathrm{keV}$ virtual monoenergetic images. The rest of the paper is organized as follows. First, we give a brief treatment of how we simulated photon-counting CT data. Second, the key building blocks from the deep learning literature used in this paper are presented. Third, we present our qualitative and quantitative results. Finally, we discuss our results and suggest future avenues of research.

\section{Methods}
\subsection{Photon-counting spectral CT}
\subsubsection{Material decomposition}
Consider a system with $B>2$ energy bins and, for simplicity, a 2-dimensional image space. Generating CT images from the measured counts in a photon-counting detector can be done in several ways. Here, we consider the projection-based two step approach. The first step is to map the measured counts to the basis material line integrals. We solve this by making the ansatz that the X-ray linear attenuation coefficient $\mu(x,y;E)$ can be approximated by a linear combination of $K$ basis materials 
\begin{equation}
    \mu(x,y;E) \approx \sum_{k=1}^K a_k (x,y) \tau_k(E),
\end{equation}
where $a_k$ and $\tau_k(E)$ denote the basis coefficients and basis functions, respectively. The projection-based approach resolves the material basis decomposition in the sinogram domain and thus our target variables are the material line integrals
\begin{equation}
    A_k(\ell) = \int_\ell a_k(x,y)ds = \mathcal{R}(a_k),
    \label{linint}
\end{equation}
where $\mathcal{R}$ denotes the Radon transform operator. The forward model is the polychromatic Beer-Lambert law, which relates the expected number of photons in bin $j$ to the material line integrals, 
\begin{equation}
    \lambda_j(\bm{A}) = \int_0^\infty \omega_j(E) \exp \left( - \sum_{k=1}^K A_m \tau_k(E) \right) dE,
\end{equation}
where $\omega_j(E)$ models the joint effect of the detector efficiency, the X-ray source, and the energy response from bin $j$ \cite{roessl2007}. We stack the measured counts in the $B$ energy bins into the vector $\bm{y} := [y_1,...,y_B].$ We will assume that $y_j$ are independent Poisson random variables, that is, for each $j$  
\begin{equation}
    y_j \sim \Poisson(\lambda_j(\bm{A})).
\end{equation}
Thus, the material decomposition can be formulated as the, non-linear, inverse problem of mapping the measured photon counts $\bm{y}$ to the material line integrals $\bm{A} := [A_1,...,A_K].$ The most common approach to this problem is the maximum likelihood method\cite{gronberg2020,alvarez2011,drucos2017}. More formally, setting up the likelihood and simplifying yields the following program 
\begin{mini}
		{\textbf{A}}{\sum_{j=1}^B \left( \lambda_j (\bm{A}) - y_j \log(\lambda_j (\bm{A}))\right)}
		{}{}
		\addConstraint{A_k}{\geq 0 \quad \forall k = 1,...,K},
\end{mini}
which is usually solved using some iterative optimization algorithm such as the logarithmic barrier method \cite{boyd2004}. The second step of this image reconstruction chain is tomographic reconstruction, for which we use Filtered Back Projection (FBP). 

\subsubsection{Simulation}
\label{section_sim}
The generation of photon-counting data is similar to the approach taken in \cite{eguizabal2021}. First, a dataset of numerical basis phantoms was created by thresholding CT images from the KiTS19 \cite{heller2021} and NSCLC \cite{bakr2018} datasets. Let $\bm{z} \in \mathbb{R}^{N \times N}$ denote the upsampled\footnote{We use bilinear upsampling to transform the original $512\times512$ images to $1024\times1024.$} CT image in Hounsfield units (HU) and $\tau \in \mathbb{R}$ a threshold in HU for bone. To avoid the issue of differentiating bone from contrast agent, contrast-enhanced images were excluded. Bone phantoms $\bm{z}_{i,j}^{\mathrm{bone}}$ are defined as $\bm{z}_{i,j}$ for $\bm{z}_{i,j}>\tau$ and $0$ for $\bm{z}_{i,j}\leq \tau$ and soft tissue phantoms $\bm{z}_{i,j}^{\mathrm{soft}}$ as  $\bm{z}_{i,j}$ for $\bm{z}_{i,j}\leq \tau$ and $0$ for $\bm{z}_{i,j}> \tau$. We found that $\tau=200$ gave reasonable results. The next step was to transform the basis phantoms into relative unitless terms. We did this by adding and dividing by 1000 HU. Finally, the basis phantoms were weighted by their relative linear attenuation coefficients at 70 $\mathrm{keV}$ $\mu^{\mathrm{water}}/\mu^{\mathrm{soft}}$ and $\mu^{\mathrm{water}}/\mu^{\mathrm{bone}}$, respectively. 

\begin{figure}
    \centering
    \includegraphics[width=\columnwidth]{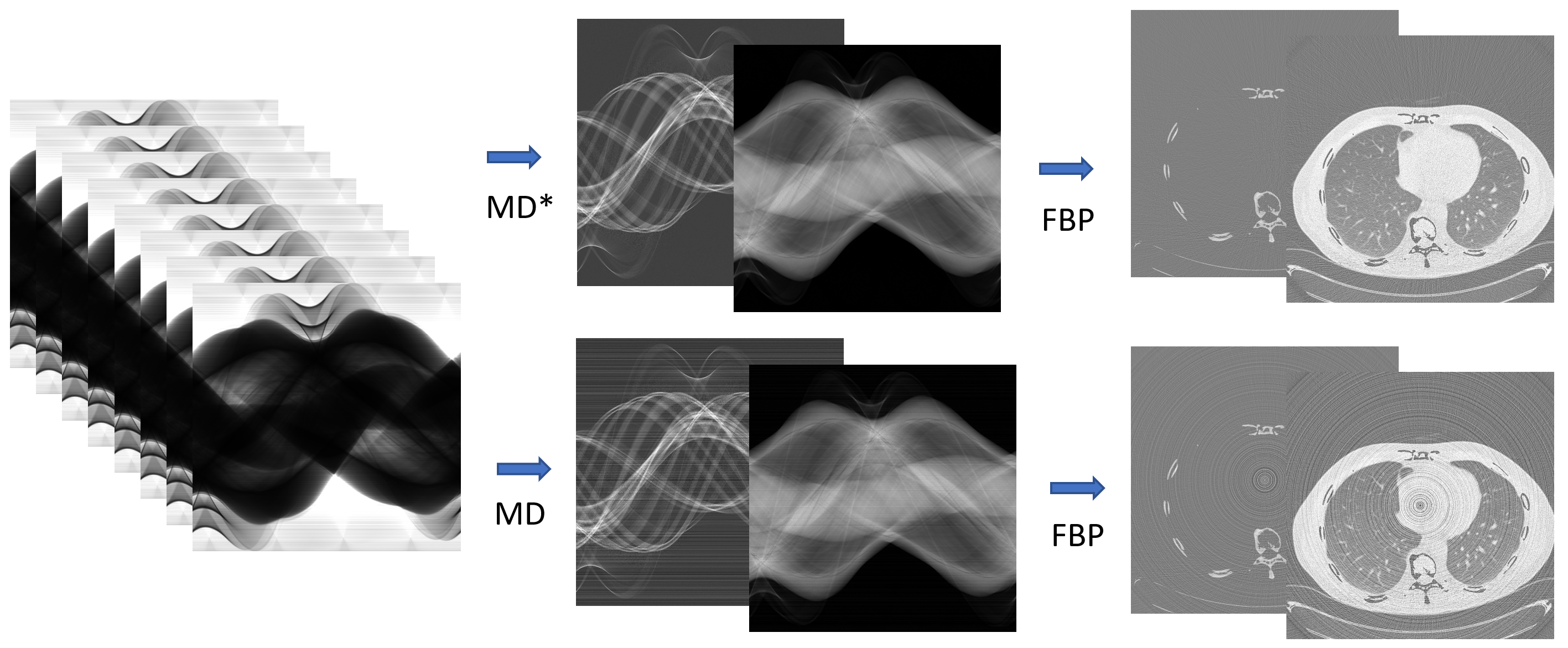}
    \caption{Overview of data generation pipeline. Here MD* denotes material decomposition using the energy bin thresholds from the simulation, including random threshold shifts, and MD material decomposition with the nominal bin thresholds.}
    \label{data_gen}
\end{figure}

Second, after generating the numerical basis phantoms, we simulated photon-counting images by using the \texttt{fanbeam} function in Matlab and a spectral response model of a photon-counting silicon detector \cite{persson2020} with $0.5\times0.5 \; \mathrm{mm}^2$ pixels for 120 $\mathrm{kVp}$ and 200 $\mathrm{mAs}$ with 2000 detector pixels and 2000 view angles. We subsequently simulated Poisson noise and used the maximum likelihood method to decompose the simulated energy bin sinograms into soft tissue and bone material basis sinograms. To simulate detector inhomogeneity, we model threshold variations in the simulation by applying a random threshold shift ($\sigma=0.5 \; \mathrm{keV}$) applied independently to each of the eight energy bin thresholds of each detector pixel and subsequently two material decompositions were performed: one with the thresholds used in the simulation, including the random shift, and one with the nominal bin thresholds. This latter configuration yields sinograms with streak artifacts and, after reconstruction, images with ring artifacts. Finally, images were reconstructed on a $1024 \times 1024$ pixel grid using FBP. The data simulation pipeline is illustrated in Fig. \ref{data_gen}.

\subsection{Deep learning}
\subsubsection{Problem statement}
This paper proposes an image processing technique for ring artifact correction in photon-counting spectral CT using deep neural networks. For $K$ basis materials and a $H\times W$ pixel grid, we formulate this problem as learning the map
\begin{equation}
    f : \bm{x} \rightarrow \bm{y},
    \label{fmap}
\end{equation}
where $\bm{x} \in \mathbb{R}^{K \times H \times W}$ denotes the ring corrupted material basis images and $\bm{y} \in \mathbb{R}^{K \times H \times W}$ their artifact free counterpart.  We let $f$ be a convolutional neural network (CNN) parameterized by $\theta$ and learn the map (\ref{fmap}) by learning parameters $\theta.$

\subsubsection{Loss function}
Common loss functions in biomedical imaging, and deep learning more generally, include L2-loss, L1-loss, and perceptual loss. The L1-loss
\begin{equation}
    \ell_1(f) = \frac{1}{KHW} ||f(\bm{x})-\bm{y}||_1
    \label{l1_loss}
\end{equation}
and the L2-loss
\begin{equation}
    \ell_2(f) = \frac{1}{KHW} ||f(\bm{x})-\bm{y}||_2^2
    \label{l2_loss}
\end{equation}
both compare output and target pixel-by-pixel. This low-level per-pixel comparison is known to lead to over-smoothing and a loss of fine-grained details that are important to the perceptual quality of the image \cite{ledig2016,johnson2016}. To prevent these types of issues, one may instead employ a so-called perceptual loss function \cite{johnson2016} which considers differences in high level feature representations rather than pixel-by-pixel. These feature representations are usually extracted from the output and target images using a pretrained CNN as feature extractor, or loss network. In this paper we use VGG16 \cite{simonyan2014} pretrained on ImageNet \cite{deng2009} as feature extractor. VGG16 was trained on RGB images and it accepts input with three channels ($C=3$). For $C=1$ we simply duplicate the main channel three times before passing it to the VGG network. For the $j$-th layer of VGG16, let this map be denoted $\phi_j.$ Then the perceptual loss is defined as 
\begin{equation}
    \ell_{VGG}(f) = \frac{1}{C_jH_jW_j} ||\phi_j(f(\bm{x}))-\phi_j(\bm{y})||_2^2,
    \label{vgg_loss}
\end{equation}
where $C_j$ is the number of channels in layer $j.$ For $C=2$ we apply this map to each basis image and subsequently concatenate the output. We found that $j=9$ yields good results and will use this throughout\footnote{Note that this layer is denoted ``relu2\_2'' in \cite{johnson2016}.}. The perceptual loss correlates well with the, somewhat ambiguously defined, perceptual quality of the resulting image. Humans also extract and compare salient features from images \cite{nixon2008}, rather than comparing pixel-per-pixel. Hence, the perceptual loss will align better with human perception than a pixel-wise loss. For instance, if two images are exactly the same save for one pixel which has a very large absolute error, then the L1- and L2-loss, would be large despite the fact that the images would essentially be perceived as the same to the human eye. Thus, we can think of the perceptual loss akin to a mathematical observer, it is a way to formally quantify the discrepancy between two images in a way that closely resembles human vision. At an early stage of this project we noticed that although most approaches were able to correct for the rings in the material basis images such that they were no longer visible, when forming virtual monoenergetic images these rings reemerged for certain energy levels. Correcting for the rings such that they are no longer directly visible in the 70 $\mathrm{keV}$ virtual monoenergetic images proved the most difficult. These images have the lowest relative noise level, making the ring artifacts more discernable. To ensure good performance even at this energy level, we consider a ``task-aware'' reconstruction loss that explicitly incorporates the fact that we are working with spectral CT by combining a L1-loss operating on the material basis images with a perceptual loss operating on 70 $\mathrm{keV}$ virtual monoenergetic images. More formally, for $K$ basis materials, our spectral loss is defined as 
\begin{equation}
    \ell_{\mathrm{VGG}_{70}\ndash \ell_1}(f) :=  \omega_1 \ell_{\mathrm{VGG}_{70}} (f) +\omega_2\ell_1(f), 
     \label{spectral_loss} 
\end{equation}
where 
\begin{equation}
    \ell_{\mathrm{VGG}_{70}} (f) :=  \ell_{\mathrm{VGG}}\left(\sum_{k=1}^K \mu_{k}f_k(\bm{x}),\sum_{k=1}^K \mu_{k}\bm{y}_{k}\right), 
    \label{vgg_70_loss}
\end{equation}
is the perceptual loss operating on virtual monoenergetic images, $\mu_k$ is the linear attenuation coefficient for basis image $k$ at 70 $\mathrm{keV}$, and $\omega_1, \omega_2$ are hyperparameters used to trade-off the two objectives. The intuition is that it will work similarly to introducing a prior: we are, to anthropomorphize, telling the network that we are in the end interested in virtual monoenergetic images despite working on material basis images. As shown below, this spectral reconstruction loss vastly improves the performance of the network. We denote the perceptual loss operating on the basis images VGG and the perceptual loss operating on virutal monoenergetic images defined in (\ref{vgg_70_loss}) $\mathrm{VGG}_{70}.$ The corresponding two loss configurations combining the perceptual loss with the L1-loss will be denoted VGG-L1 and $\mathrm{VGG}_{70}$-L1, respectively. 

\subsubsection{Network architecture}
\begin{figure}
    \centering
    \includegraphics[width=\columnwidth]{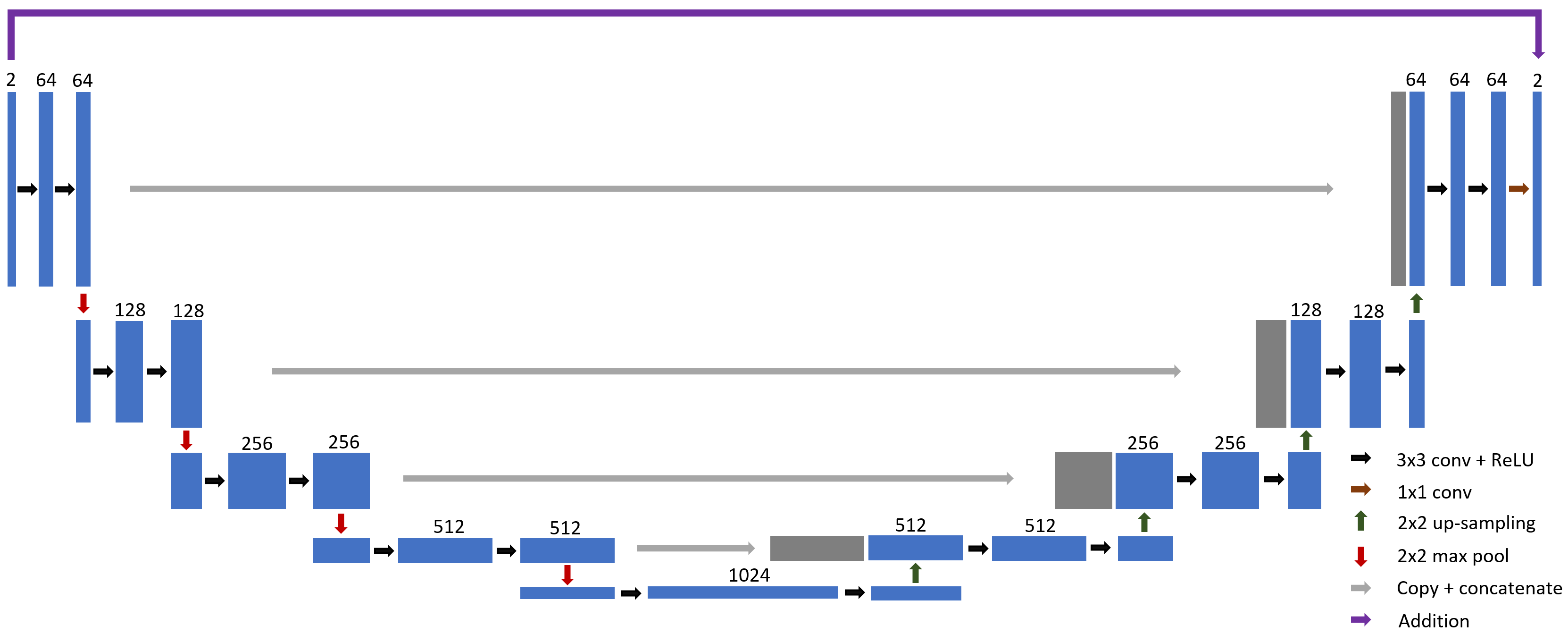}
    \caption{Illustration of the version of UNet used in this paper.}
    \label{unet}
\end{figure}

We use a modified version of the UNet \cite{ronneberger2015}, which was originally developed for image segmentation tasks in biomedical imaging. Our implementation was inspired by \cite{jin2017} and \cite{kim2019}. The key differences from the original UNet are the skip connection from input to output and that the up-convolutions have been replaced with bilinear upsampling to avoid checkerboard artifacts \cite{odena2016}. Note by including this skip connection we repurpose the actual network to learn the residual map. The version of UNet used in this paper is illustrated in Fig. \ref{unet}.

\subsection{Data}
A total of 2250 slices from 90 patients in the KiTS19 dataset, which contains chest and abdomen scans, were simulated and split 70/20/10 into train, validation and test sets. This split was conducted such that slices from any given patient ended up in only one of the three datasets. To evaluate the generalizability of the network, we simulated an additional 450 slices from the NSCLC dataset, which contains full-body scans. Evaluating on this additional dataset gives us an indication of whether the network has learned something general that can be applied successfully to datasets with different anatomies.

\subsection{Training details}
Each loss configuration is trained using Adam \cite{kingma2014} with $\beta_1=0.5,\beta_2=0.9$ and learning rate $\alpha = 10^{-4}$ for 100 epochs on a NVIDIA A100-SXM4-40GB GPU. $\omega_1$ and $\omega_2$ were tuned to get a good trade-off between the different objectives. This was achieved by setting $\omega_1 = 10$ and $\omega_2=1$ for $\mathrm{VGG}_{70}$-L1. Note that these values will put the two terms in the spectral loss on approximately the same scale. Since the scale of VGG and $\mathrm{VGG}_{70}$ are different, we opted to use $\omega_1 = 1$ and $\omega_2=10$ for VGG-L1 as this, again, puts the two objectives on approximately the same scale. For all loss configurations we use batch size 2. The network is trained on randomly extracted $512 \times 512$ patches. Note that by extracting a patch of size $512 \times 512$ from images that are $1024 \times 1024$ we are ensured to always include the center of the image. Hence, each patch will include full rings with a high probability. Training on randomly extracted patches will serve as a regularizer and may thus help prevent overfitting. As an added bonus, training on patches will drastically reduce the graphics memory requirements and speed up training. The network trained with the L2-loss is included as a baseline. The remaining loss configurations are included such that we can conduct an ablation study of our proposed spectral loss: only L1, only VGG but not $\mathrm{VGG}_{70}$, only $\mathrm{VGG}_{70}$, VGG-L1 but not $\mathrm{VGG}_{70}$, and our full setup $\mathrm{VGG}_{70}$-L1. You can find the code used for training and evaluation in the following repo: \url{https://github.com/KTH-Physics-of-Medical-Imaging/deep_spectral_ring}. 

\section{Results}

\subsection{Qualitative results}
\begin{figure}
     \centering
     \begin{subfigure}[b]{0.24\columnwidth}
         \centering
         \includegraphics[width=\columnwidth]{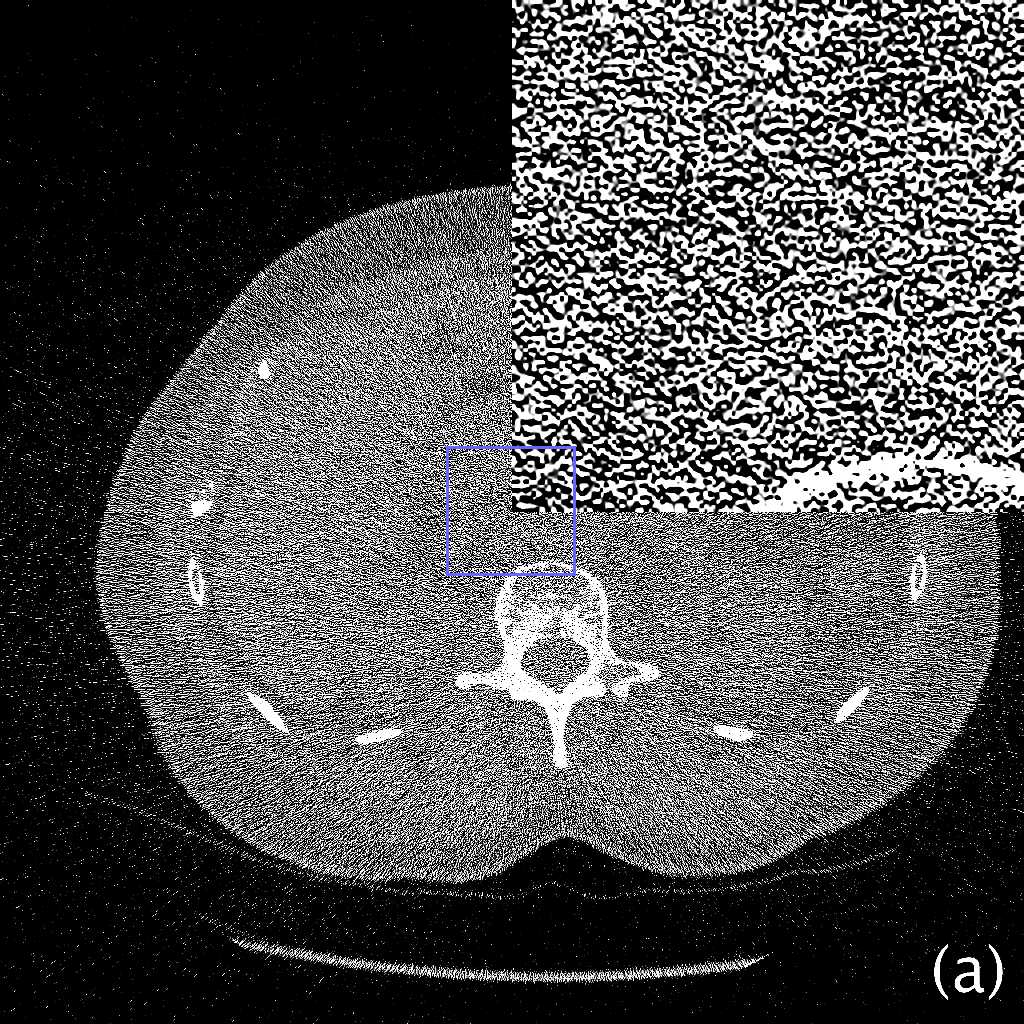}
         %\caption{}
     \end{subfigure}
     \hspace{-0.7em}
     \begin{subfigure}[b]{0.24\columnwidth}
         \centering
         \includegraphics[width=\columnwidth]{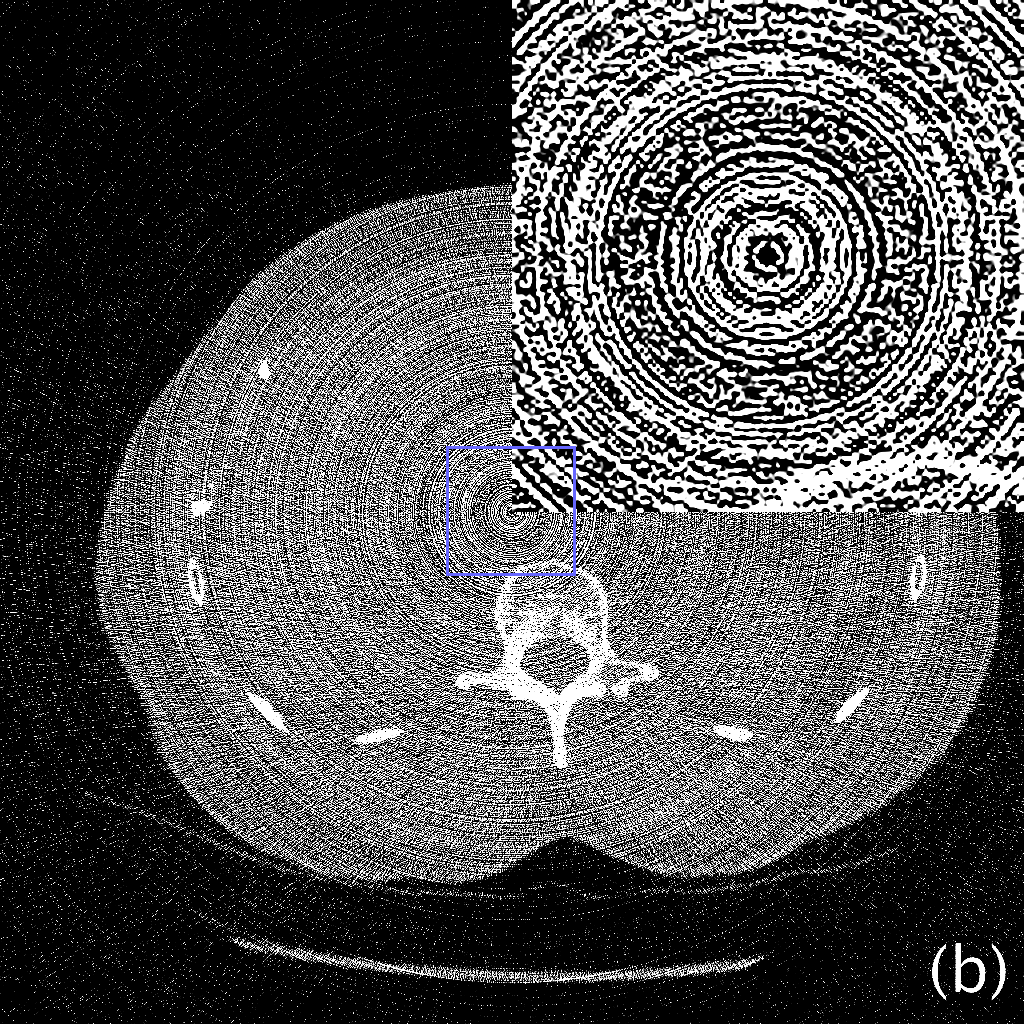}
         %\caption{}
     \end{subfigure}
     \begin{subfigure}[b]{0.24\columnwidth}
         \centering
         \includegraphics[width=\columnwidth]{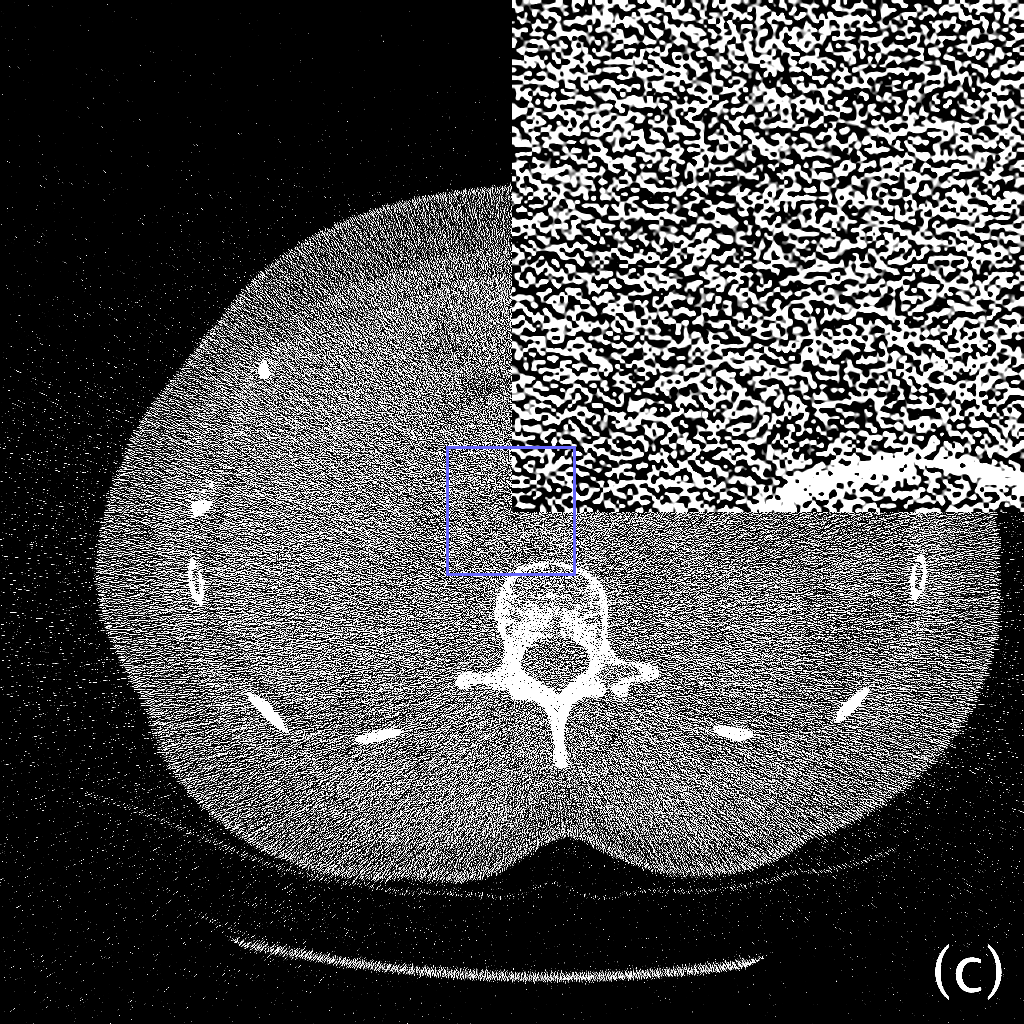}
         %\caption{}
     \end{subfigure}
     \hspace{-0.7em}
       \begin{subfigure}[b]{0.24\columnwidth}
         \centering
         \includegraphics[width=\columnwidth]{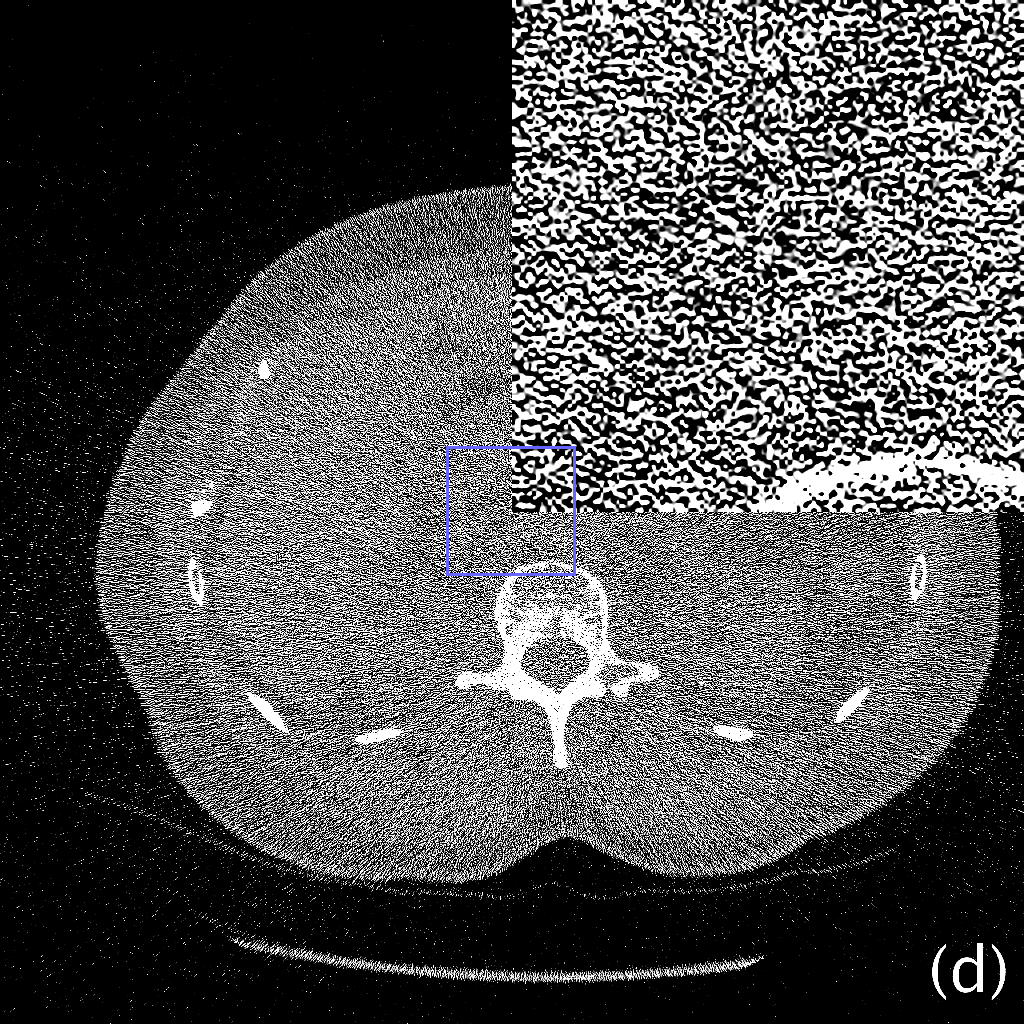}
         %\caption{}
     \end{subfigure}
    \vfill
 \begin{subfigure}[b]{0.24\columnwidth}
         \centering
         \includegraphics[width=\columnwidth]{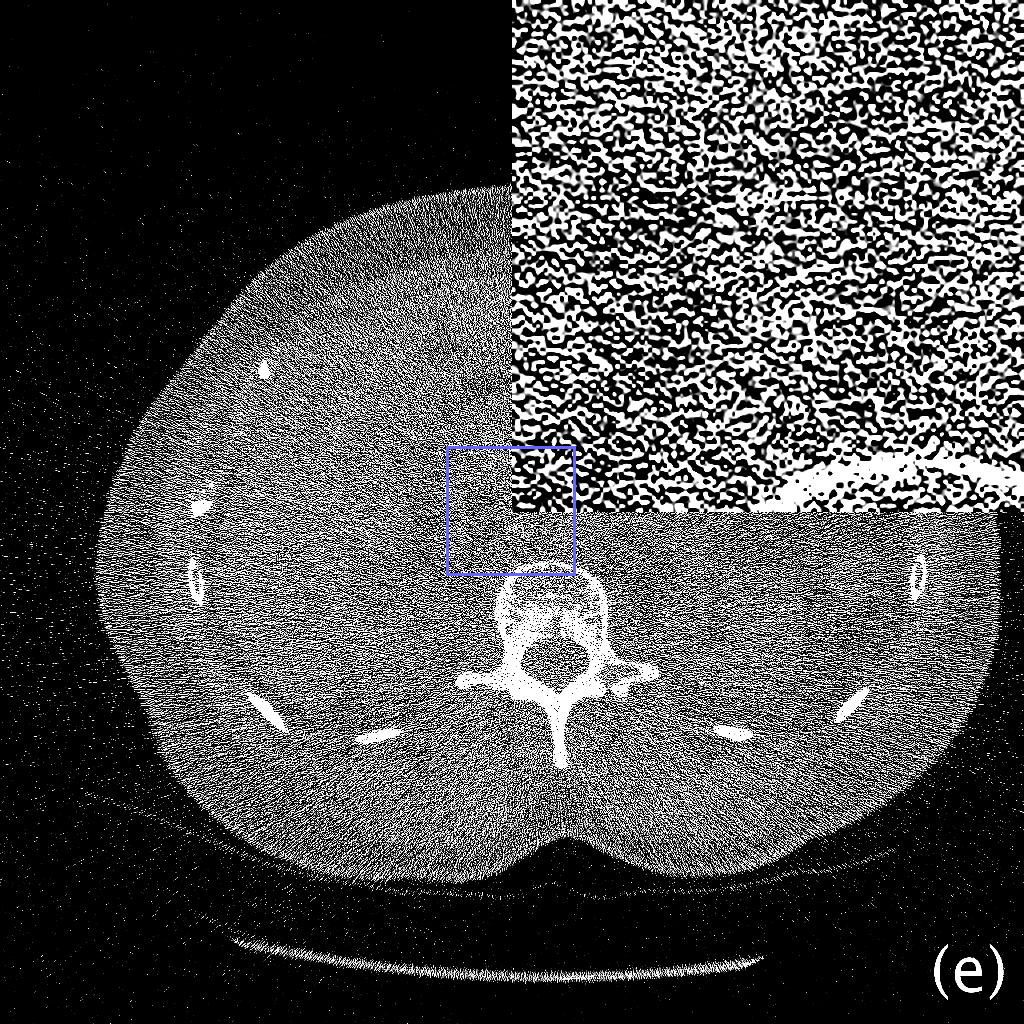}
         %\caption{}
     \end{subfigure}
     \hspace{-0.7em}
 \begin{subfigure}[b]{0.24\columnwidth}
         \centering
         \includegraphics[width=\columnwidth]{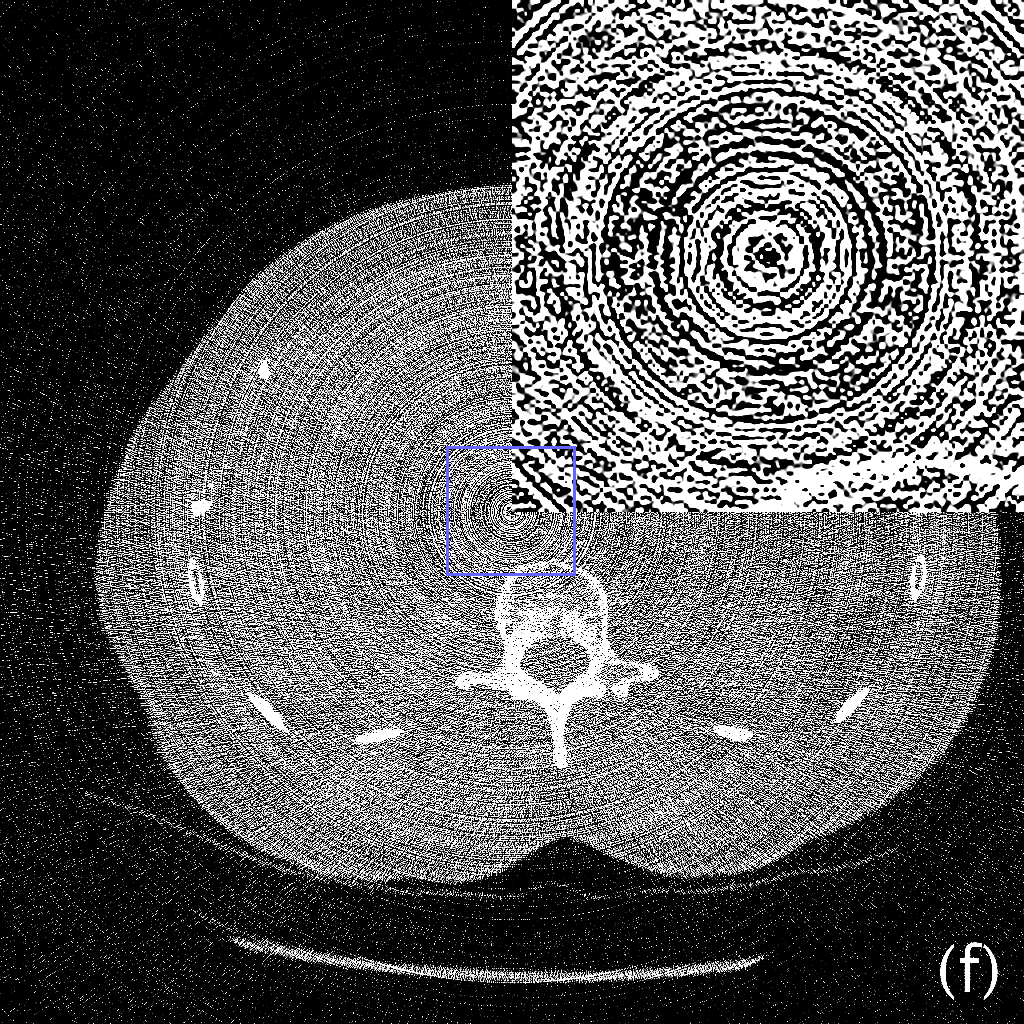}
         %\caption{}
     \end{subfigure}
     \begin{subfigure}[b]{0.24\columnwidth}
         \centering
         \includegraphics[width=\columnwidth]{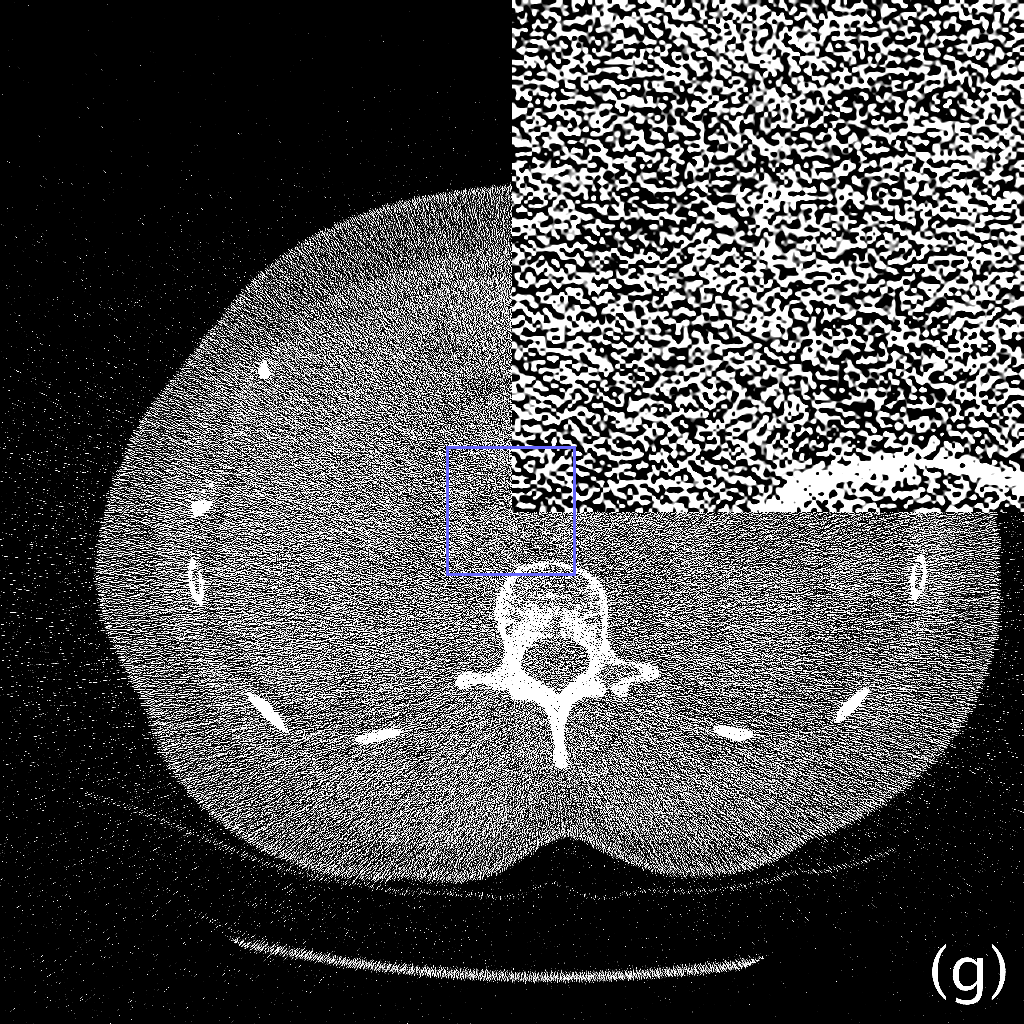}
         %\caption{}
     \end{subfigure}
     \hspace{-0.7em}
     \begin{subfigure}[b]{0.24\columnwidth}
         \centering
         \includegraphics[width=\columnwidth]{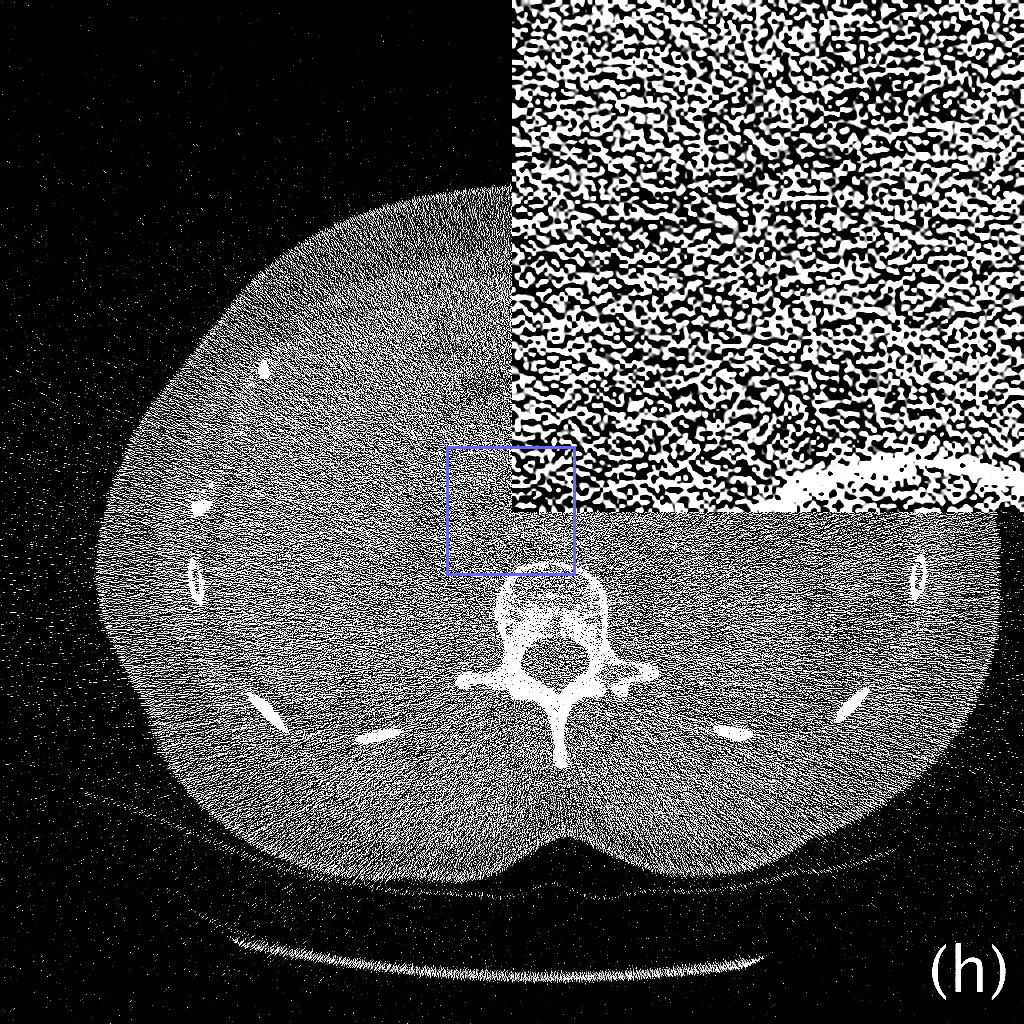}
         %\caption{}
     \end{subfigure}
        \caption{Example slice from the KiTS19 test set. 40 $\mathrm{keV}$ virtual monoenergetic images. Display window [-160,240] HU. (a) ring artifact free (truth), (b) ring corrupted (observed), (c) L2, (d)  L1, (e) VGG, (f) $\mathrm{VGG}_{70}$, (g) VGG-L1, (h) $\mathrm{VGG}_{70}$-L1.}
        \label{qual_40_kits}
\end{figure}

\begin{figure}
     \centering
     \begin{subfigure}[b]{0.24\columnwidth}
         \centering
         \includegraphics[width=\columnwidth]{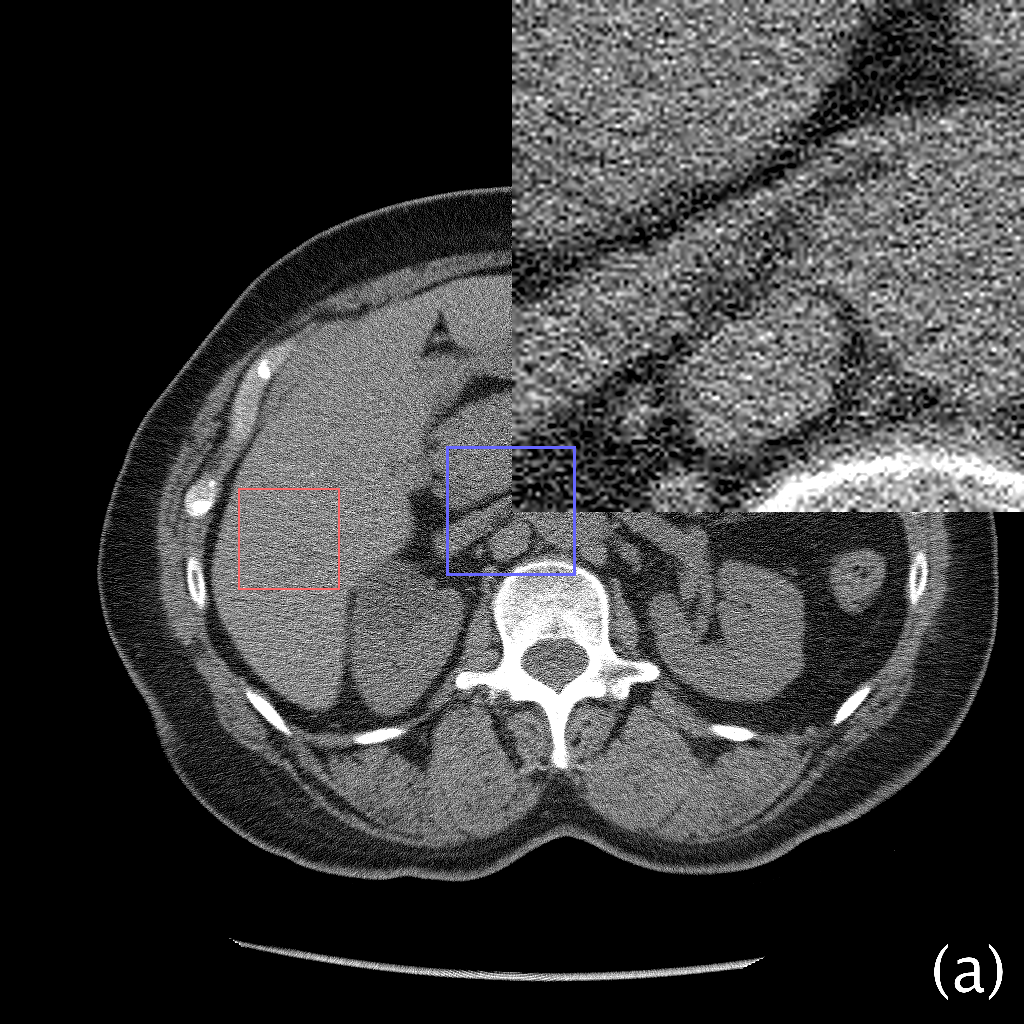}
         %\caption{}
     \end{subfigure}
     \hspace{-0.7em}
     \begin{subfigure}[b]{0.24\columnwidth}
         \centering
         \includegraphics[width=\columnwidth]{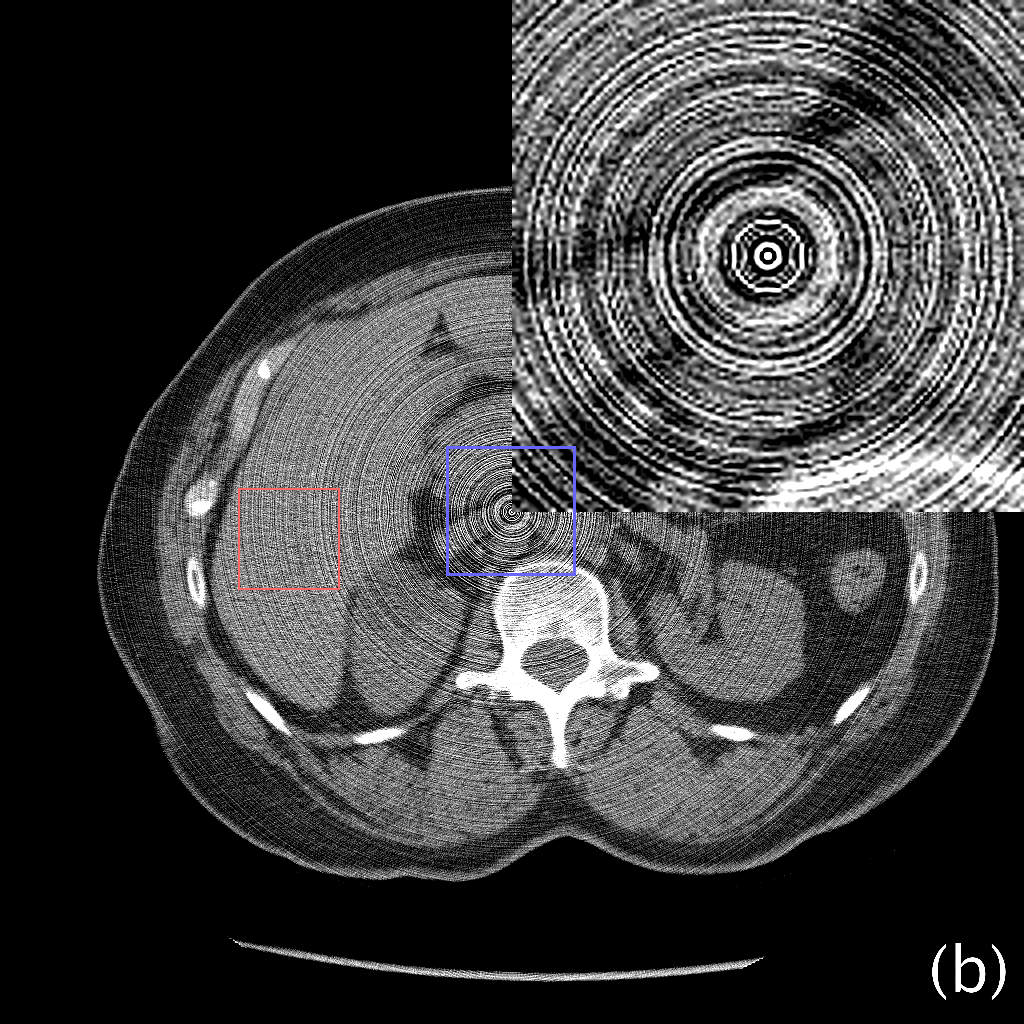}
         %\caption{}
     \end{subfigure}
     \begin{subfigure}[b]{0.24\columnwidth}
         \centering
         \includegraphics[width=\columnwidth]{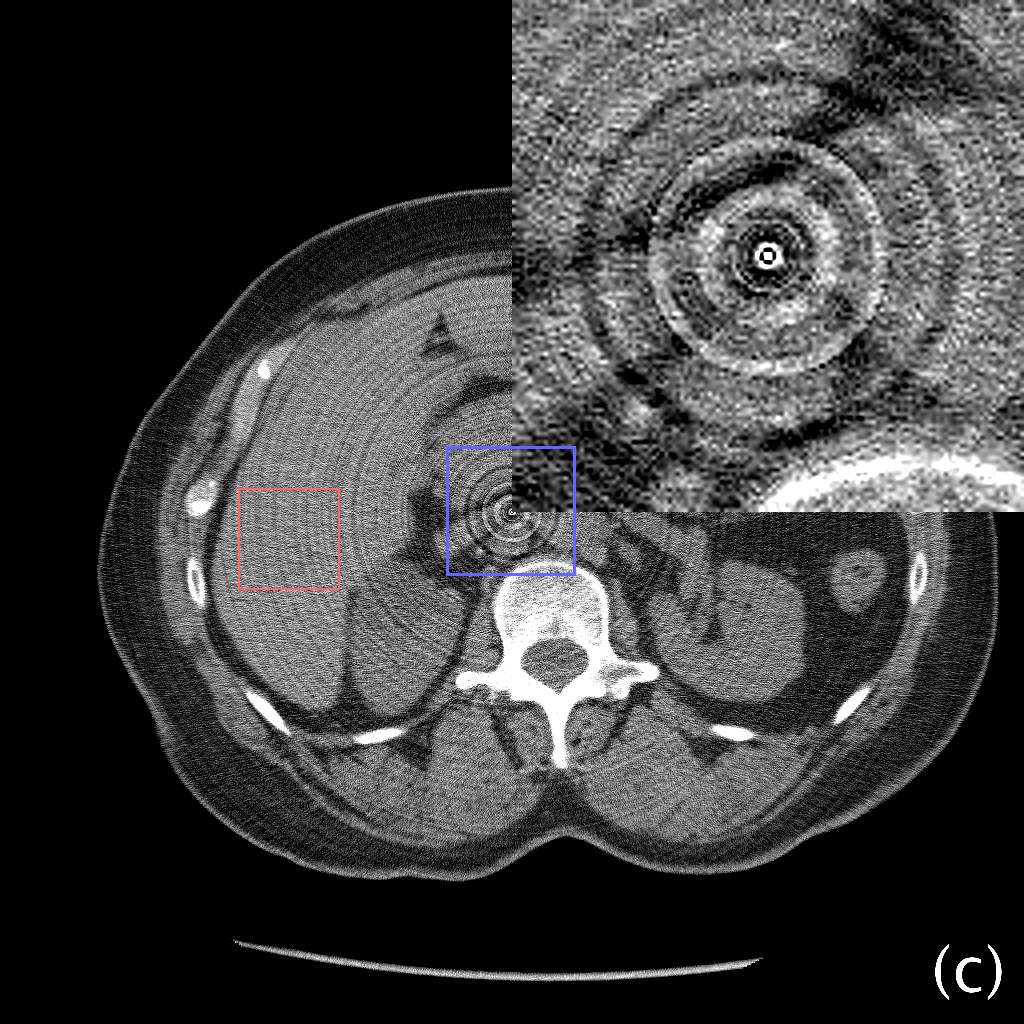}
         %\caption{}
     \end{subfigure}
     \hspace{-0.7em}
       \begin{subfigure}[b]{0.24\columnwidth}
         \centering
         \includegraphics[width=\columnwidth]{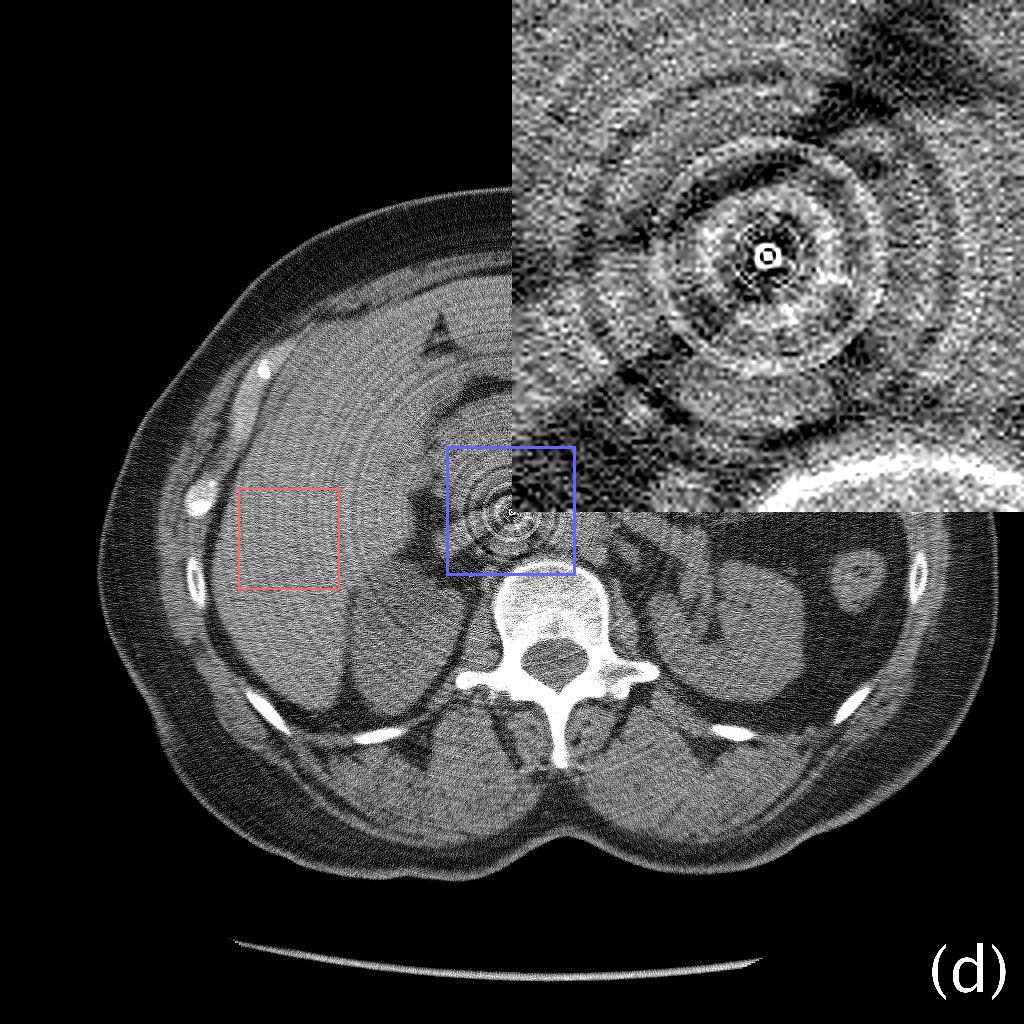}
         %\caption{}
     \end{subfigure}
    \vfill
 \begin{subfigure}[b]{0.24\columnwidth}
         \centering
         \includegraphics[width=\columnwidth]{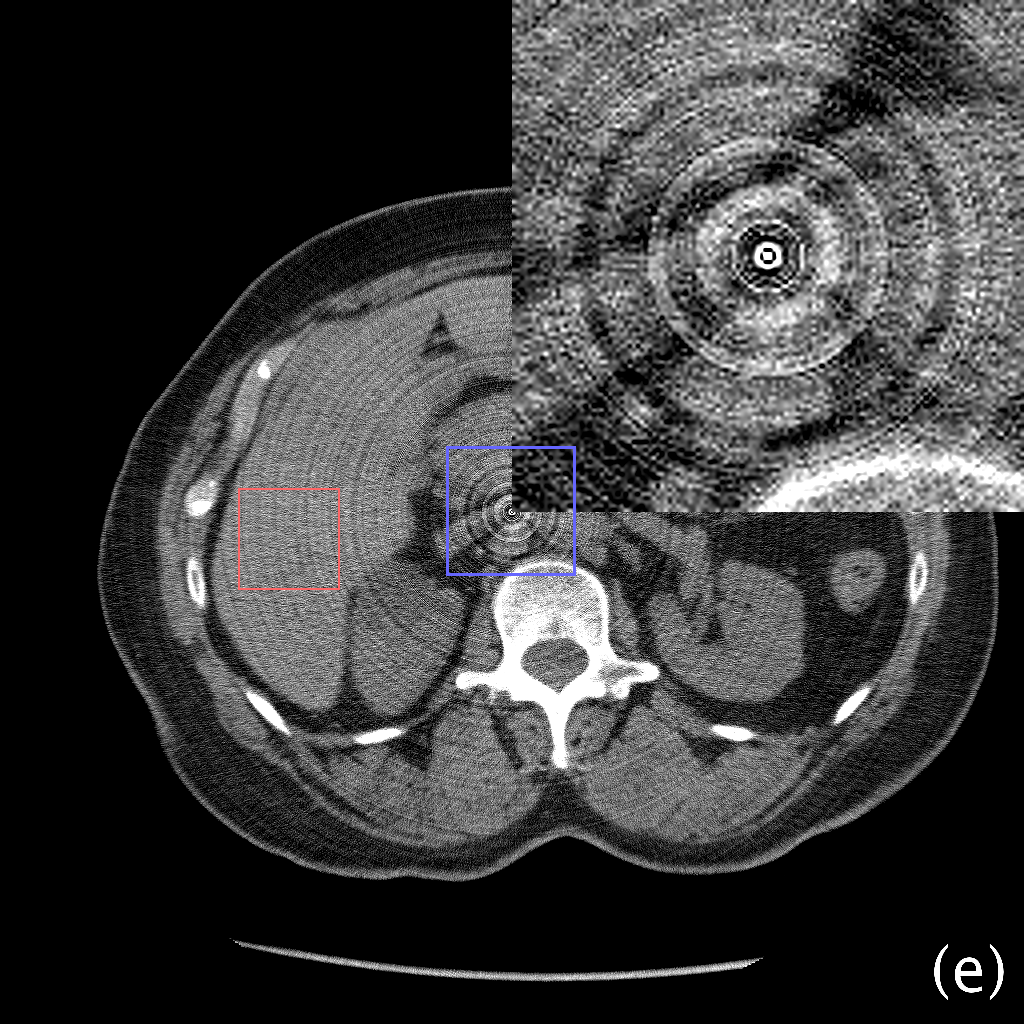}
         %\caption{}
     \end{subfigure}
     \hspace{-0.7em}
 \begin{subfigure}[b]{0.24\columnwidth}
         \centering
         \includegraphics[width=\columnwidth]{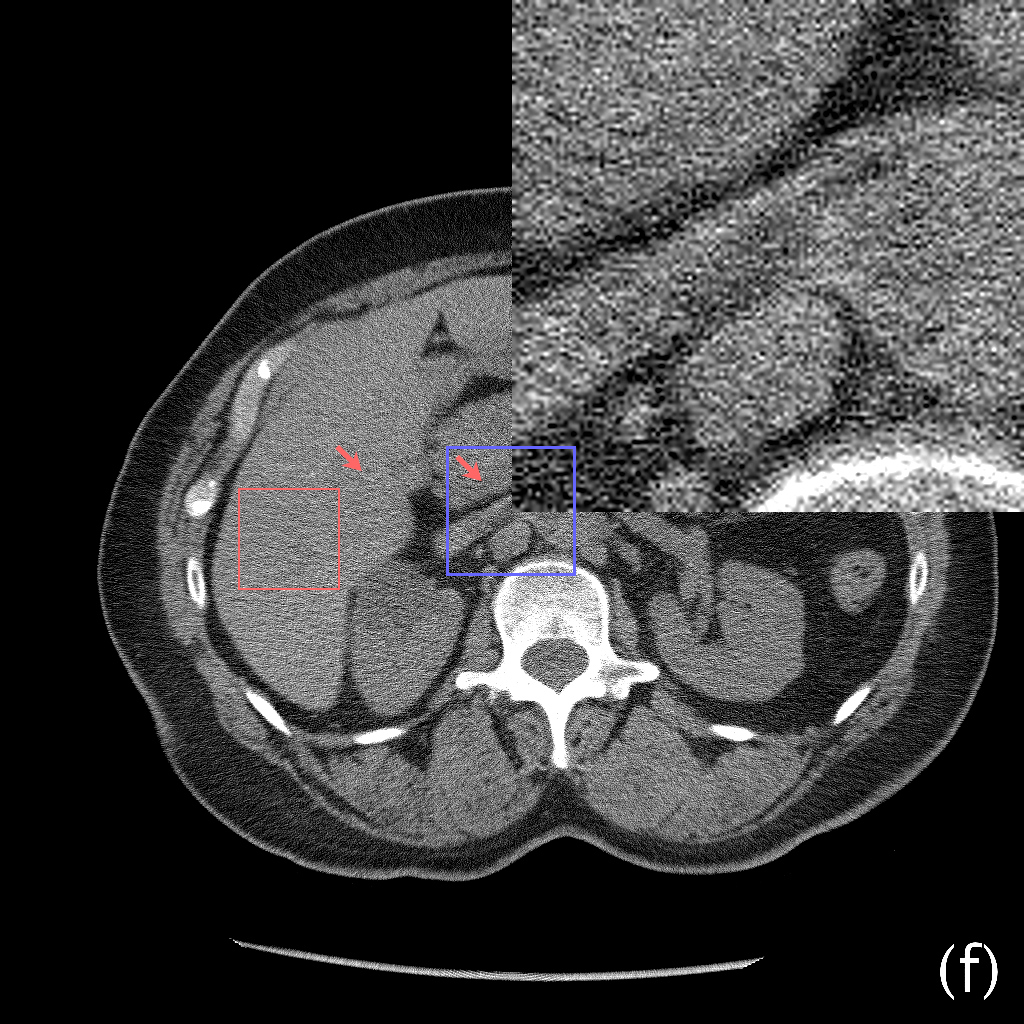}
         %\caption{}
     \end{subfigure}
     \begin{subfigure}[b]{0.24\columnwidth}
         \centering
         \includegraphics[width=\columnwidth]{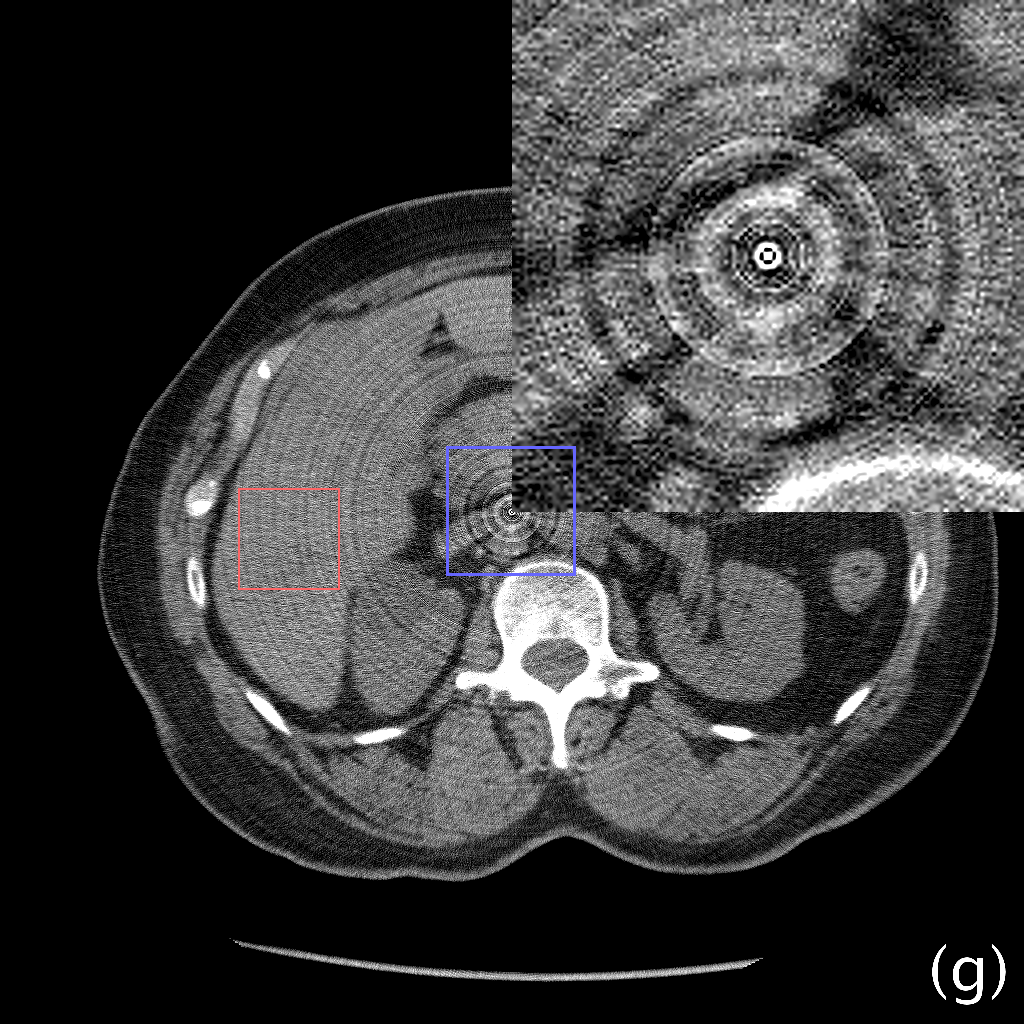}
         %\caption{}
     \end{subfigure}
     \hspace{-0.7em}
     \begin{subfigure}[b]{0.24\columnwidth}
         \centering
         \includegraphics[width=\columnwidth]{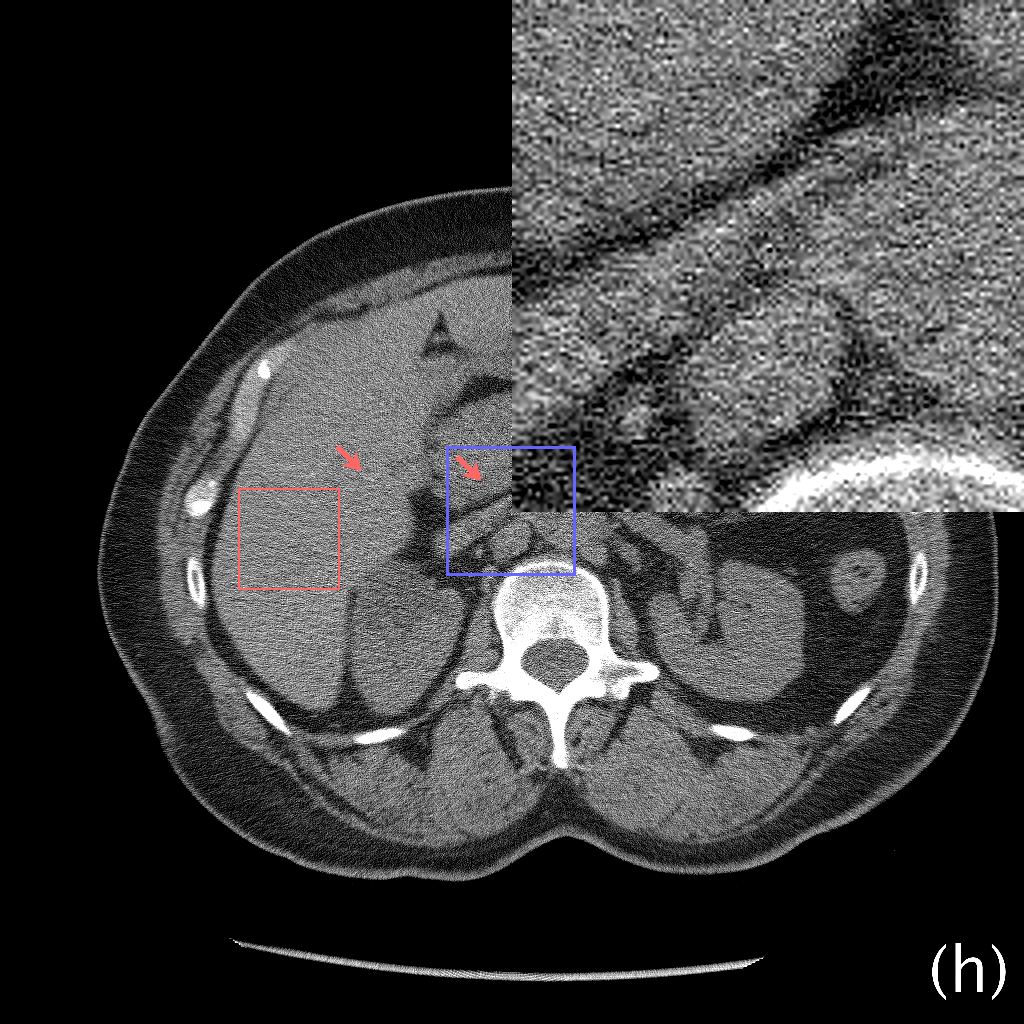}
         %\caption{}
     \end{subfigure}
        \caption{Example slice from the KiTS19 test set. 70 $\mathrm{keV}$ virtual monoenergetic images. Display window [-160,240] HU. (a) ring artifact free (truth), (b) ring corrupted (observed), (c) L2, (d)  L1, (e) VGG, (f) $\mathrm{VGG}_{70}$, (g) VGG-L1, (h) $\mathrm{VGG}_{70}$-L1.}
        \label{qual_70_kits}
\end{figure}

\begin{figure}
     \centering
     \begin{subfigure}[b]{0.24\columnwidth}
         \centering
         \includegraphics[width=\columnwidth]{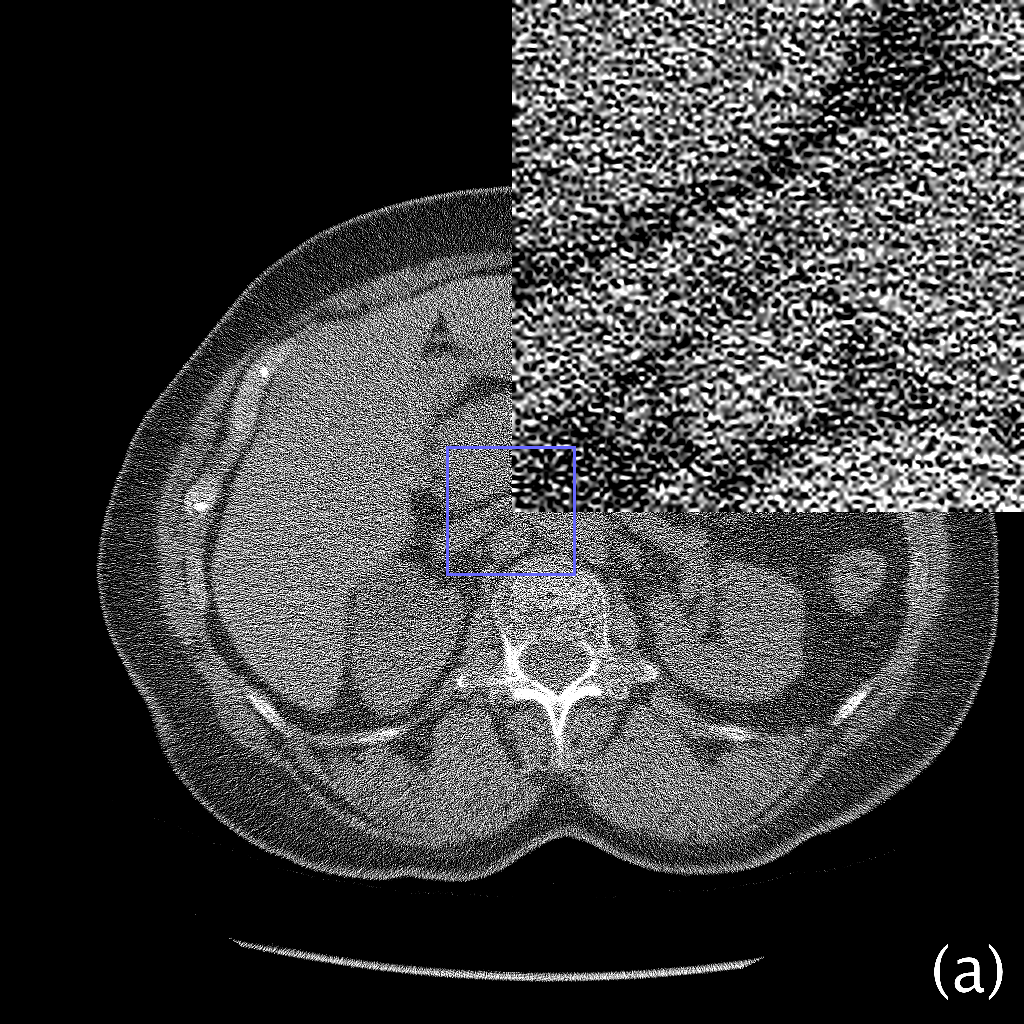}
         %\caption{}
     \end{subfigure}
     \hspace{-0.7em}
     \begin{subfigure}[b]{0.24\columnwidth}
         \centering
         \includegraphics[width=\columnwidth]{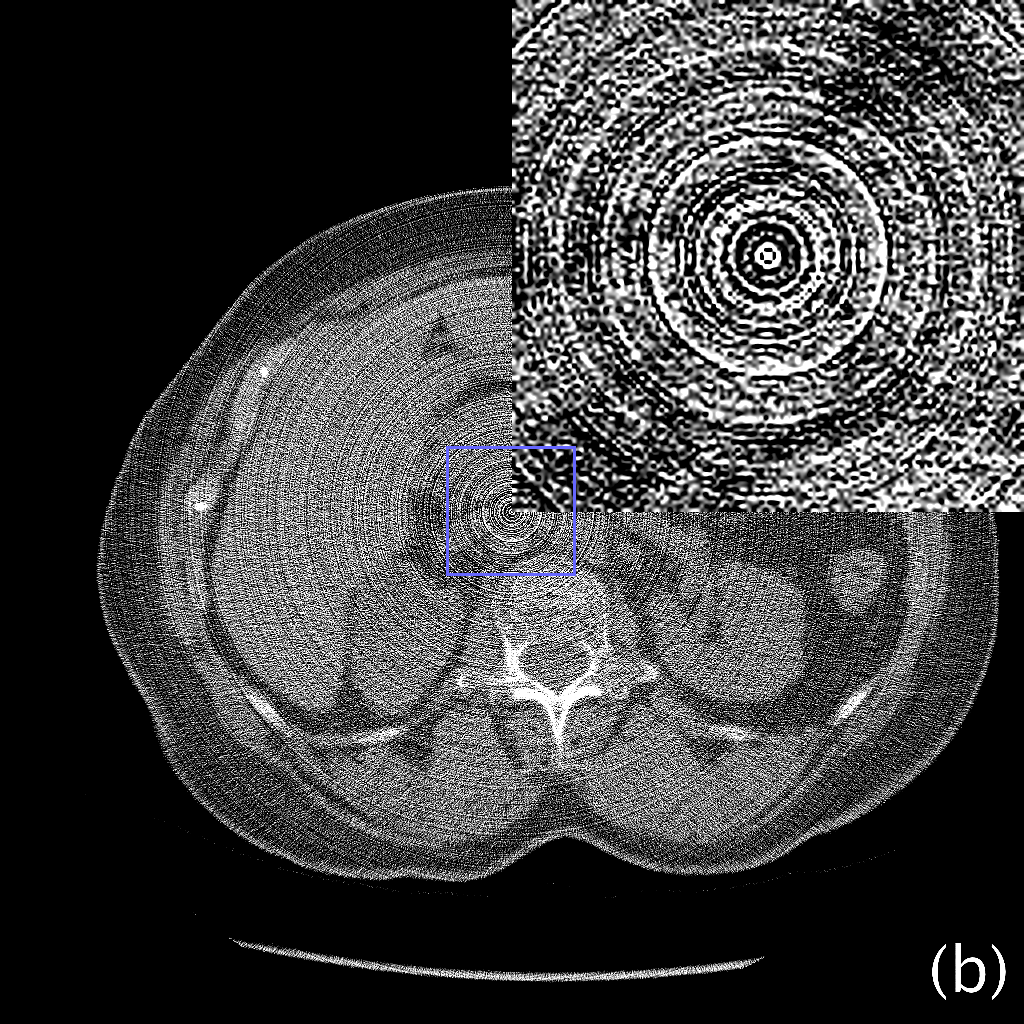}
         %\caption{}
     \end{subfigure}
     \begin{subfigure}[b]{0.24\columnwidth}
         \centering
         \includegraphics[width=\columnwidth]{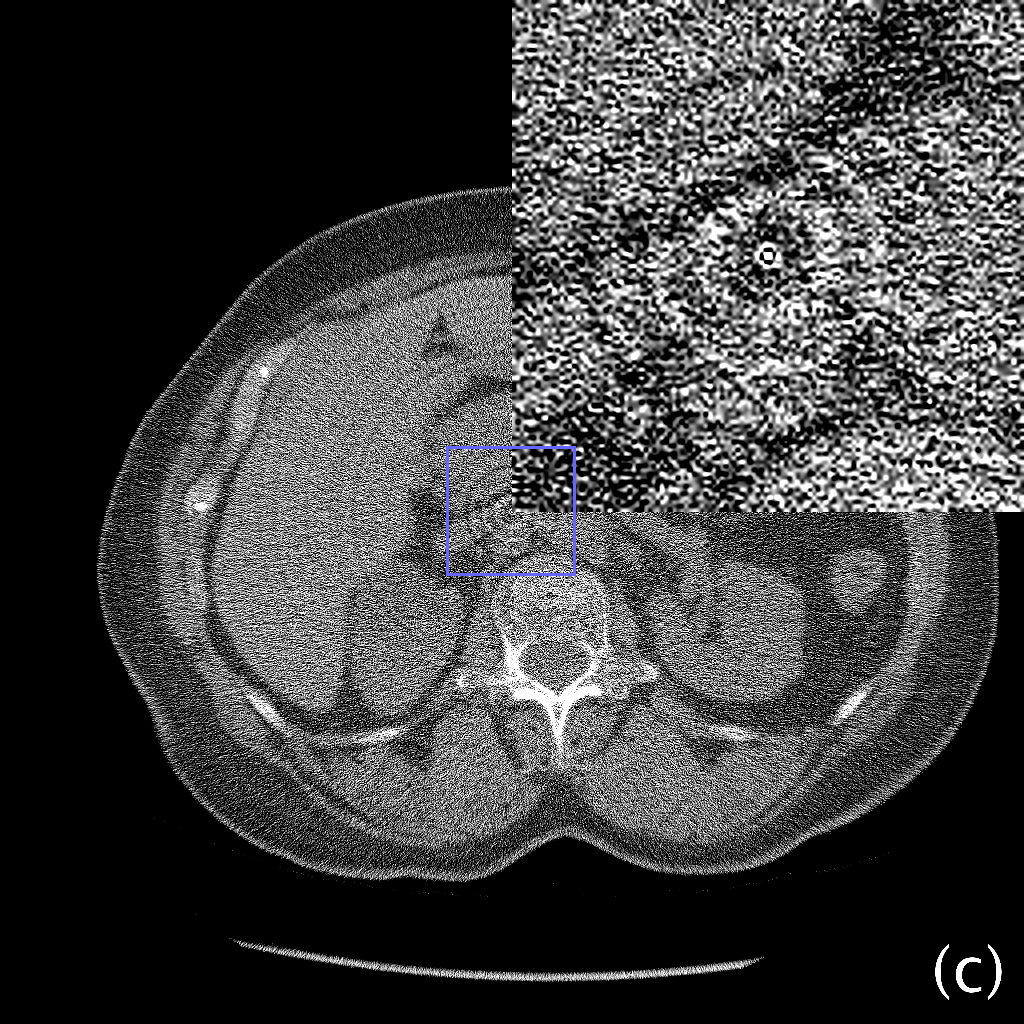}
         %\caption{}
     \end{subfigure}
     \hspace{-0.7em}
       \begin{subfigure}[b]{0.24\columnwidth}
         \centering
         \includegraphics[width=\columnwidth]{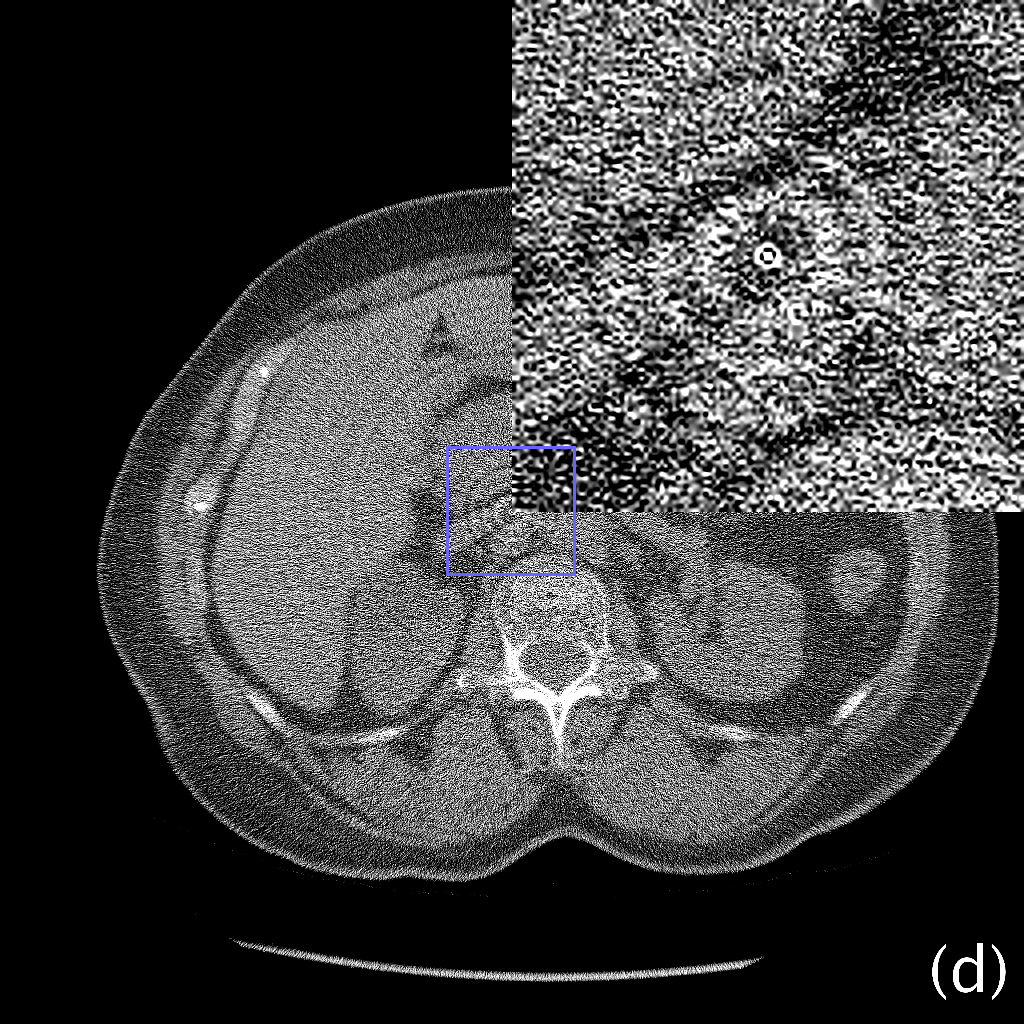}
         %\caption{}
     \end{subfigure}
    \vfill
 \begin{subfigure}[b]{0.24\columnwidth}
         \centering
         \includegraphics[width=\columnwidth]{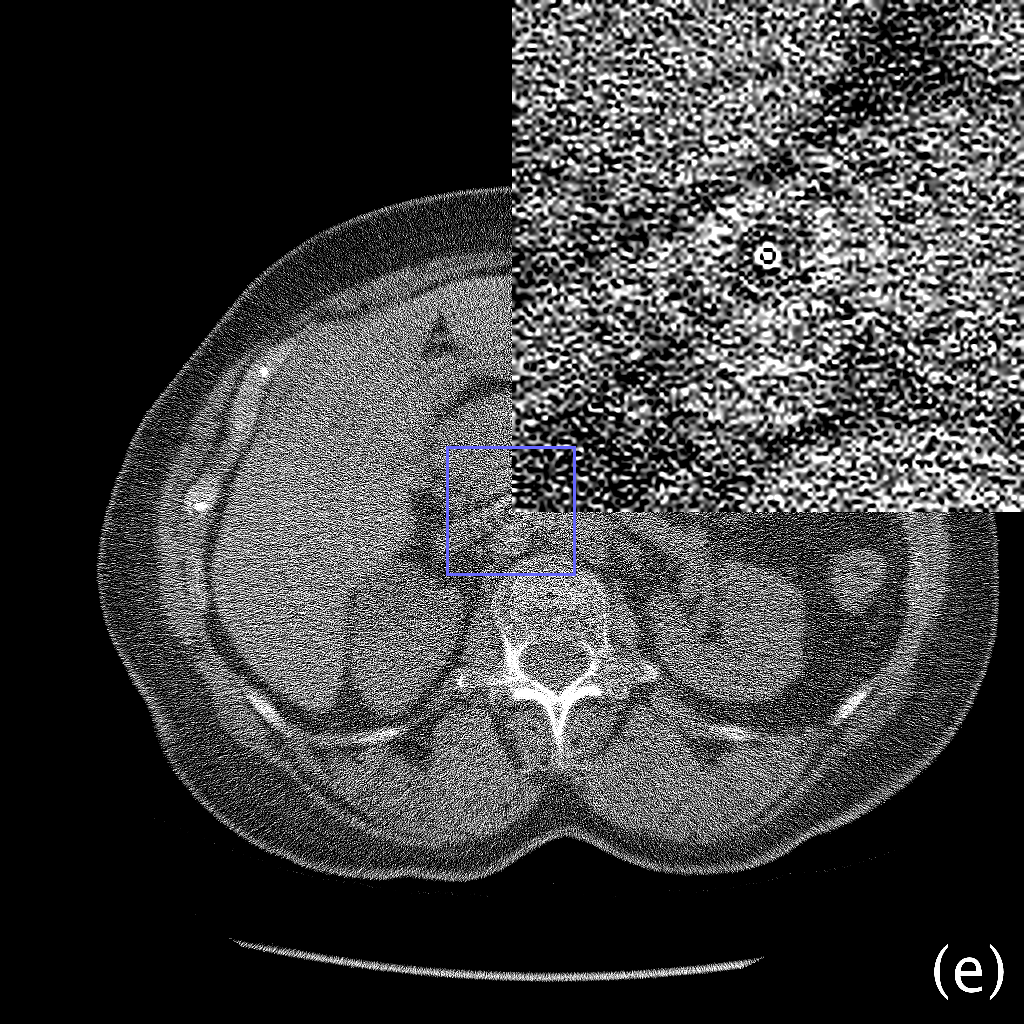}
         %\caption{}
     \end{subfigure}
     \hspace{-0.7em}
 \begin{subfigure}[b]{0.24\columnwidth}
         \centering
         \includegraphics[width=\columnwidth]{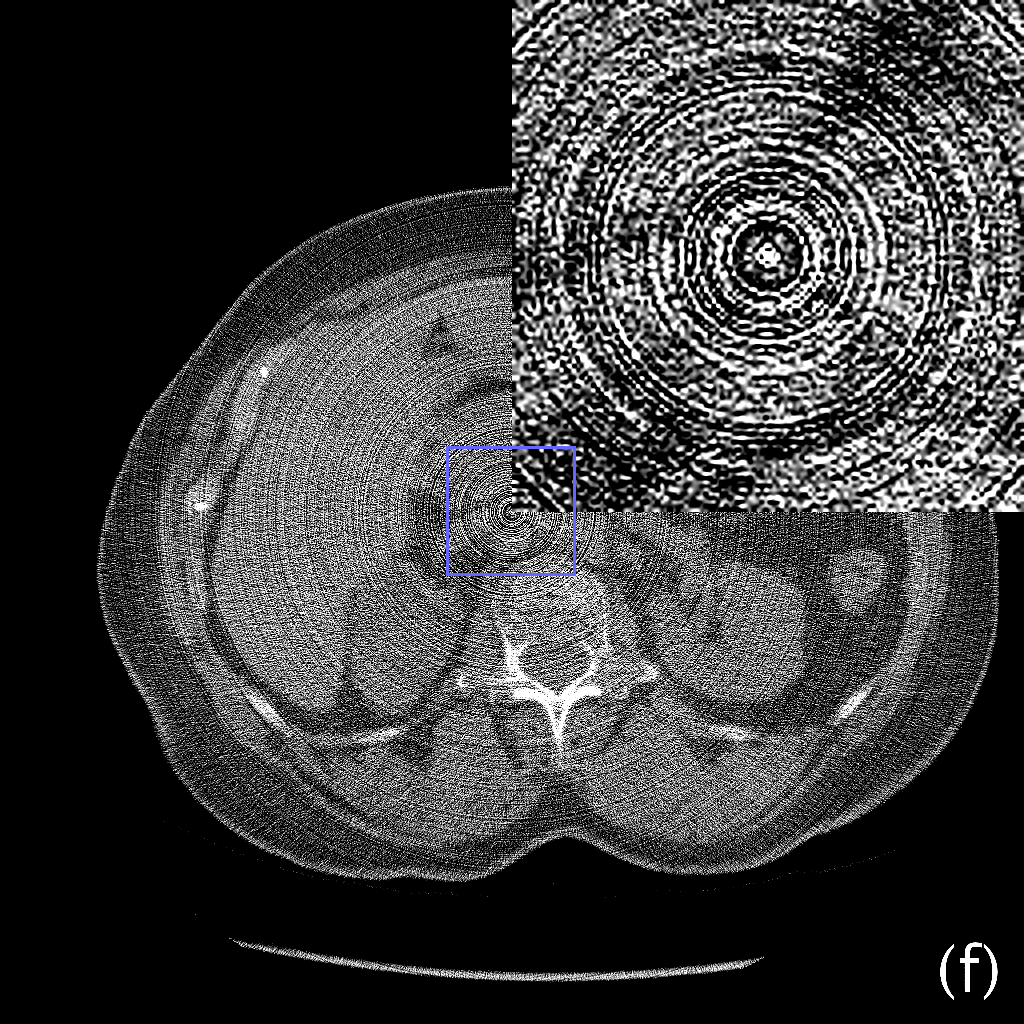}
         %\caption{}
     \end{subfigure}
     \begin{subfigure}[b]{0.24\columnwidth}
         \centering
         \includegraphics[width=\columnwidth]{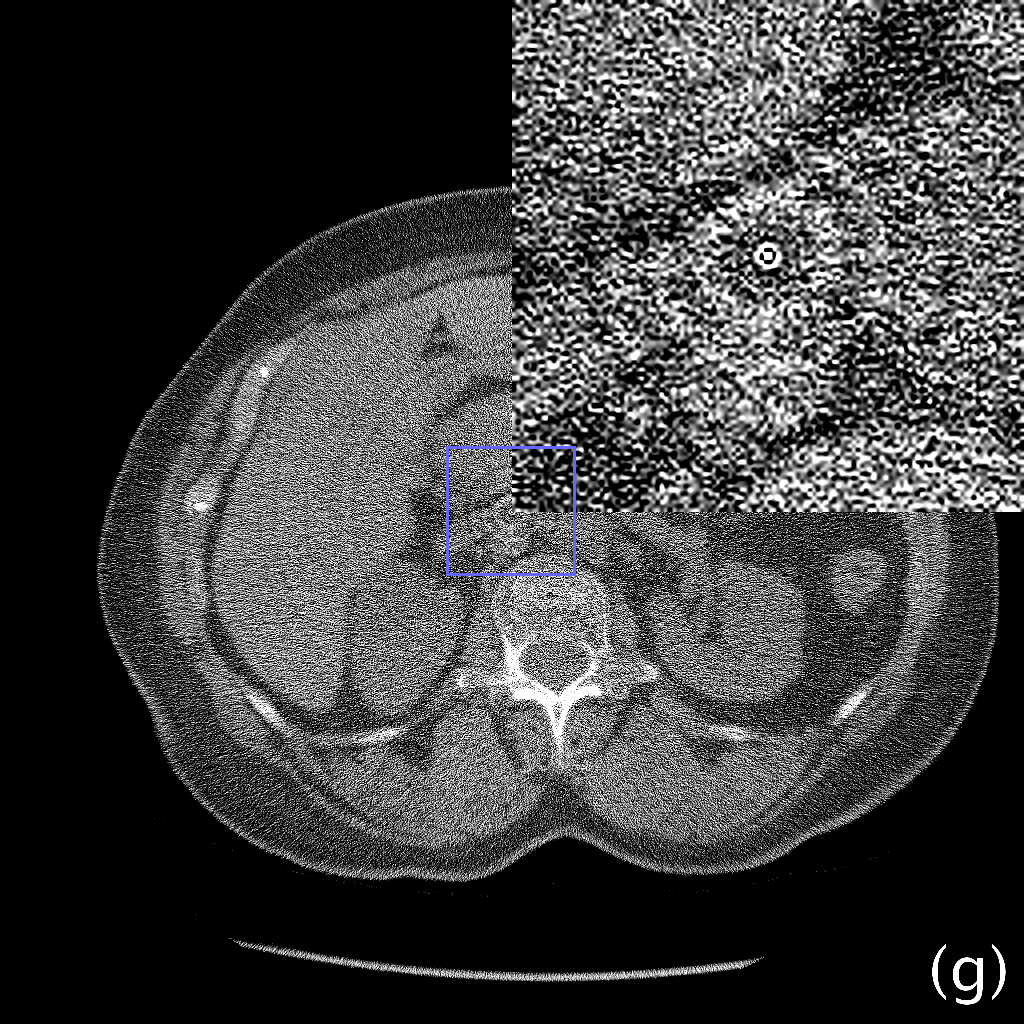}
         %\caption{}
     \end{subfigure}
     \hspace{-0.7em}
     \begin{subfigure}[b]{0.24\columnwidth}
         \centering
         \includegraphics[width=\columnwidth]{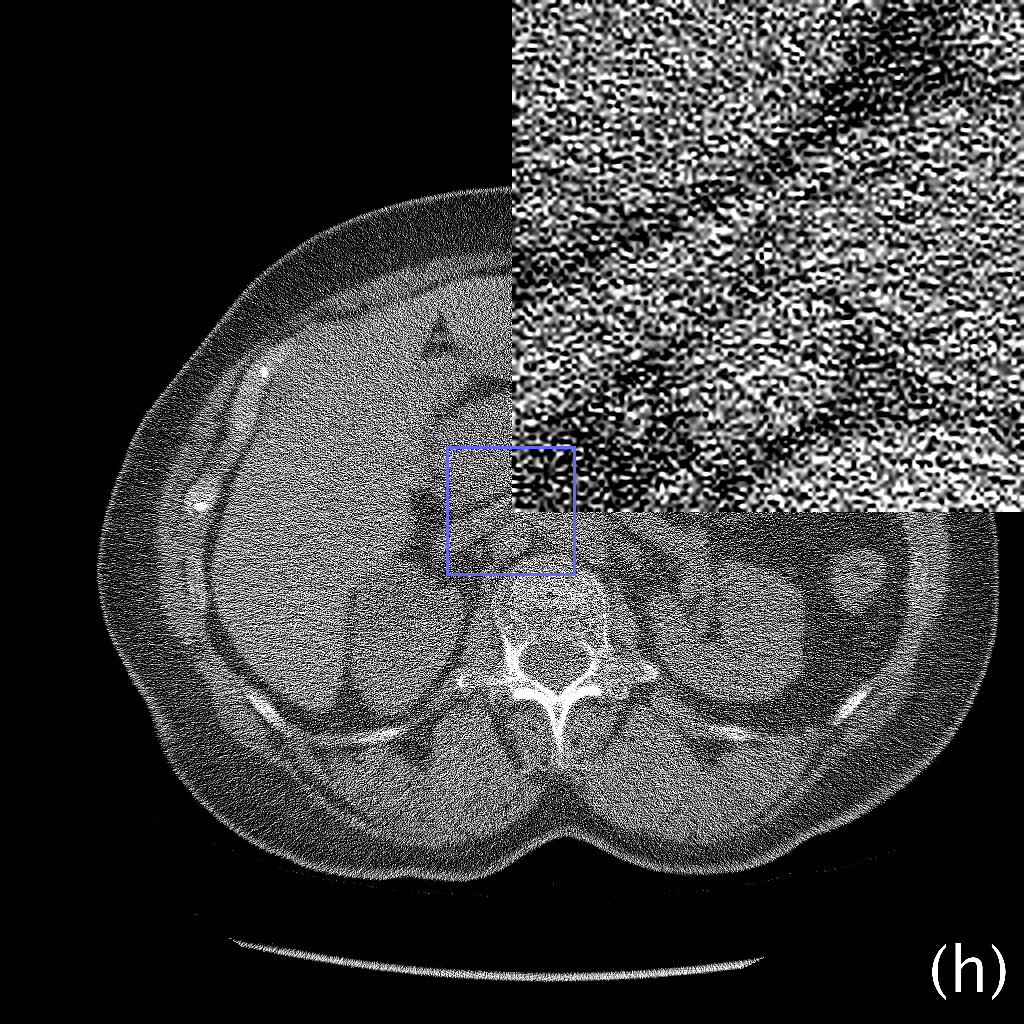}
         %\caption{}
     \end{subfigure}
        \caption{Example slice from the KiTS19 test set. 100 $\mathrm{keV}$ virtual monoenergetic images. Display window [-160,240] HU. (a) ring artifact free (truth), (b) ring corrupted (observed), (c) L2, (d)  L1, (e) VGG, (f) $\mathrm{VGG}_{70}$, (g) VGG-L1, (h) $\mathrm{VGG}_{70}$-L1.}
        \label{qual_100_kits}
\end{figure}

\begin{figure}
     \centering
     \begin{subfigure}[b]{0.24\columnwidth}
         \centering
         \includegraphics[width=\columnwidth]{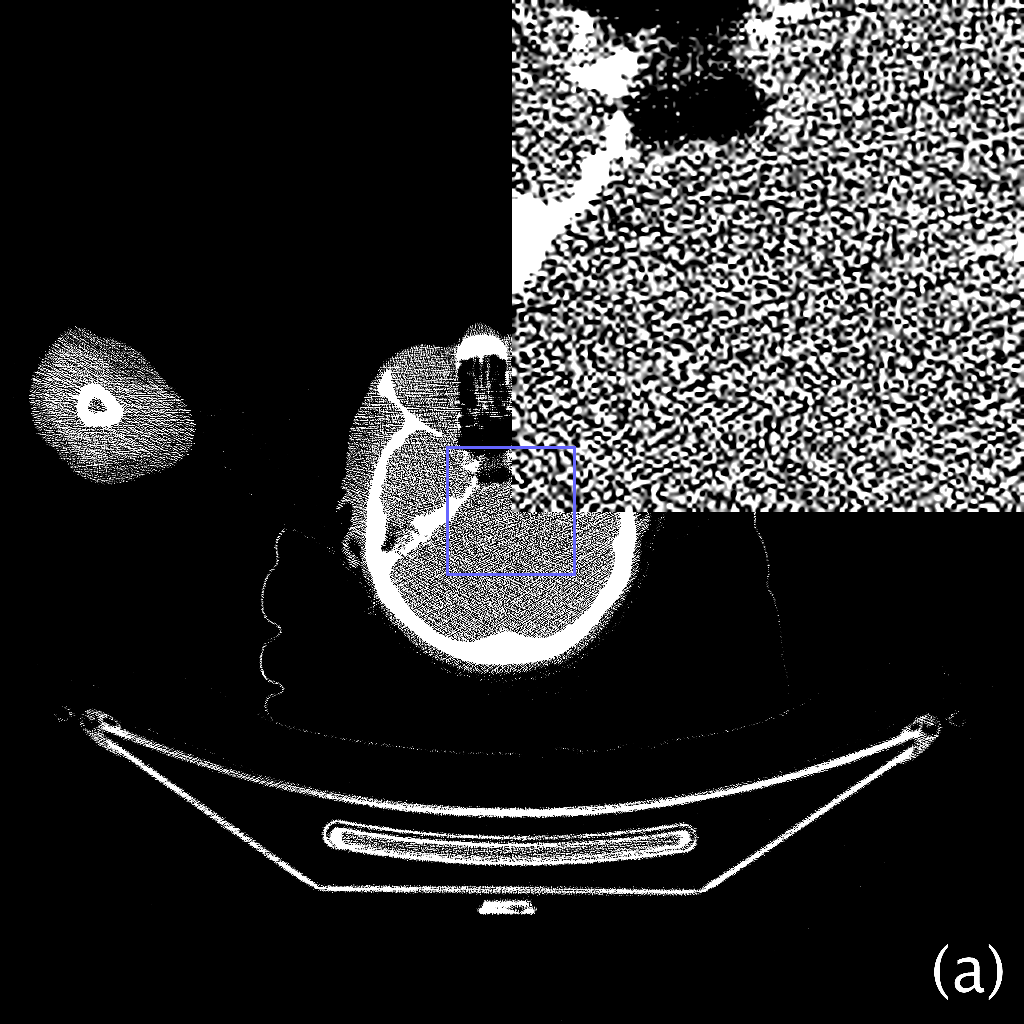}
         %\caption{}
     \end{subfigure}
     \hspace{-0.7em}
     \begin{subfigure}[b]{0.24\columnwidth}
         \centering
         \includegraphics[width=\columnwidth]{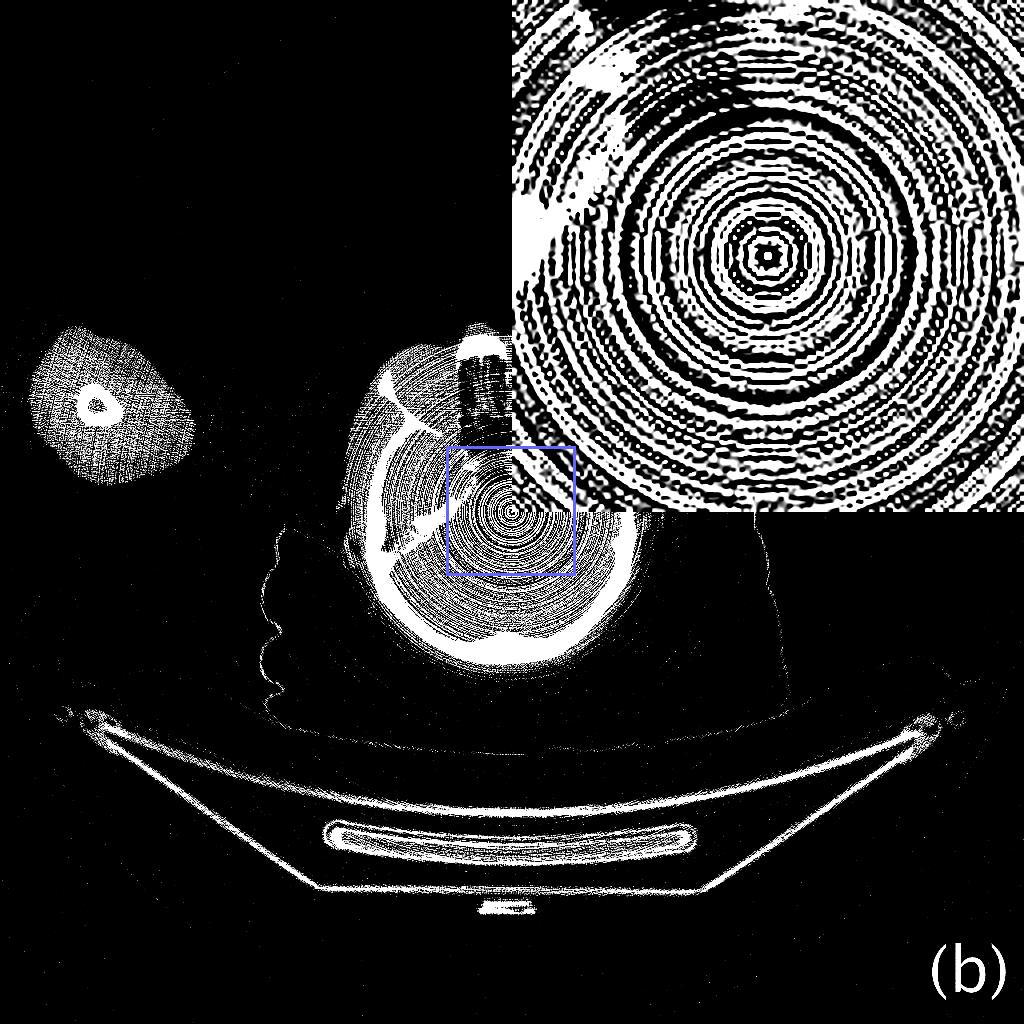}
         %\caption{}
     \end{subfigure}
     \begin{subfigure}[b]{0.24\columnwidth}
         \centering
         \includegraphics[width=\columnwidth]{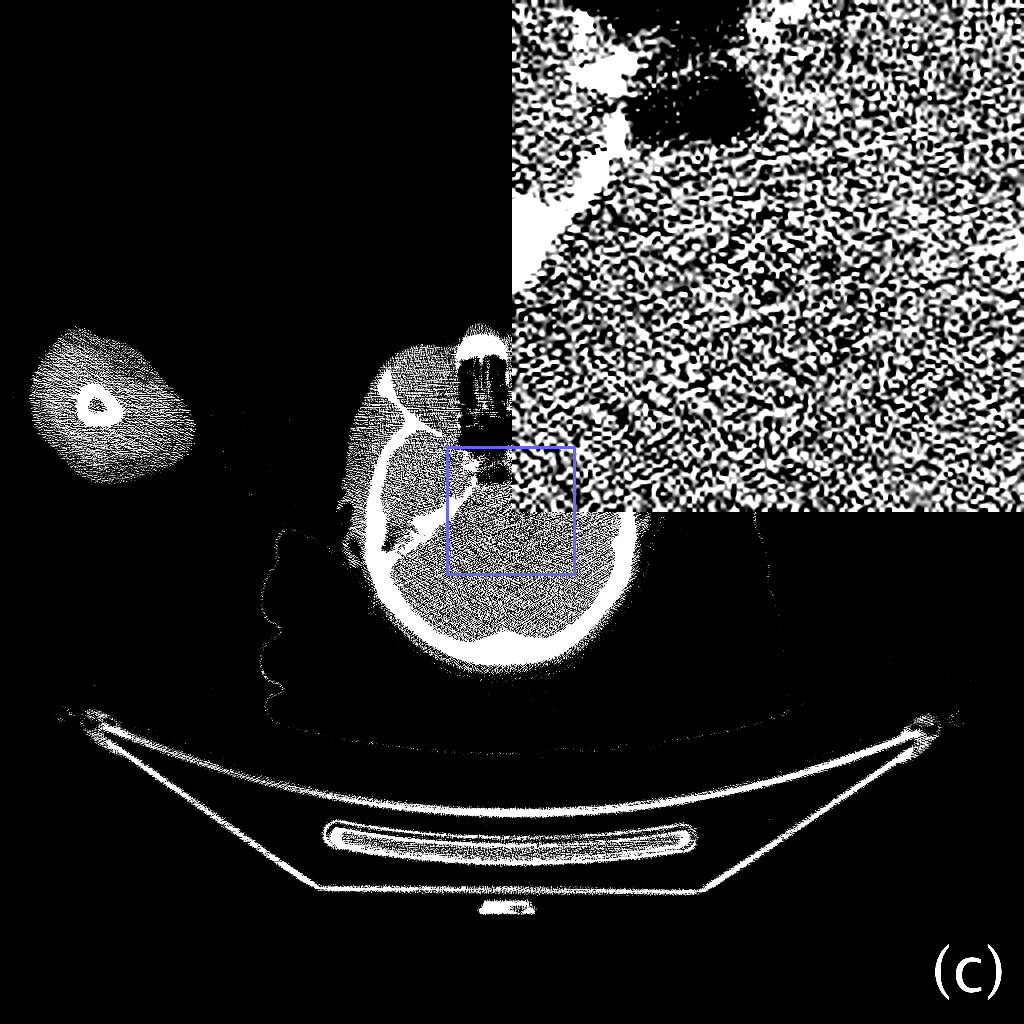}
         %\caption{}
     \end{subfigure}
     \hspace{-0.7em}
       \begin{subfigure}[b]{0.24\columnwidth}
         \centering
         \includegraphics[width=\columnwidth]{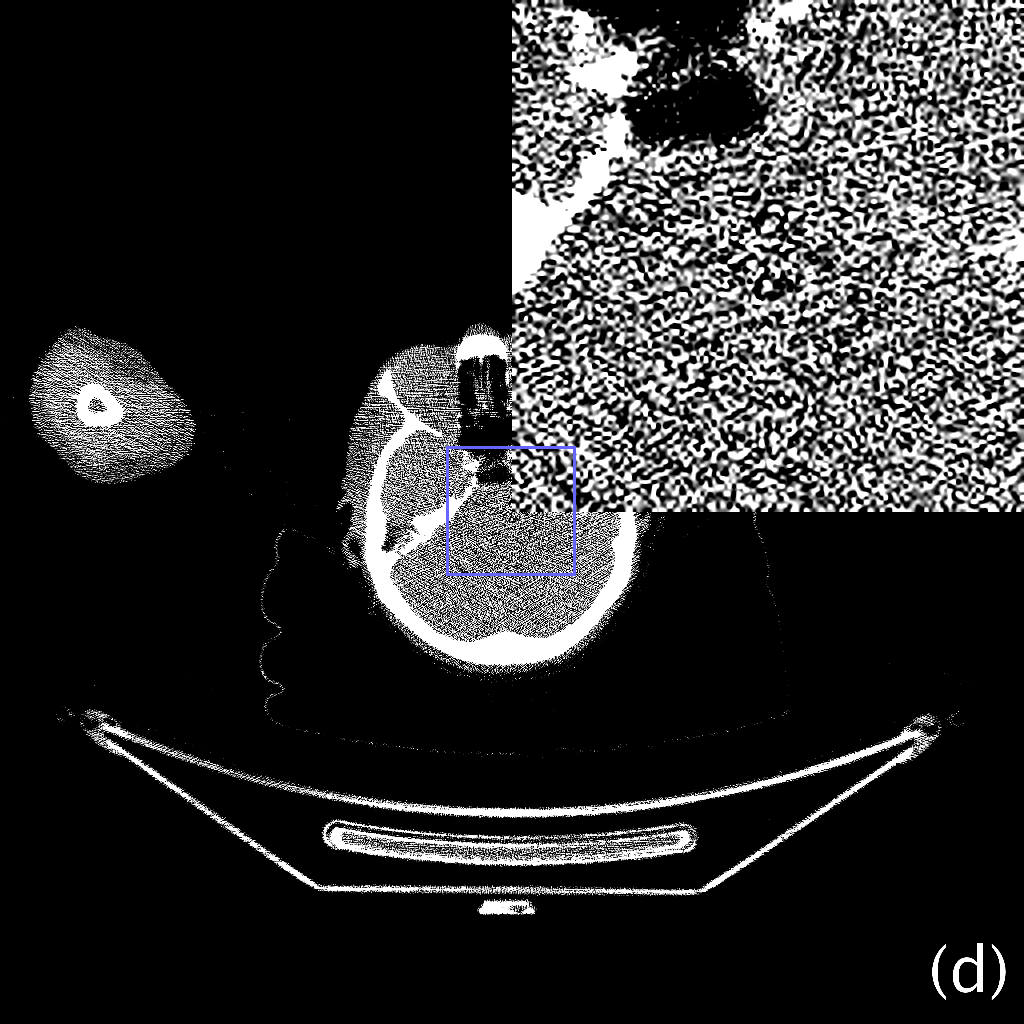}
         %\caption{}
     \end{subfigure}
      \vfill
 \begin{subfigure}[b]{0.24\columnwidth}
         \centering
         \includegraphics[width=\columnwidth]{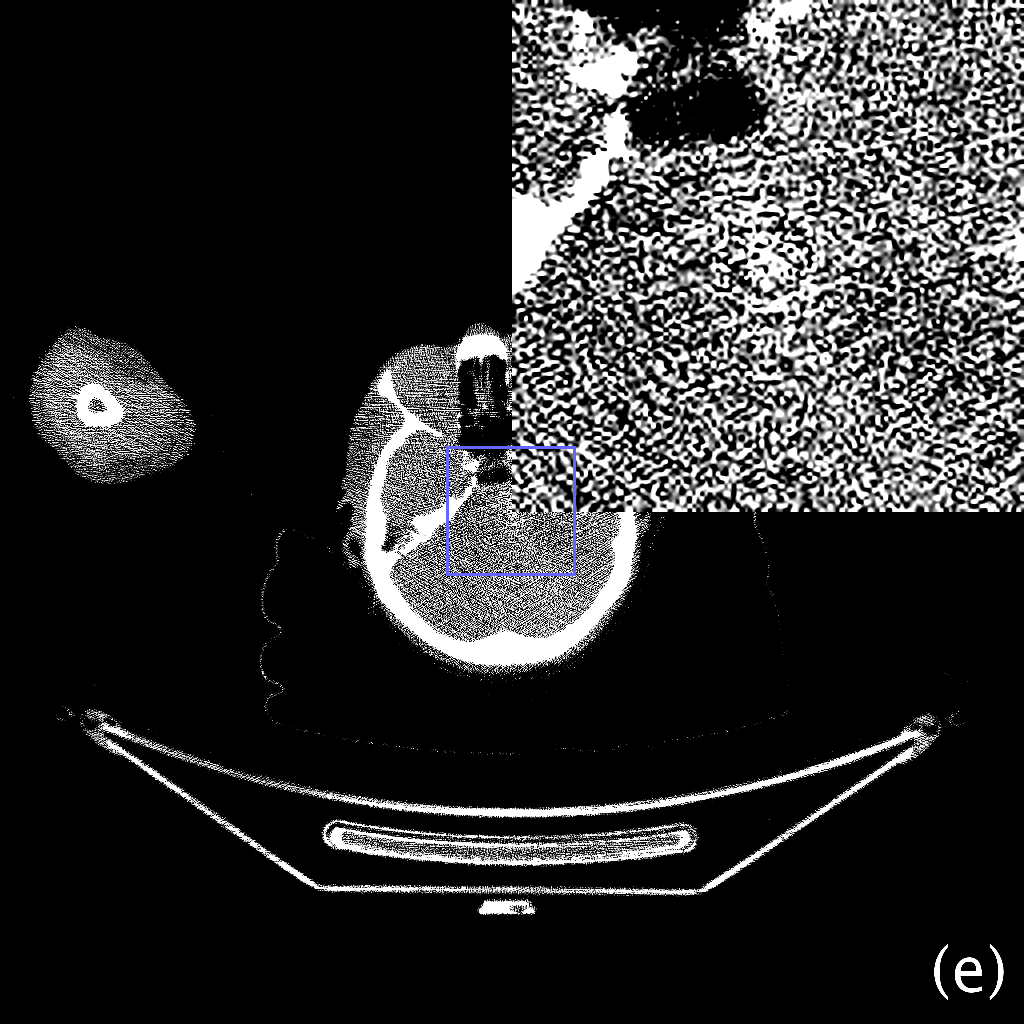}
         %\caption{}
     \end{subfigure}
     \hspace{-0.7em}
 \begin{subfigure}[b]{0.24\columnwidth}
         \centering
         \includegraphics[width=\columnwidth]{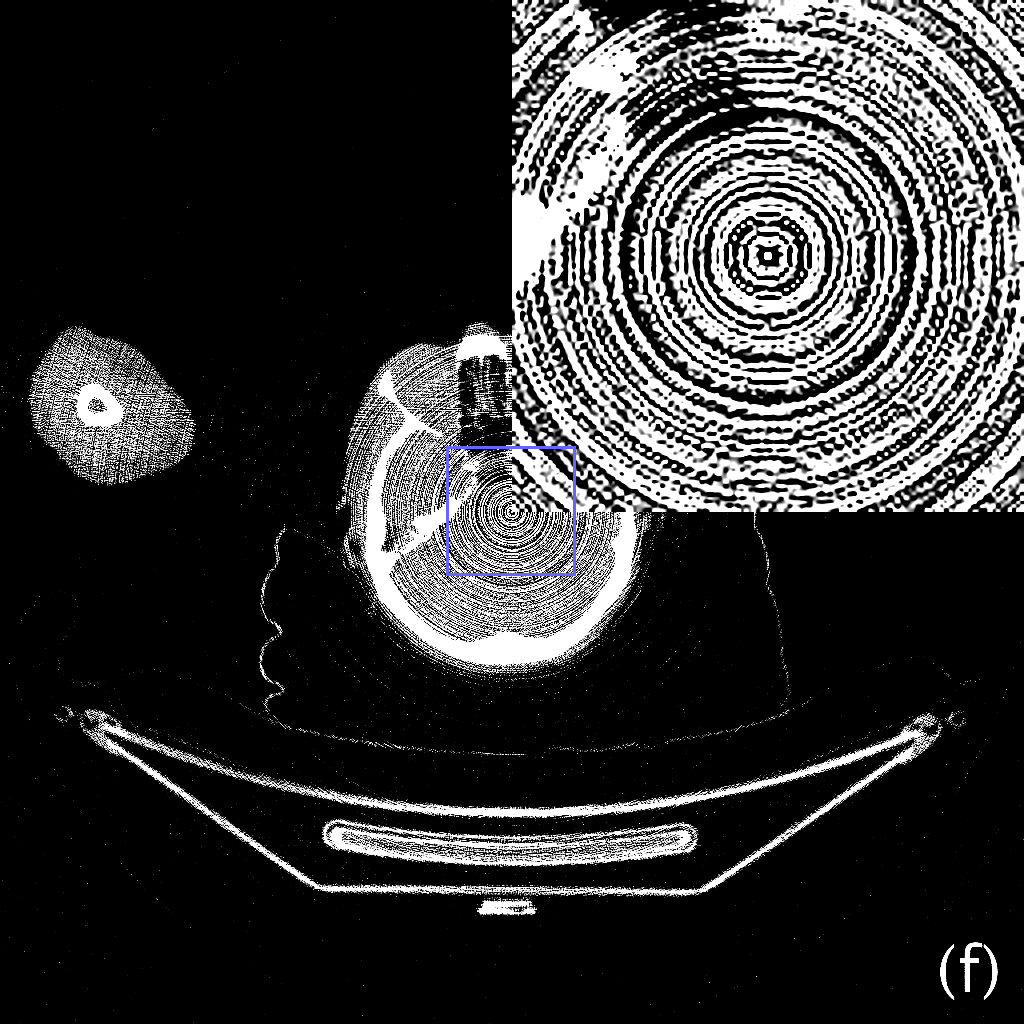}
         %\caption{}
     \end{subfigure}
     \begin{subfigure}[b]{0.24\columnwidth}
         \centering
         \includegraphics[width=\columnwidth]{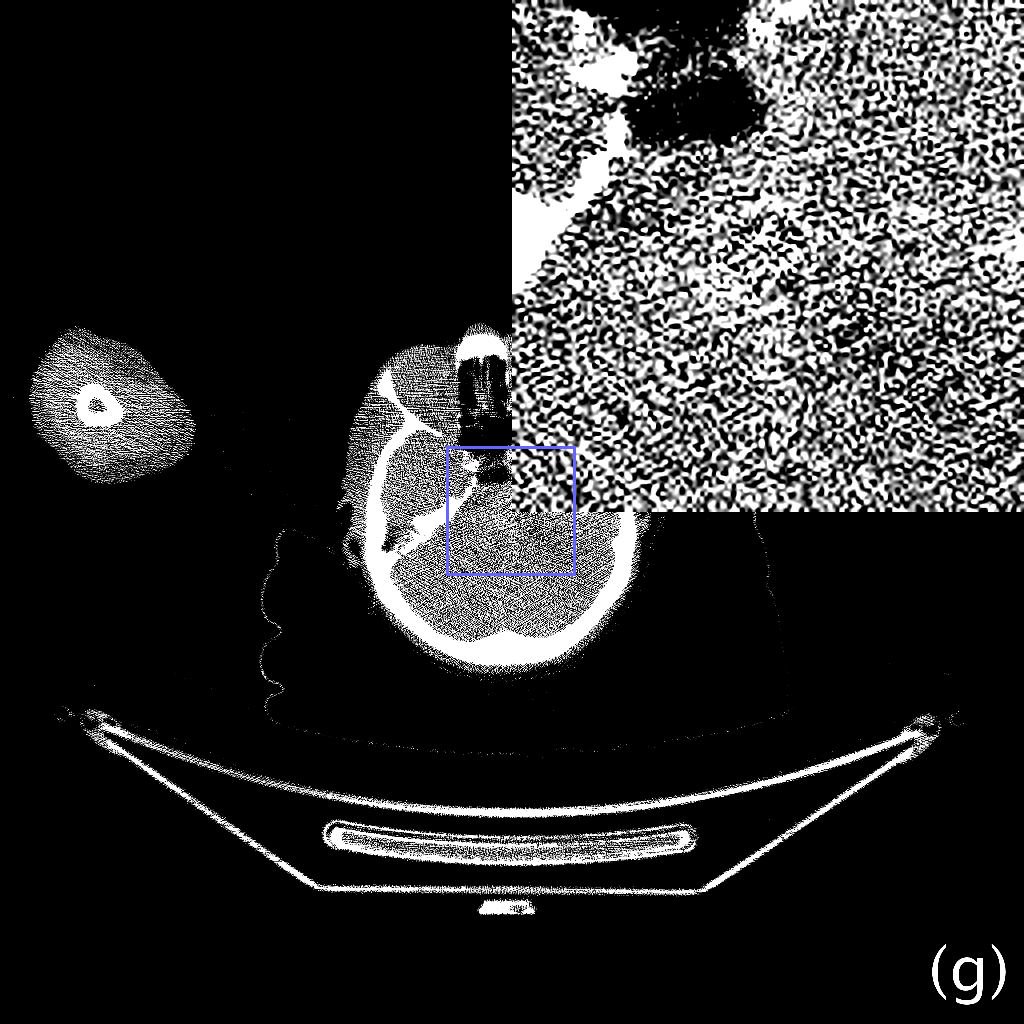}
         %\caption{}
     \end{subfigure}
     \hspace{-0.7em}
     \begin{subfigure}[b]{0.24\columnwidth}
         \centering
         \includegraphics[width=\columnwidth]{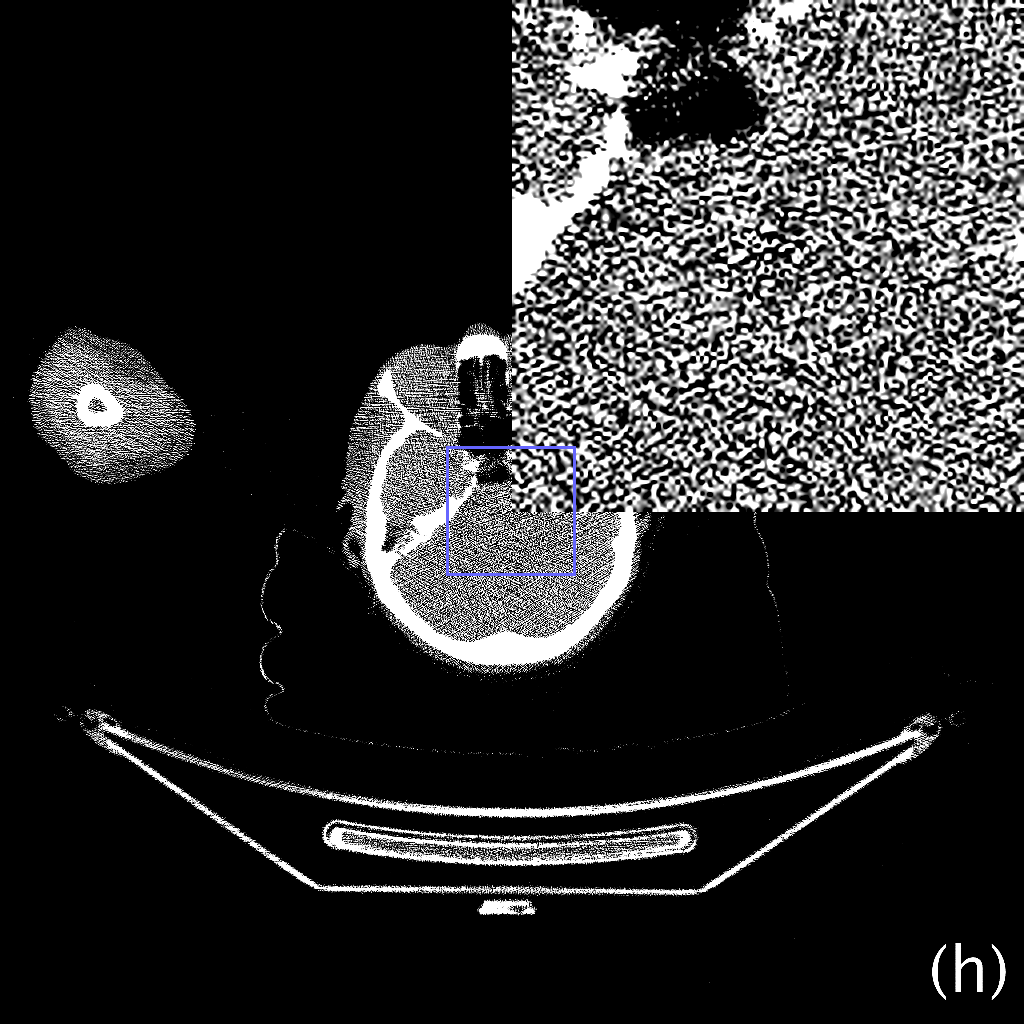}
         %\caption{}
     \end{subfigure}
        \caption{Example slice from the NSCLC test set. 40 $\mathrm{keV}$ virtual monoenergetic images. Display window [-160,240] HU. (a) ring artifact free (truth), (b) ring corrupted (observed), (c) L2, (d)  L1, (e) VGG, (f) $\mathrm{VGG}_{70}$, (g) VGG-L1, (h) $\mathrm{VGG}_{70}$-L1.}
        \label{qual_40_nsclc}
\end{figure}

\begin{figure}
     \centering
     \begin{subfigure}[b]{0.24\columnwidth}
         \centering
         \includegraphics[width=\columnwidth]{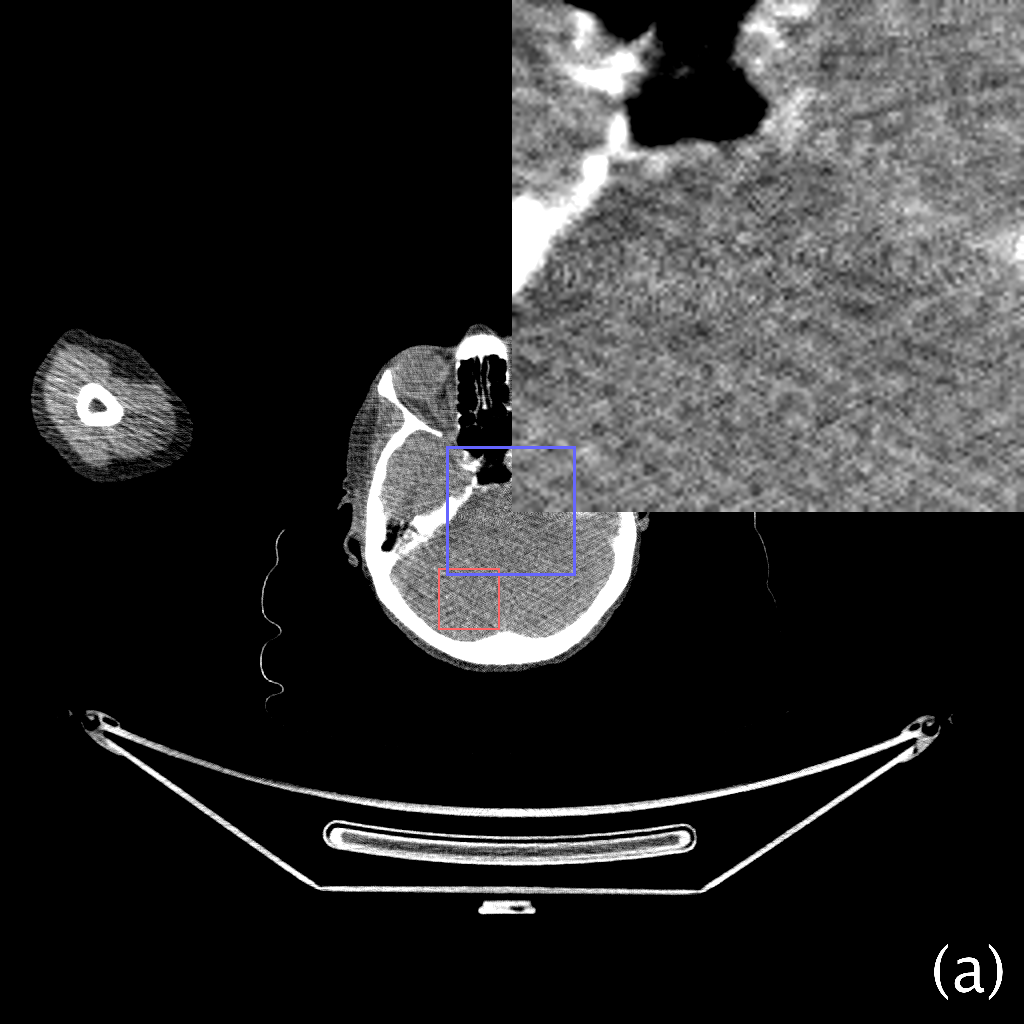}
         %\caption{}
     \end{subfigure}
     \hspace{-0.7em}
     \begin{subfigure}[b]{0.24\columnwidth}
         \centering
         \includegraphics[width=\columnwidth]{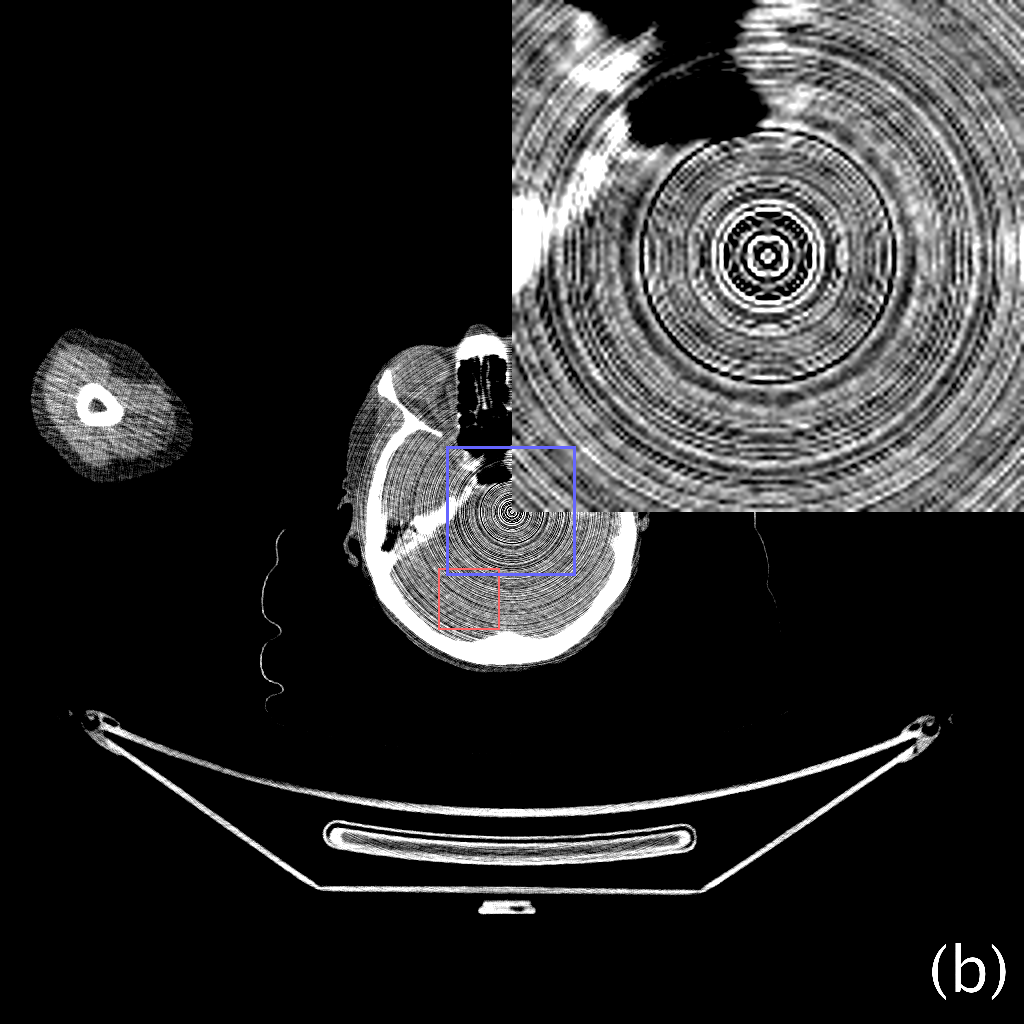}
         %\caption{}
     \end{subfigure}
     \begin{subfigure}[b]{0.24\columnwidth}
         \centering
         \includegraphics[width=\columnwidth]{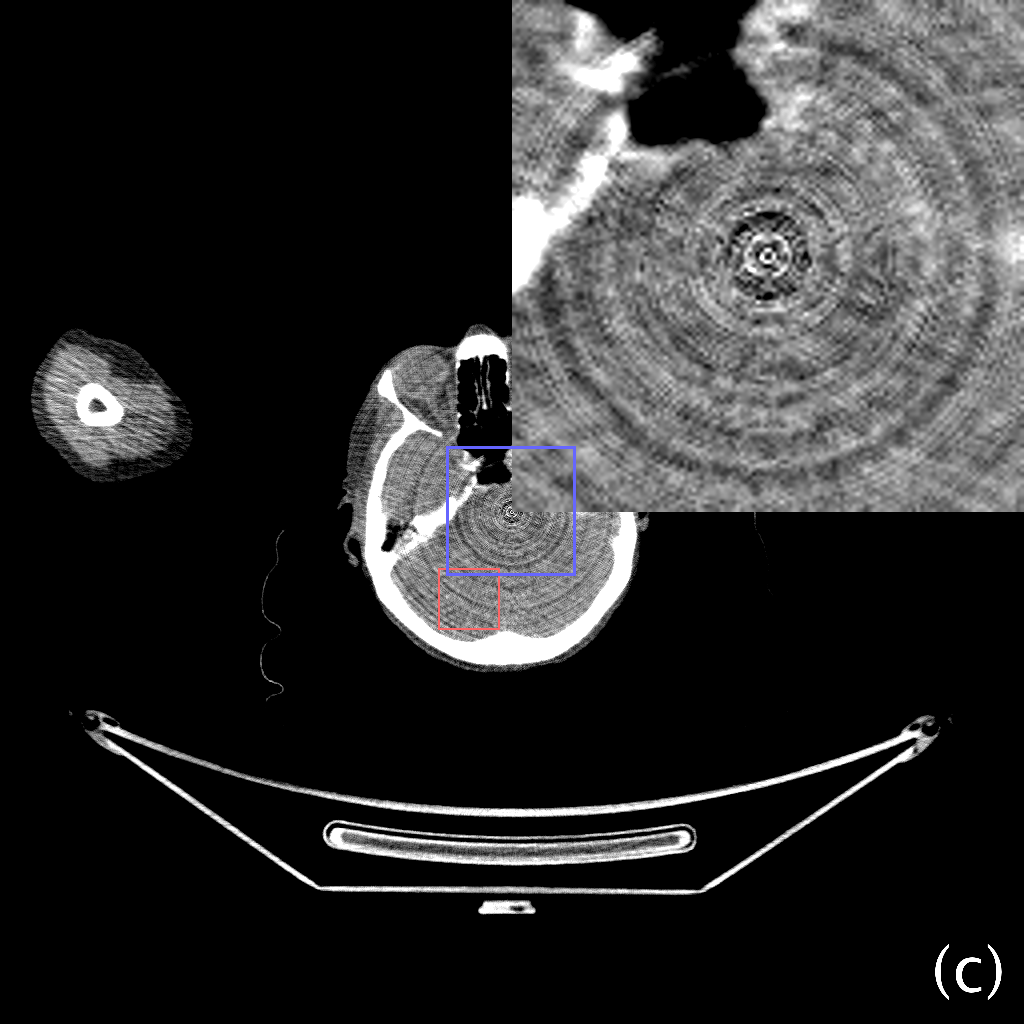}
         %\caption{}
     \end{subfigure}
     \hspace{-0.7em}
       \begin{subfigure}[b]{0.24\columnwidth}
         \centering
         \includegraphics[width=\columnwidth]{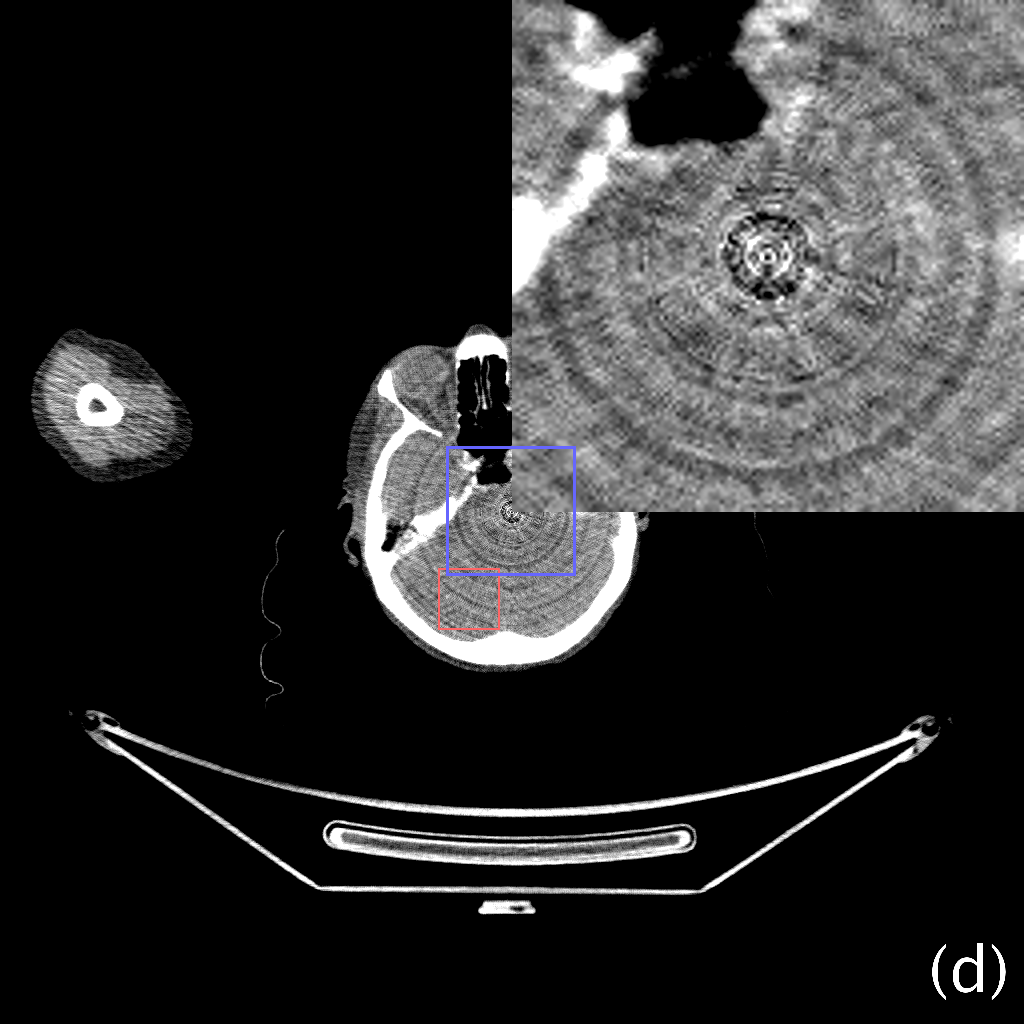}
         %\caption{}
     \end{subfigure}
      \vfill
 \begin{subfigure}[b]{0.24\columnwidth}
         \centering
         \includegraphics[width=\columnwidth]{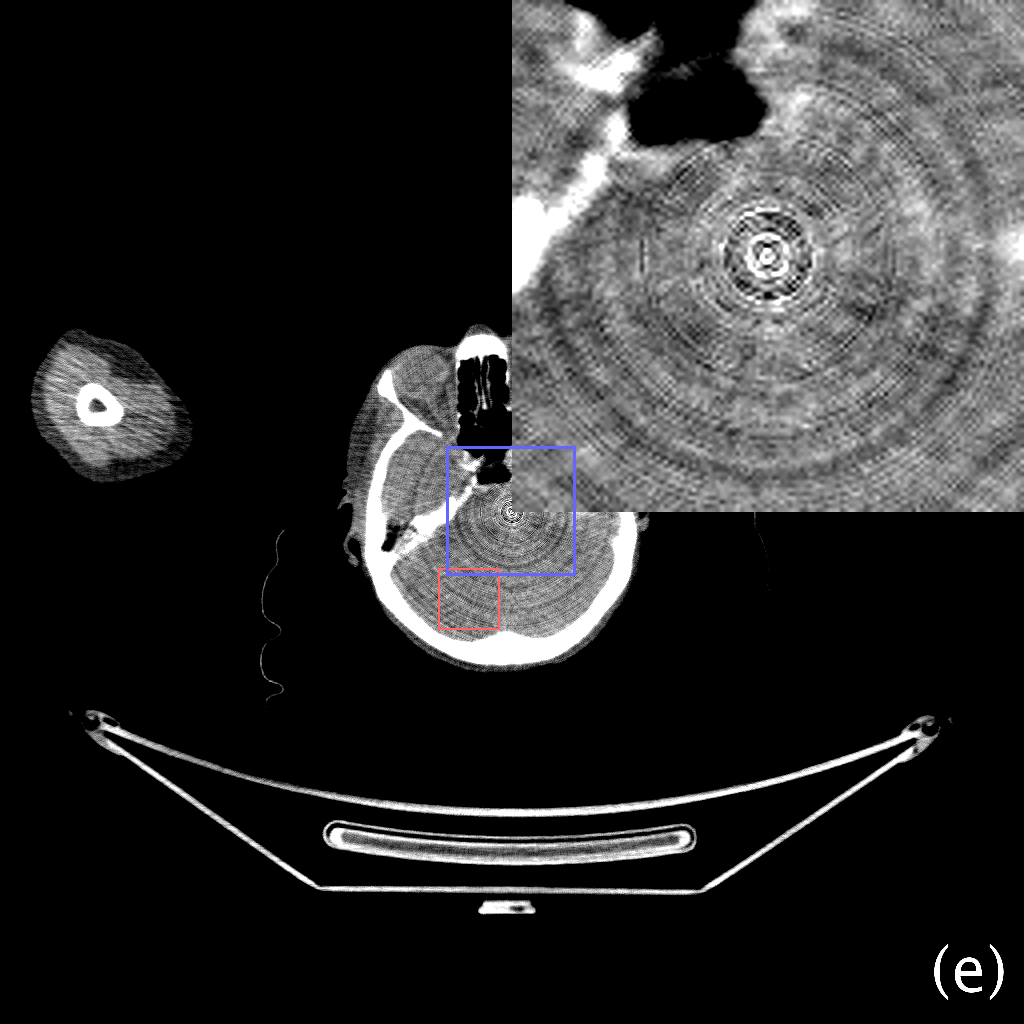}
         %\caption{}
     \end{subfigure}
     \hspace{-0.7em}
 \begin{subfigure}[b]{0.24\columnwidth}
         \centering
         \includegraphics[width=\columnwidth]{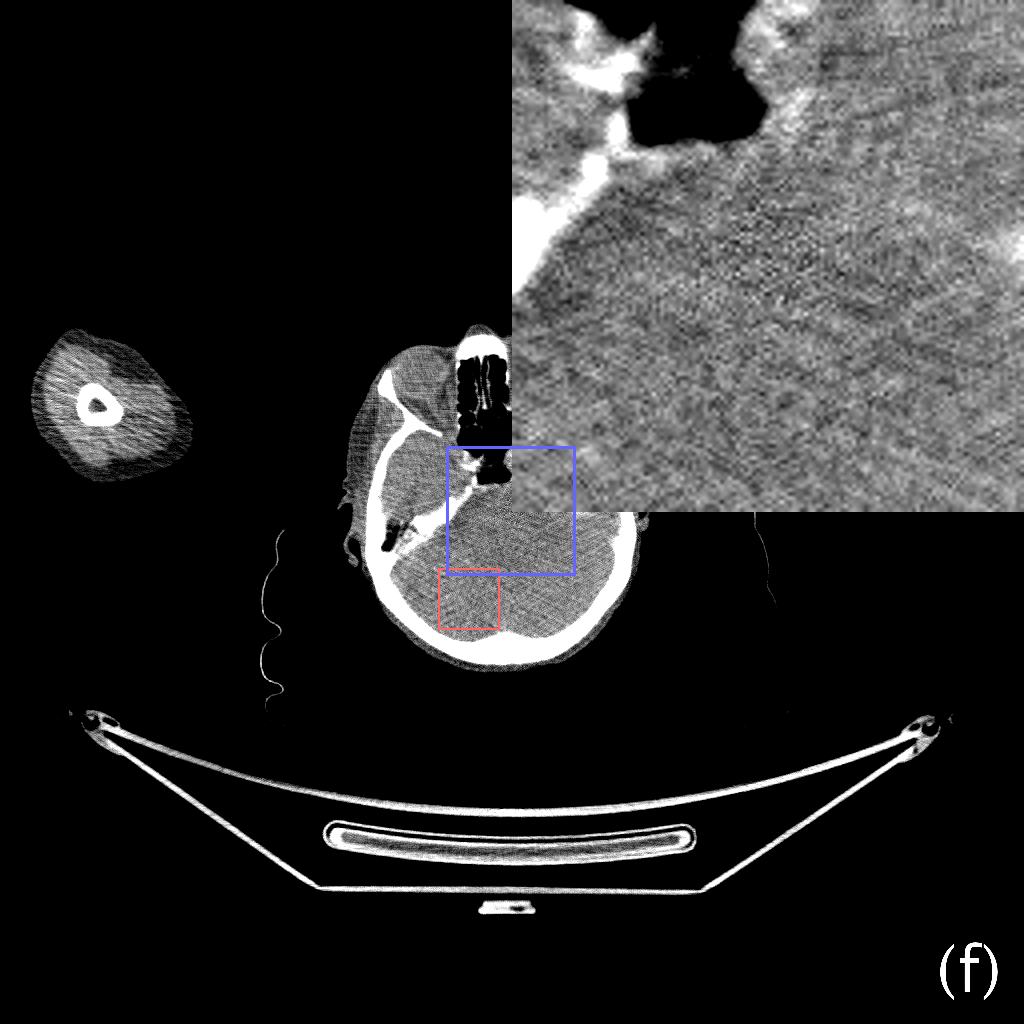}
         %\caption{}
     \end{subfigure}
     \begin{subfigure}[b]{0.24\columnwidth}
         \centering
         \includegraphics[width=\columnwidth]{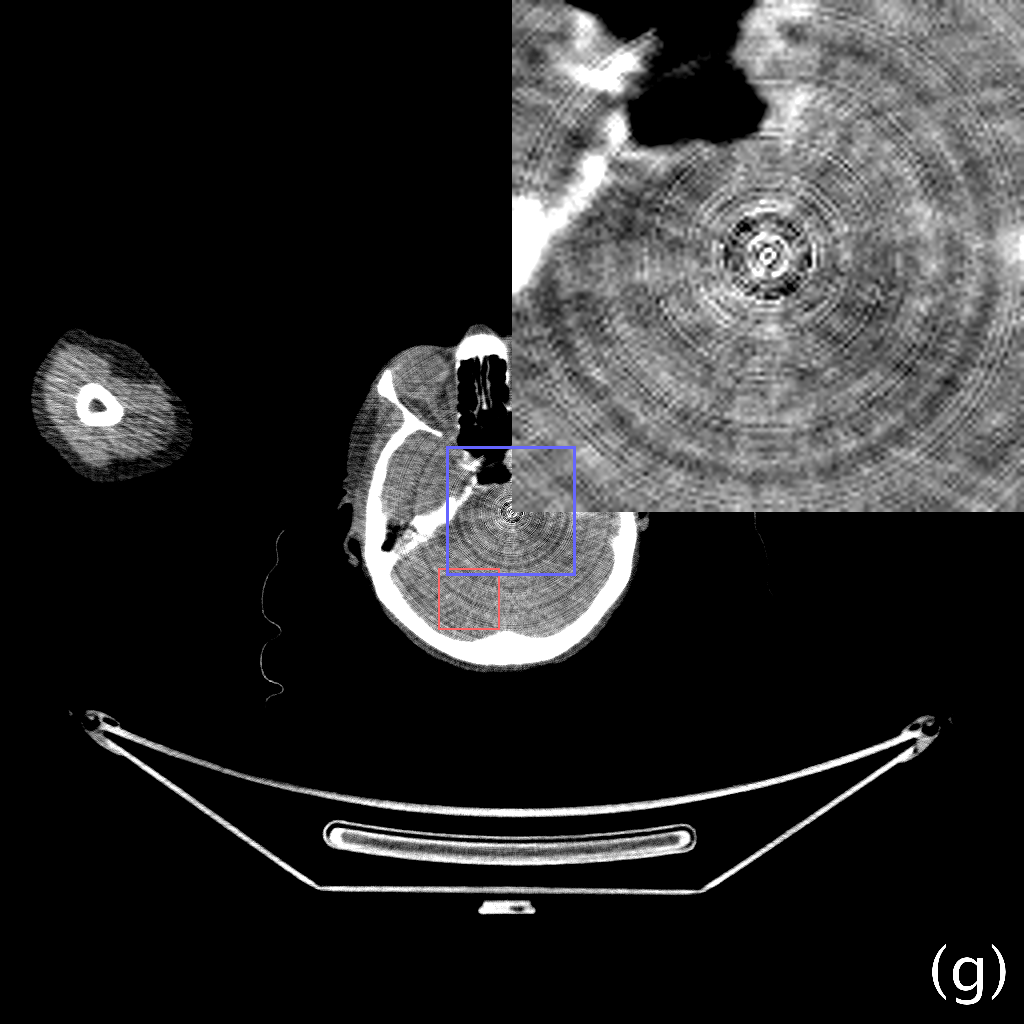}
         %\caption{}
     \end{subfigure}
     \hspace{-0.7em}
     \begin{subfigure}[b]{0.24\columnwidth}
         \centering
         \includegraphics[width=\columnwidth]{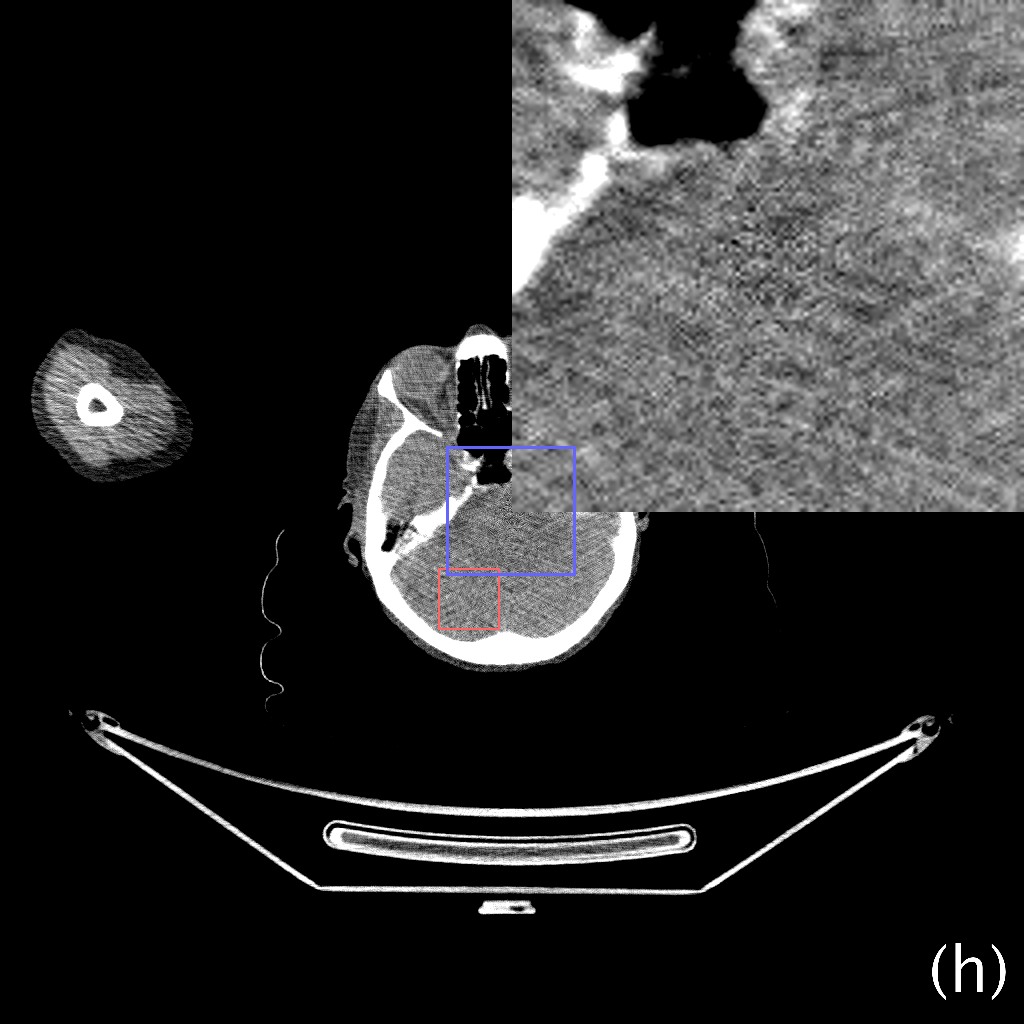}
         %\caption{}
     \end{subfigure}
        \caption{Example slice from the NSCLC test set. 70 $\mathrm{keV}$ virutal monoenergetic images. Display window [-160,240] HU. (a) ring artifact free (truth), (b) ring corrupted (observed), (c) L2, (d)  L1, (e) VGG, (f) $\mathrm{VGG}_{70}$, (g) VGG-L1, (h) $\mathrm{VGG}_{70}$-L1.}
        \label{qual_70_nsclc}
\end{figure}

\begin{figure}
     \centering
     \begin{subfigure}[b]{0.24\columnwidth}
         \centering
         \includegraphics[width=\columnwidth]{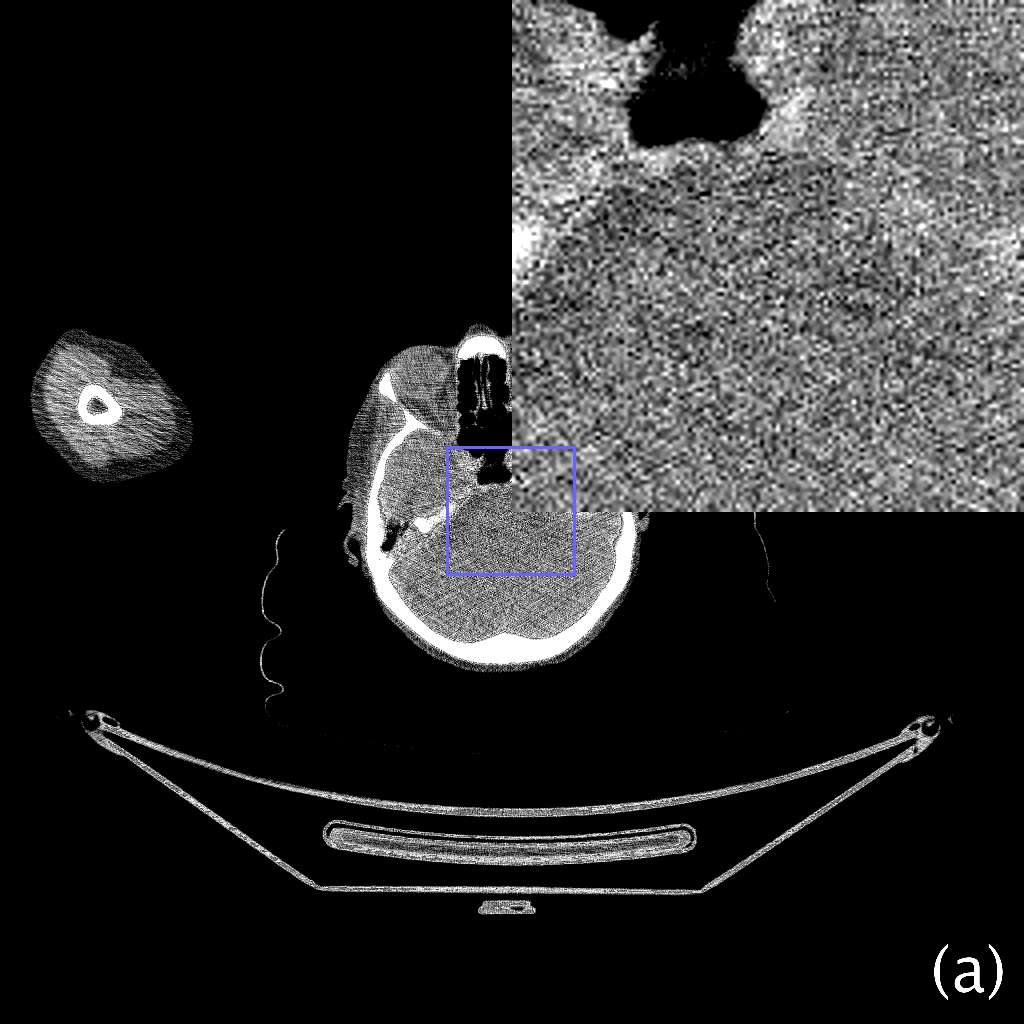}
         %\caption{}
     \end{subfigure}
     \hspace{-0.7em}
     \begin{subfigure}[b]{0.24\columnwidth}
         \centering
         \includegraphics[width=\columnwidth]{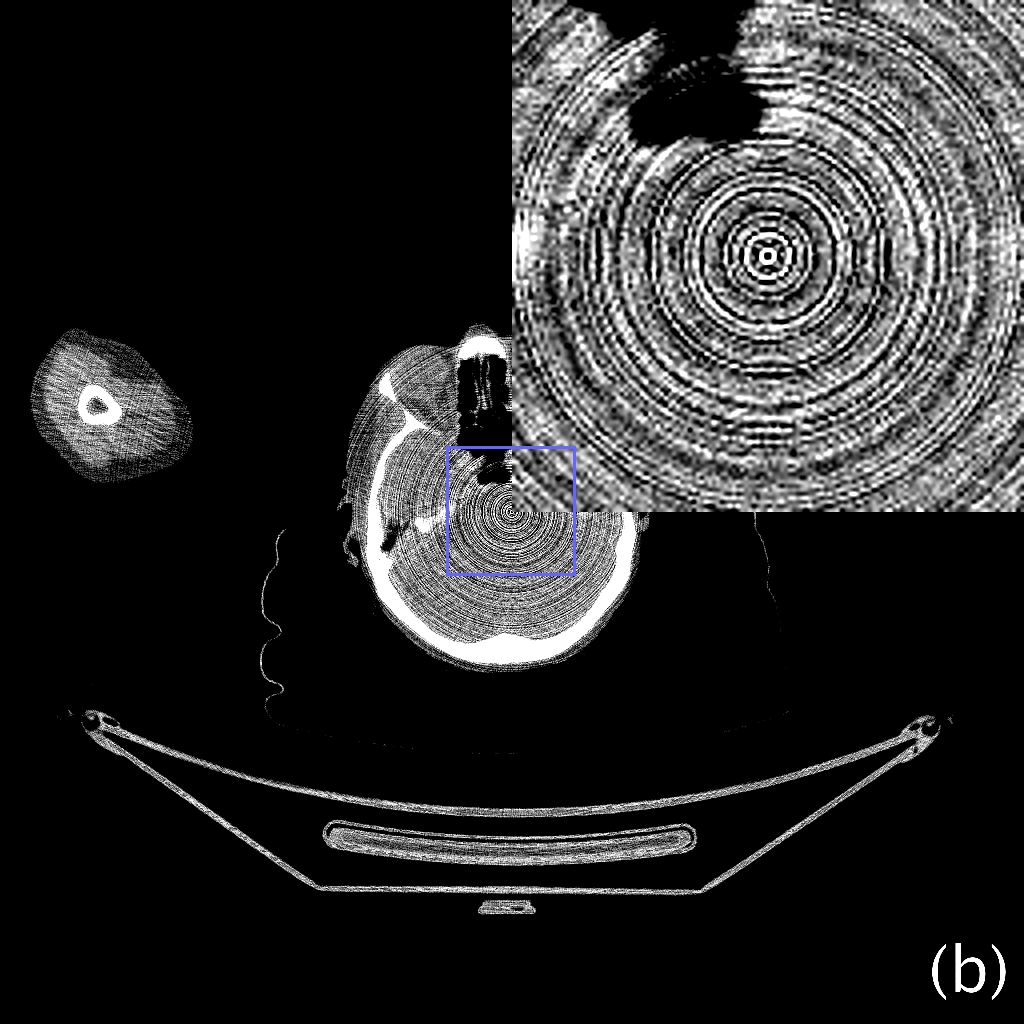}
         %\caption{}
     \end{subfigure}
     \begin{subfigure}[b]{0.24\columnwidth}
         \centering
         \includegraphics[width=\columnwidth]{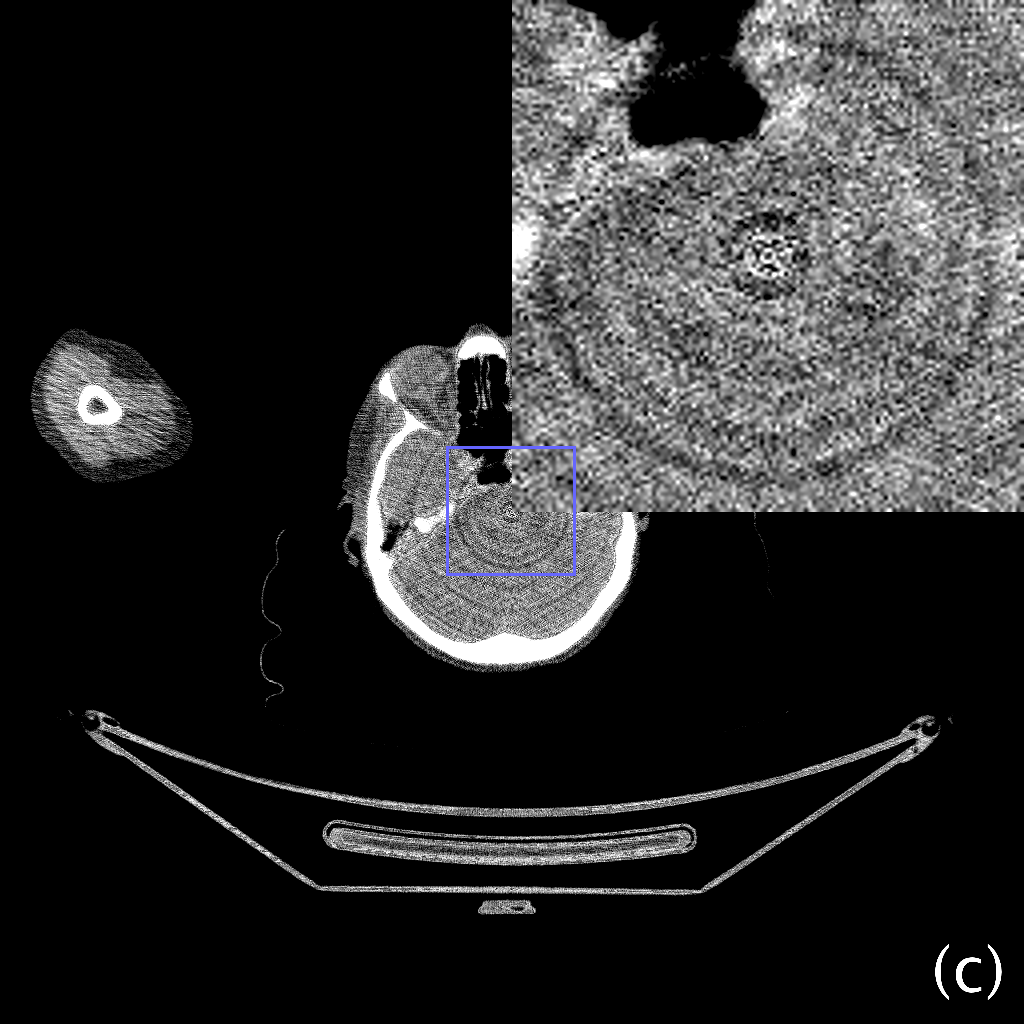}
         %\caption{}
     \end{subfigure}
     \hspace{-0.7em}
       \begin{subfigure}[b]{0.24\columnwidth}
         \centering
         \includegraphics[width=\columnwidth]{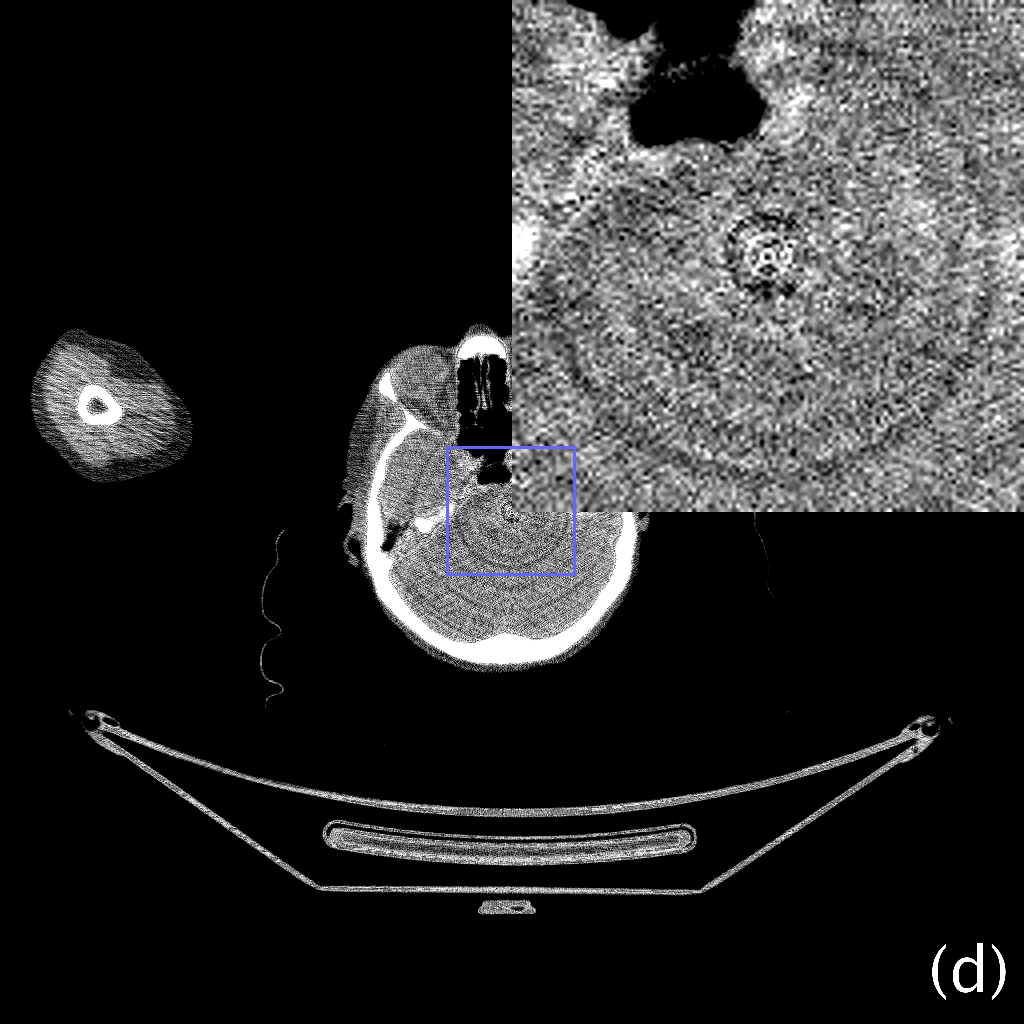}
         %\caption{}
     \end{subfigure}
      \vfill
 \begin{subfigure}[b]{0.24\columnwidth}
         \centering
         \includegraphics[width=\columnwidth]{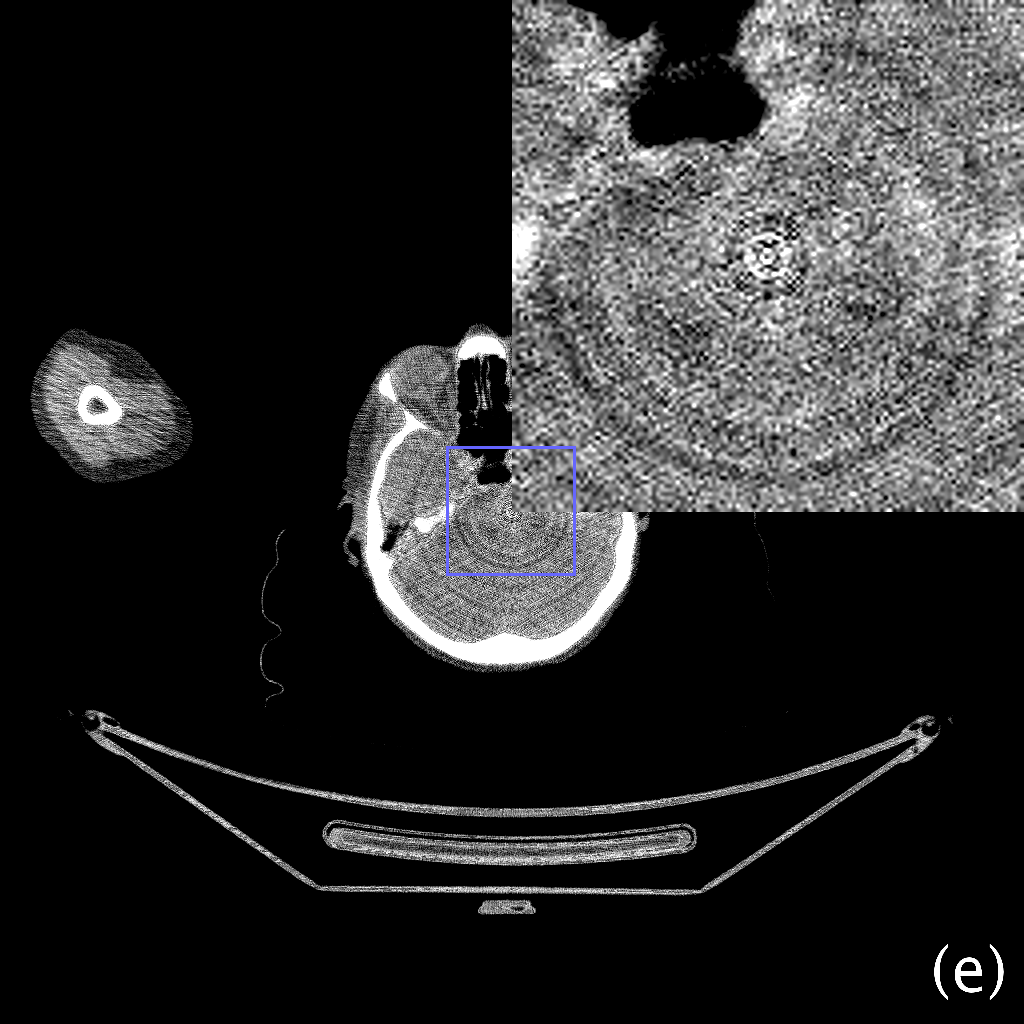}
         %\caption{}
     \end{subfigure}
     \hspace{-0.7em}
 \begin{subfigure}[b]{0.24\columnwidth}
         \centering
         \includegraphics[width=\columnwidth]{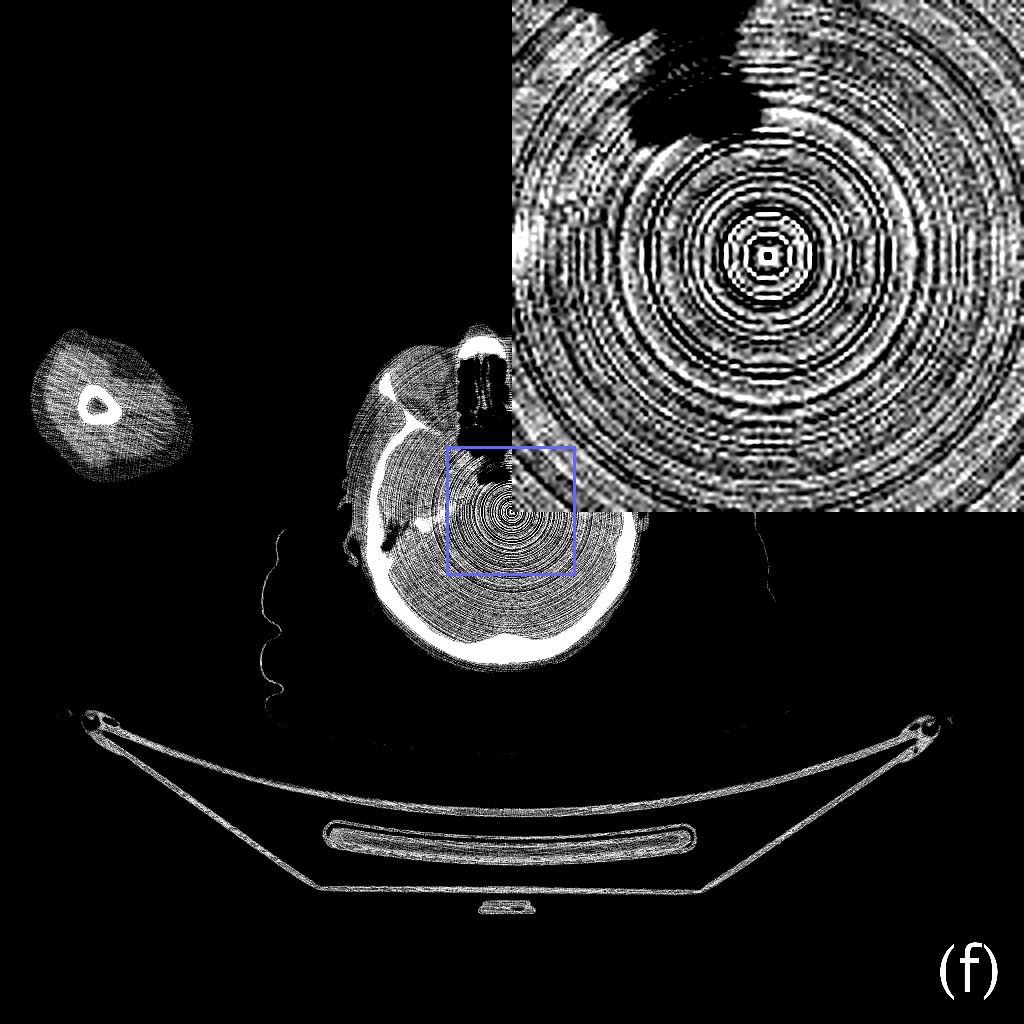}
         %\caption{}
     \end{subfigure}
     \begin{subfigure}[b]{0.24\columnwidth}
         \centering
         \includegraphics[width=\columnwidth]{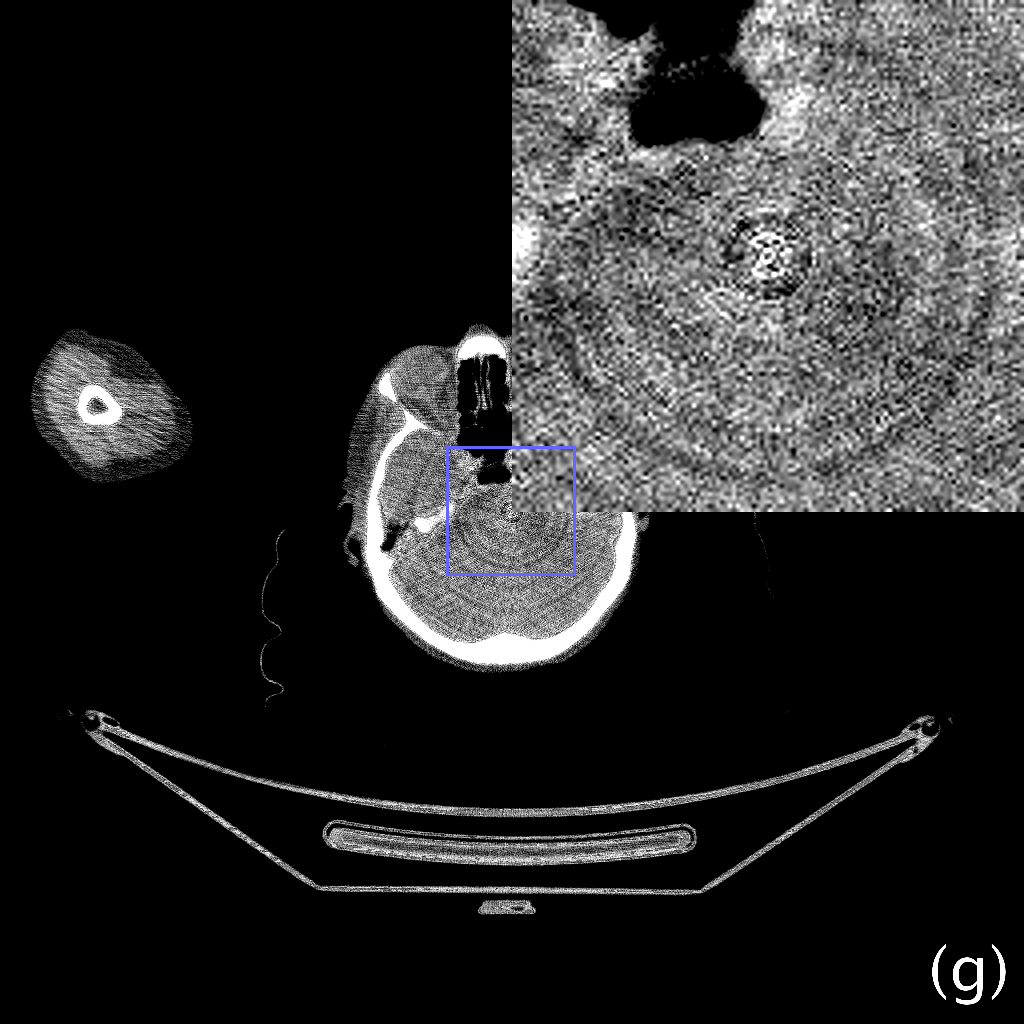}
         %\caption{}
     \end{subfigure}
     \hspace{-0.7em}
     \begin{subfigure}[b]{0.24\columnwidth}
         \centering
         \includegraphics[width=\columnwidth]{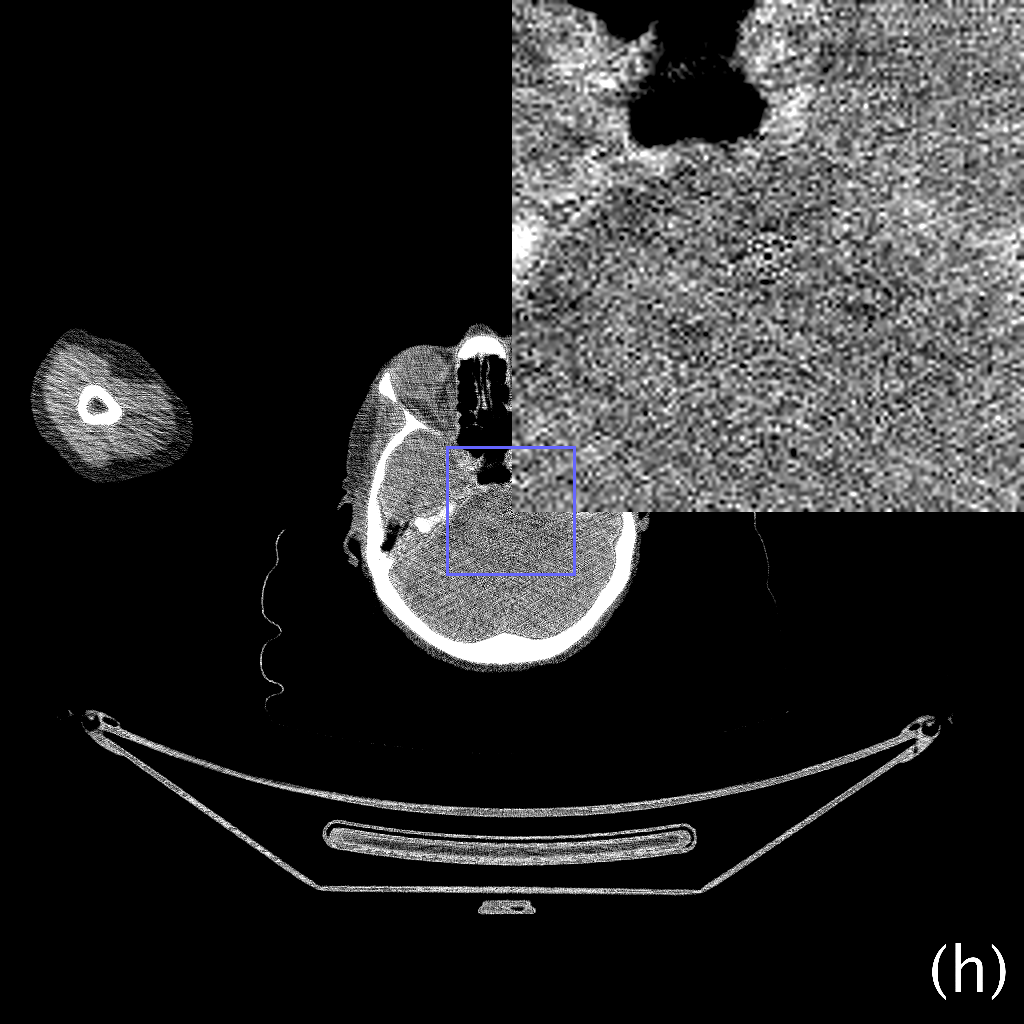}
         %\caption{}
     \end{subfigure}
        \caption{Example slice from the NSCLC test set. 100 $\mathrm{keV}$ virtual monoenergetic images. Display window [-160,240] HU. (a) ring artifact free (truth), (b) ring corrupted (observed), (c) L2, (d)  L1, (e) VGG, (f) $\mathrm{VGG}_{70}$, (g) VGG-L1, (h) $\mathrm{VGG}_{70}$-L1.}
        \label{qual_100_nsclc}
\end{figure}

Qualitative results are available in Fig. \ref{qual_40_kits}-\ref{qual_100_nsclc} where we show 40, 70, and 100 $\mathrm{keV}$ virtual monoenergetic images from the KiTS19 and the NSCLC test sets. Despite the fact that the network is operating on a pair of material basis images, we show virtual monoenergetic images. This is mainly due to the fact that the rings are correlated between the two material basis images. Hence, although the rings may no longer be visible in the ring corrected basis images, it is possible that the rings reemerge for certain linear combinations considered when forming virtual monoenergetic images. In addition, and perhaps more importantly, the virtual monoenergetic images are often what is in the end of interest to the radiologist. In Fig. \ref{qual_40_kits} we have formed 40 $\mathrm{keV}$ virtual monoenergetic images from an example slice in the KiTS19 test set. This figure contains the ring artifact free (truth), ring corrupted (observed), and processed (predicted) images for each loss configuration considered. For each, we plot a magnification of the center region of interest (ROI) in the upper right corner to emphasize details. We can see that all loss configurations except $\mathrm{VGG}_{70}$ have done a good job correcting for the rings at this energy level. That $\mathrm{VGG}_{70}$ fails at 40 $\mathrm{keV}$ is rather expected since this loss function is completely agnostic to the spectral nature of the problem and only ``sees'' 70 $\mathrm{keV}$ virtual monoenergetic images. In Fig. \ref{qual_70_kits} we show the 70 $\mathrm{keV}$ virtual monoenergetic image. Comparing to observed, we can see that all loss configurations have reduced the degree of ring artifacts. However, only the loss configurations including $\mathrm{VGG}_{70}$ adequately correct for the ring artifacts such that they are no longer obviously visible. In addition, we note that the processed images for $\mathrm{VGG}_{70}$ and $\mathrm{VGG}_{70}$-L1 seem to accurately reproduce truth and preserve fine details. This is particularly evident when comparing the magnified version of the center ROI. The ability to faithfully reproduce truth is explored further below by considering the statistical properties of the indicated red ROIs in addition to considering profile plots. Despite doing a good job correcting for the rings, some residual rings are still visible. However, they now appear more as thicker, band-like, partial circles. We have annotated the processed image for $\mathrm{VGG}_{70}$ and $\mathrm{VGG}_{70}$-L1 in Fig. \ref{qual_70_kits} with red arrows to draw the reader's attention to these remaining artifacts. We can also note that these artifacts are slightly stronger for $\mathrm{VGG}_{70}$-L1 than for $\mathrm{VGG}_{70}$. Finally, for the KiTS19 example, we have Fig. \ref{qual_100_kits}, where we show the results for the 100 $\mathrm{keV}$ virtual monoenergetic image. The results closely parallel that of the 70 $\mathrm{keV}$ case. For all loss configurations, save $\mathrm{VGG}_{70},$ there is a clear reduction in the ring artifacts. However, in contrast to the 70 $\mathrm{keV}$ case, it is now only $\mathrm{VGG}_{70}$-L1 that suppresses the rings sufficiently such that they are no longer visible. $\mathrm{VGG}_{70}$ fails here for the same reason as in the 40 $\mathrm{keV}$ case. In Fig. \ref{qual_40_nsclc}-\ref{qual_100_nsclc} we can see that the network is doing a good job generalizing from chest and abdomen scans to a head scan from the NSCLC test set. The results in Fig. \ref{qual_40_nsclc}-\ref{qual_100_nsclc} closely parallels what we saw for the KiTS19 example. All loss configurations, save $\mathrm{VGG}_{70},$ adequately correct for the rings such that they are no longer visible in the 40 $\mathrm{keV}$ case. For the 70 $\mathrm{keV}$ case in Fig. \ref{qual_70_nsclc} only the two loss configurations including $\mathrm{VGG}_{70}$ sufficiently correct for the rings. In Fig. \ref{qual_100_nsclc} only our proposed spectral loss $\mathrm{VGG}_{70}$-L1 remains successful. The fact that the network seems to perform fairly well on head scans, despite being trained on chest and abdomen scans, is indicating that the network has learned some general features that are seemingly robust to the specific anatomy considered. Out of the loss configurations considered, only our proposed spectral loss, $\mathrm{VGG}_{70}$-L1, is able to produce ring corrected 40, 70, and 100 $\mathrm{keV}$ virtual monoenergetic images for both the KiTS19 and NSCLC case. Removing any of the components of $\mathrm{VGG}_{70}$-L1, the L1-loss, the $\mathrm{VGG}_{70}$-loss, our using VGG instead of $\mathrm{VGG}_{70}$, produces some failure to correct for the rings at the energy levels considered.

\begin{figure}
     \centering
     \begin{subfigure}[b]{0.33\columnwidth}
         \centering
         \includegraphics[width=\columnwidth]{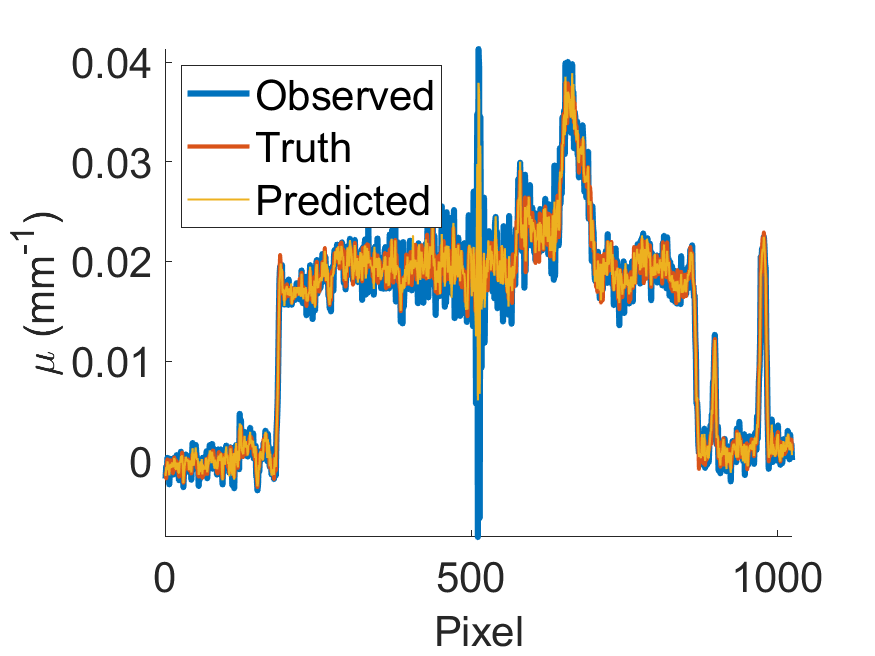}
         \caption{}
     \end{subfigure}
       \begin{subfigure}[b]{0.33\columnwidth}
         \centering
         \includegraphics[width=\columnwidth]{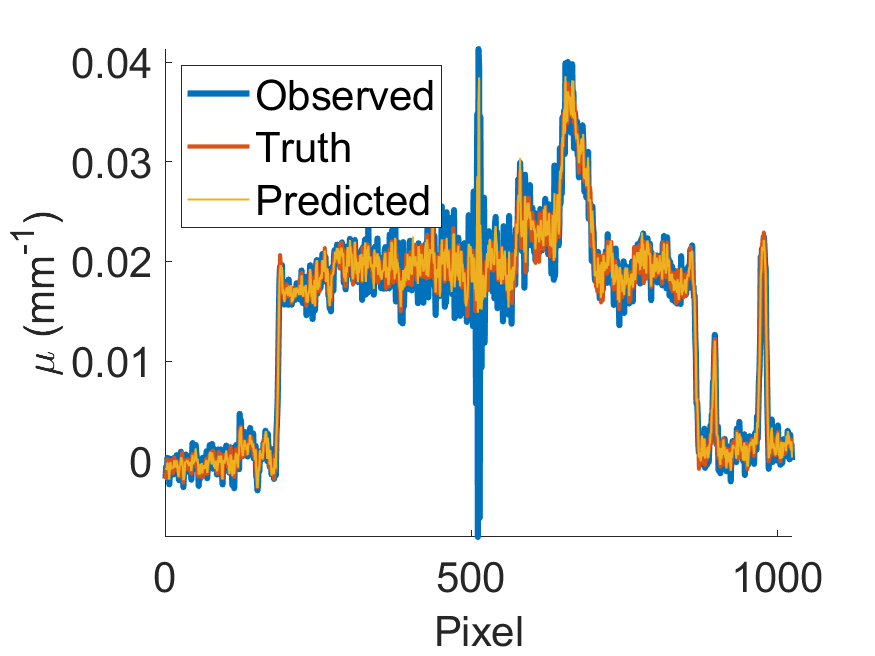}
         \caption{}
     \end{subfigure}
     \begin{subfigure}[b]{0.33\columnwidth}
         \centering
         \includegraphics[width=\columnwidth]{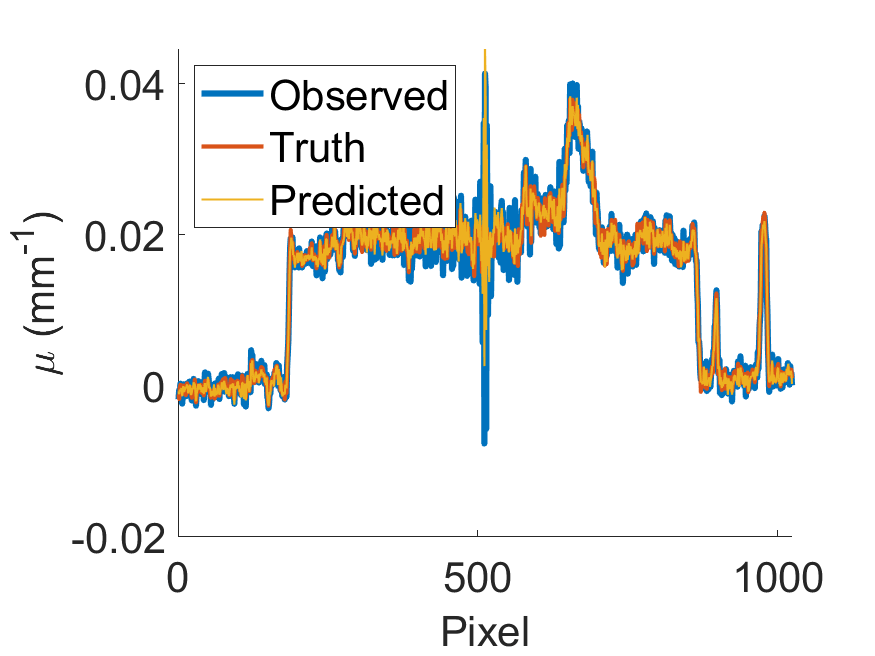}
         \caption{}
     \end{subfigure}
     \vfill
     \begin{subfigure}[b]{0.33\columnwidth}
         \centering
         \includegraphics[width=\columnwidth]{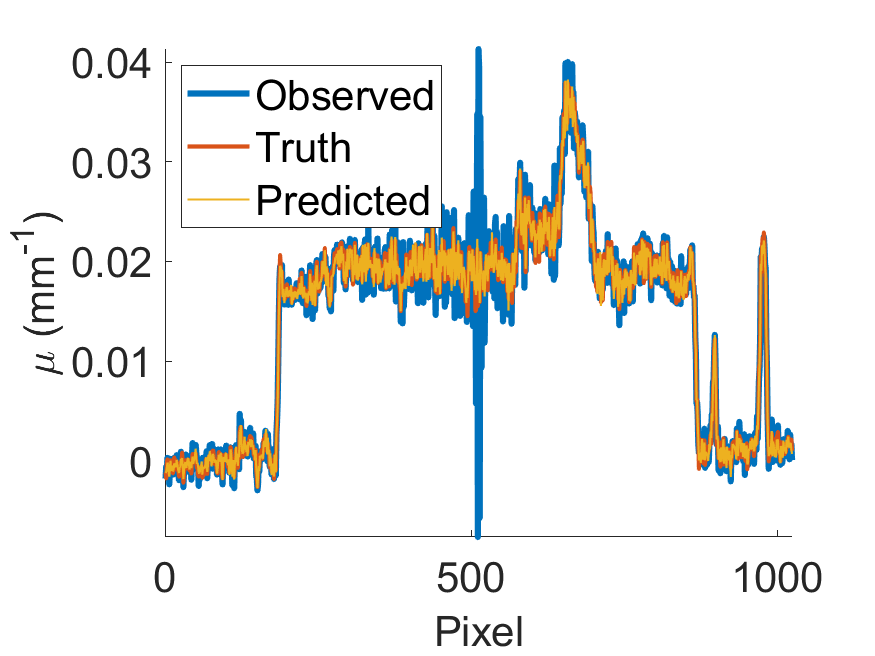}
         \caption{}
     \end{subfigure}
     \begin{subfigure}[b]{0.33\columnwidth}
         \centering
         \includegraphics[width=\columnwidth]{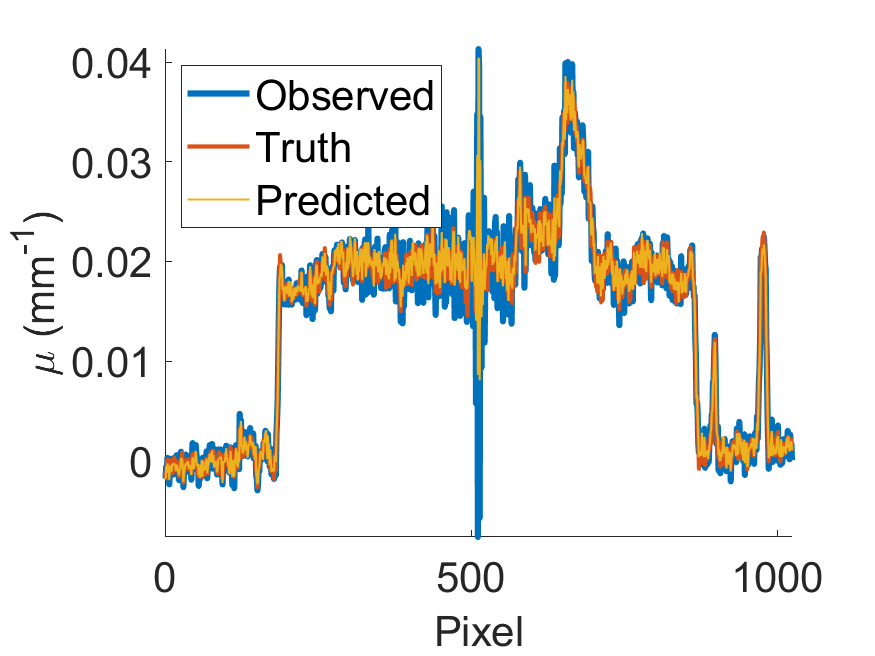}
         \caption{}
     \end{subfigure}
     \begin{subfigure}[b]{0.33\columnwidth}
         \centering
         \includegraphics[width=\columnwidth]{unet_alt_64_vgg16_alt_9_100_2_sc_512_1_1_1_2_train_kits_img_test_kits_img_34_profile.png}
         \caption{}
     \end{subfigure}
        \caption{Profile of a vertical line through spine in the 70 $\mathrm{keV}$ virtual monoenergetic image for the KiTS19 case shown in Fig. \ref{qual_70_kits}. (a) L2, (b)  L1, (c) VGG, (d) $\mathrm{VGG}_{70}$, (e) VGG-L1, (f) $\mathrm{VGG}_{70}$-L1.}
        \label{profile_plots_kits}
\end{figure}

\begin{figure}
     \centering
     \begin{subfigure}[b]{0.33\columnwidth}
         \centering
         \includegraphics[width=\columnwidth]{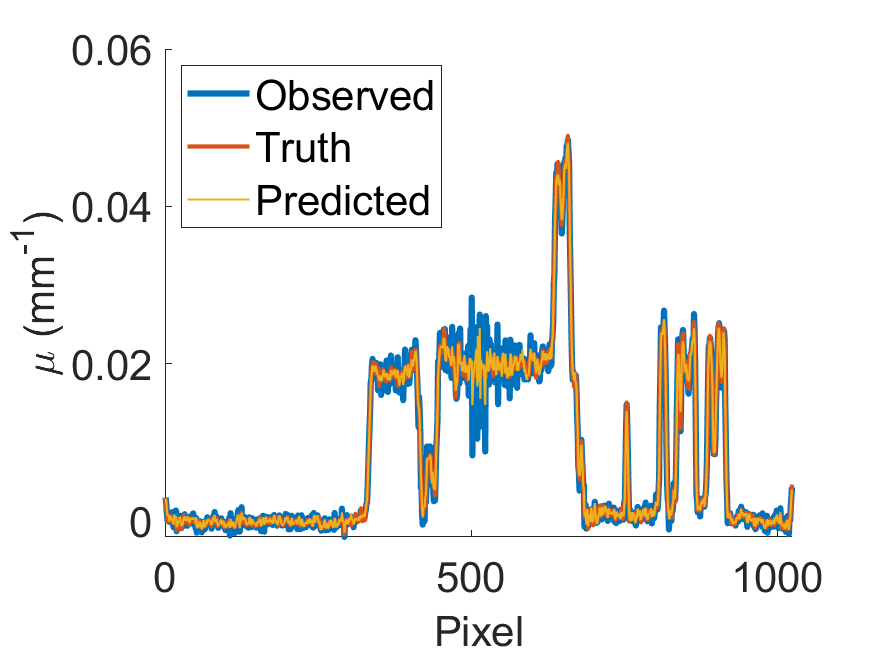}
         \caption{}
     \end{subfigure}
       \begin{subfigure}[b]{0.33\columnwidth}
         \centering
         \includegraphics[width=\columnwidth]{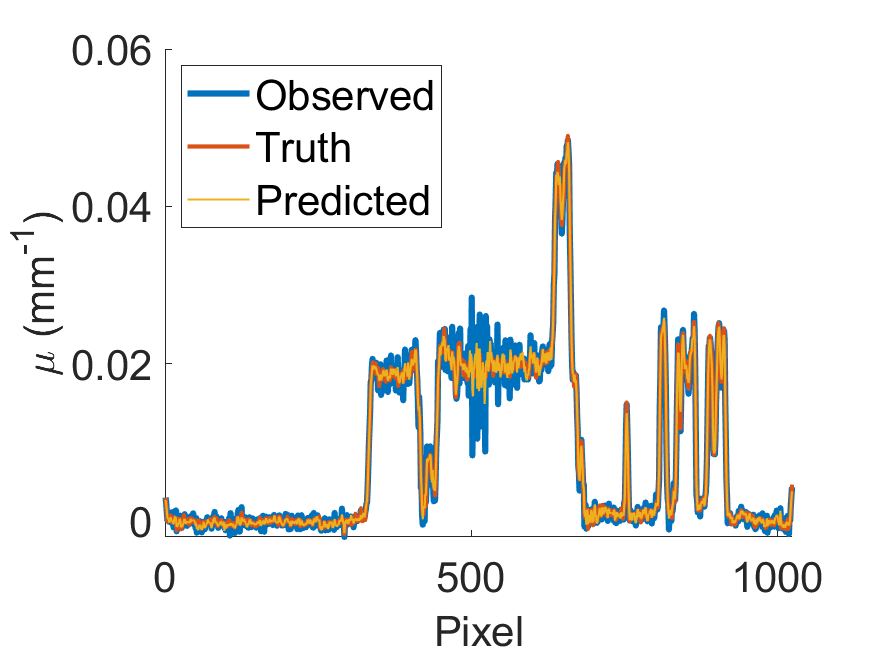}
         \caption{}
     \end{subfigure}
     \begin{subfigure}[b]{0.33\columnwidth}
         \centering
         \includegraphics[width=\columnwidth]{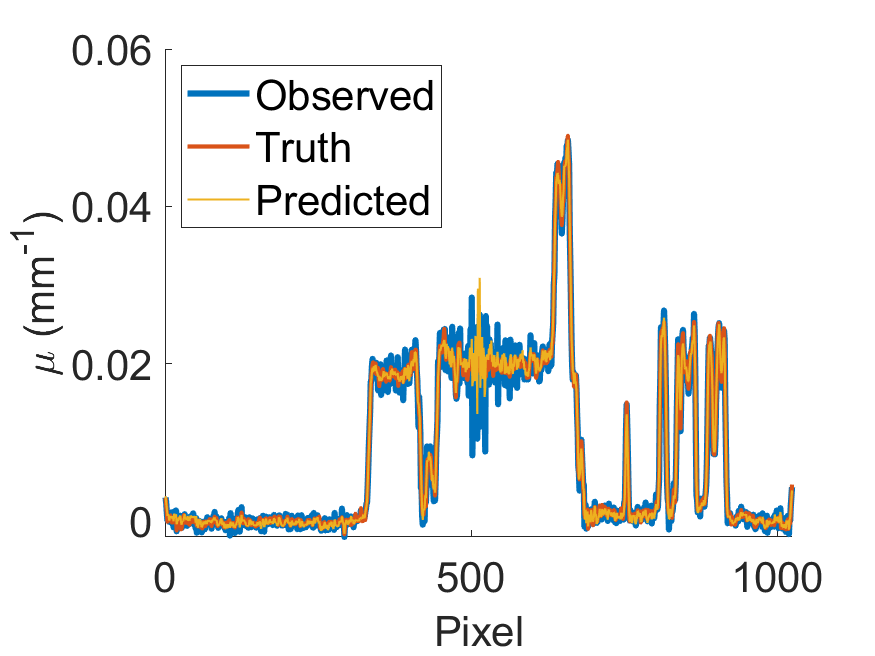}
         \caption{}
     \end{subfigure}
     \vfill
     \begin{subfigure}[b]{0.33\columnwidth}
         \centering
         \includegraphics[width=\columnwidth]{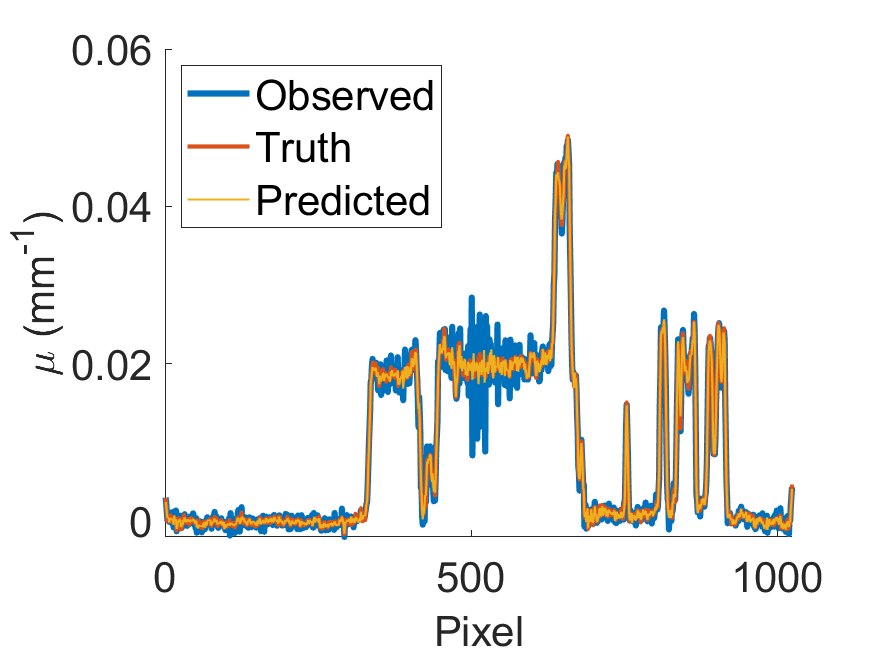}
         \caption{}
     \end{subfigure}
     \begin{subfigure}[b]{0.33\columnwidth}
         \centering
         \includegraphics[width=\columnwidth]{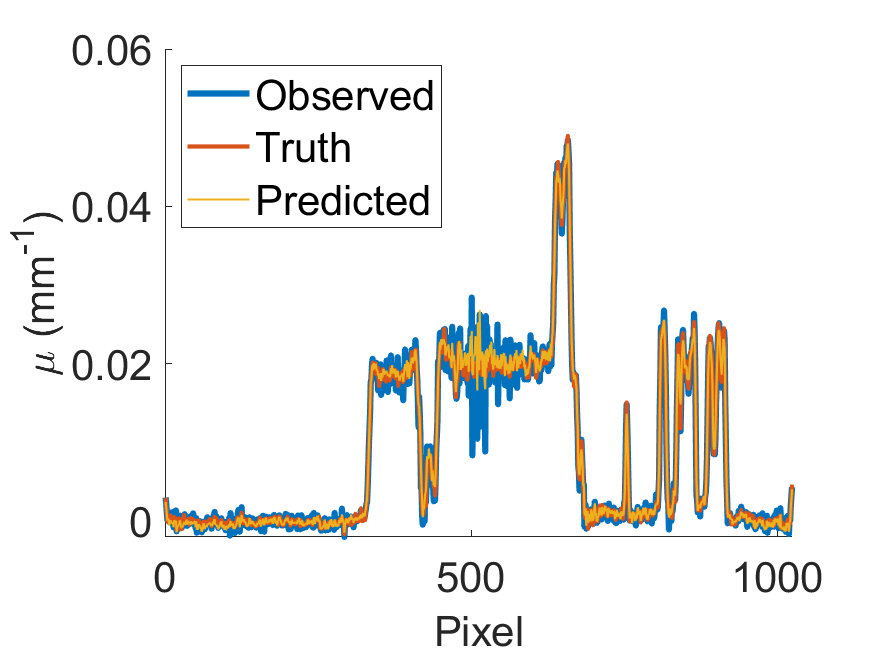}
         \caption{}
     \end{subfigure}
     \begin{subfigure}[b]{0.33\columnwidth}
         \centering
         \includegraphics[width=\columnwidth]{unet_alt_64_vgg16_alt_9_100_2_sc_512_1_1_1_2_train_kits_img_val_nsclc_img_175_profile.png}
         \caption{}
     \end{subfigure}
        \caption{Profile of a vertical line through spine in the 70 $\mathrm{keV}$ virtual monoenergetic image for the NSCLC case shown in Fig. \ref{qual_70_nsclc}. (a) L2, (b)  L1, (c) VGG, (d) $\mathrm{VGG}_{70}$, (e) VGG-L1, (f) $\mathrm{VGG}_{70}$-L1.}
        \label{profile_plots_nsclc}
\end{figure}

To further analyze the processed images, we consider the profile of a vertical line at pixel 512 going through the spine in the KiTS19 case and the head in the NSCLC case for the 70 $\mathrm{keV}$ monoenergetic images in Fig. \ref{qual_70_kits} and Fig. \ref{qual_70_nsclc}. The results can be found in Fig. \ref{profile_plots_kits} and Fig. \ref{profile_plots_nsclc}. In Fig. \ref{profile_plots_kits}, we can see that $\mathrm{VGG}_{70}$ and $\mathrm{VGG}_{70}$-L1 seem to be doing a very good job reproducing truth. There is no visible introduction of bias or change in standard deviation. For the remaining loss configurations we can see that they are mostly doing a good job, save for large deviations near the center. This is likely due to the remaining ring artifacts. The conclusions from Fig. \ref{profile_plots_nsclc} are largely the same. Again, although slightly less obviously so, $\mathrm{VGG}_{70}$ and $\mathrm{VGG}_{70}$-L1 are the best performing setups. Between the two only our proposed spectral loss, $\mathrm{VGG}_{70}$-L1, is capable of correcting for ring artifacts at a range of energy levels. 

\subsection{Quantitative results}
To quantitatively assess the results, we first use two standard metrics in the literature: Structural Similarity Index (SSIM) \cite{wang2004} and Peak Signal-to-Noise Ratio (PSNR). The average of these two metrics are available for the KiTS19 and NSCLC test sets in Table \ref{table_quant}. We note, however, that these metrics are not necessarily optimal when it comes to assessing the perceptual quality of the processed images\cite{johnson2016}---which is ultimately what we are interested in. The top performers are L1 and L2. Note also that L1 outperforms L2 in terms of PSNR. This is a bit odd since it is equivalent to the network trained with the L1-loss achieving lower L2-loss than a network trained with the L2-loss. One possible explanation for this is that the L2-loss is more prone to get stuck in local minima due to its smoothness and convexity properties \cite{zhao2017}. Another possibility, since the difference in PSNR between the two loss configurations is very small, is that it might simply be due to stochastic variation in the optimization procedure. Another thing that sticks out in Table \ref{table_quant} is that $\mathrm{VGG}_{70}$ achieves lower SSIM and PSNR than observed for both test sets. In other words, according to these metrics, the processed images are worse than the ring corrupted ones. That $\mathrm{VGG}_{70}$ performs worse than observed is likely explained by the fact that $\mathrm{VGG}_{70}$  only operates on virtual monoenergetic images whereas these results are based on the material basis images. In particular, we can see a steady improvement for $\mathrm{VGG}_{70}$-L1 where we have added the L1-loss, which operates on the material basis images. In essence, $\mathrm{VGG}_{70}$ only works for the, non-spectral, 70 $\mathrm{keV}$ case. Comparing performance on the two test sets, it may seem at first glance as if there is a drop in performance on the NSCLC test set. However, note that the percentage improvement over observed is actually higher on average for the NSCLC test set.

To further evaluate the processed images we consider the first two moments of the distribution of the pixels in two roughly flat ROIs, indicated red, in Fig. \ref{qual_70_kits} and \ref{qual_70_nsclc}. The results are available in Table \ref{table_stat}. We treat ``truth'' as the gold standard. In other words, a perfect model would reproduce the mean and standard deviation of truth. By comparing the results for truth and observed we can see that the rings mainly impact the standard deviation. The mean, on the other hand, remains largely unaffected. L2 achieved the lowest error in reproducing the mean for the KiTS19 and the NSCLC case and $\mathrm{VGG}_{70}$ is the top performer in terms of reproducing the standard deviation. Although $\mathrm{VGG}_{70}$-L1 is not the best at reproducing the mean nor standard deviation, it does a very good job getting both approximately correct simultaneously. For the KiTS19 case, the error for both standard deviation and mean is less than one percent. For the NSCLC case, it is around two percent. Hence, there is tentatively a drop in performance when applying the network to a head scan. Overall, the results in Table \ref{table_stat}, in conjunction with the qualitative results, indicate that the network trained with our spectral loss does a good job faithfully reproducing truth. 

\begin{table}[]
    \centering
    \begin{tabular}{ccccc} \toprule
	&	KiTS19	&		&	NSCLC	&		\\ \midrule
	&	SSIM	&	PSNR	&	SSIM	&	PSNR	\\ \midrule 
Observed 	&	0.66	&	29.2	&	0.55	&	27.8	\\
L2	&	0.88	&	33.7	&	0.81	&	33.1	\\
L1	&	0.88	&	33.8	&	0.81	&	33.2	\\
VGG	&	0.83	&	33.6	&	0.75	&	32.9	\\
$\mathrm{VGG}_{70}$	&	0.45	&	28.9	&	0.25	&	27.4	\\
VGG-L1	&	0.84	&	33.5	&	0.75	&	32.8	\\
$\mathrm{VGG}_{70}$-L1	&	0.87	&	33.4	&	0.79	&	32.7	\\
 \bottomrule
    \end{tabular}
    \caption{Average SSIM and PSNR measured in the material basis images in the KiTS19 and NSCLC test sets.}
    \label{table_quant} 
\end{table}

\begin{table}[]
    \centering
    \begin{tabular}{ccccc} \toprule
	&	KiTS19	&		&	NSCLC	&		\\ \midrule
	&	Mean	&	Std	&	Mean	&	Std	\\ \midrule
Truth	&	48.8	&	67.5	&	44.9	&	41.3	\\
Observed 	&	48.6	&	83.8	&	44.5	&	76.0	\\
L2	&	49.0	&	63.7	&	44.9	&	42.6	\\
L1	&	60.7	&	63.8	&	47.6	&	42.6	\\
VGG	&	47.9	&	61.6	&	39.7	&	44.6	\\
$\mathrm{VGG}_{70}$	&	47.6	&	67.2	&	46.2	&	41.0	\\
VGG-L1	&	41.9	&	61.8	&	51.1	&	42.5	\\
$\mathrm{VGG}_{70}$-L1	&	48.5	&	66.9	&	44.2	&	40.4	\\
 \bottomrule
    \end{tabular}
    \caption{Mean and standard deviation of red ROIs in Fig. \ref{qual_70_kits} and \ref{qual_70_nsclc}. Values are in Hounsfield Units (HU).}
    \label{table_stat} 
\end{table}

\section{Discussion}
Next generation X-ray CT scanners, based on photon-counting detectors, are now clinically available. One common issue for photon-counting detectors is detector inhomogeneity, which, if left uncorrected for, leads to streak artifacts in the projection domain and corresponding ring artifacts in the image domain. The proper way to combat this issue is detector calibration. However, this calibration cause wear on the X-ray tube and may be very time-consuming. In this paper, we have proposed a deep learning-based ring correction technique for photon-counting spectral CT. We have demonstrated a proof-of-concept of how a CNN can be used to satisfactorily correct for ring artifacts whilst preserving fine details that are of clinical importance. Our proposed method is not meant to replace detector calibration, but rather to complement it. Having available an image processing technique for ring artifact correction may, for instance, allow postponing the calibration process to a more convenient time and enable higher patient throughout as it allows for longer intervals between calibration. In addition, increased robustness to detector inhomogeneity can also improve image quality in situations where the recommended calibration procedure has not been followed properly.

While our results are very promising, this method is not without its limitations. In particular, we found the network unable to fully correct for ring artifacts in some slices imaging the head. This is most likely because the anatomy of the head is very different from other parts of the body. There is bone completely encapsulating soft tissue. The first obvious step in trying to improve the performance of skull images is to include head scans in the training data. Recall that the network is currently trained on a dataset including only chest and abdomen scans. In future research, we plan to train on a larger a more diverse dataset, containing head scans, and see if we can ensure high performance even on these problematic slices.

Moreover, the network is currently trained on a continuous ring pattern, which is intuitive although not the most realistic. In a more realistic scenario, the types of ring artifacts that one may encounter are more likely in the shape of single rings, partial rings, bands, and combinations of these. In future research, we will extend the results presented in this paper to ring patterns that occur more frequently in practice. We also plan to evaluate our proposed technique on clinical data. 

Another interesting avenue, left for future research, is to further explore the properties of the processed images. Here we do an initial analysis which mainly relies on qualitative results. One could do this more formally using tools from medical imaging such as contrast-to-noise-ratio (CNR), signal-to-noise-ratio (SNR), modulation transfer function (MTF), and the noise power spectrum (NPS).

For the results in this paper we trained the network on randomly extracted $512 \times 512$ patches. As mentioned above, this has numerous benefits including a regularizing effect and reduced graphics memory requirements. It might be beneficial to train on smaller patches, say $128 \times 128$, as one could extract more patches, augmenting the dataset, within the same computational budget. However, initial results indicated that the network failed to properly generalize from smaller patches to full images. This is in contrast to the results in \cite{trapp2022}, where they train a 2D network, very similar to ours, on patches that are $64 \times 64.$ One key difference between their work and ours is that we are concerned with spectral CT and pass a pair of material basis images, instead of a single CT image, through the network. It is possible that larger patch sizes are necessary to achieve adequate performance in the spectral case. Further exploring the issue of optimal patch size if left for future research.  

Moreover, GANs have been used very successfully for ring correction \cite{wang2019} and image denoising \cite{yang2018,kim2019,wolterink2017}. Hence, a natural extension of the work in this paper is to combine our proposed spectral loss with an adversarial loss. We did some experiments, not shown in this paper, using an adversarial loss based on the WGAN-GP \cite{arjovsky2017,gulrajani2017}. An adjusted version of the discriminator used in PatchGAN \cite{isola2016} was used as critic. Initial results indicated that combining the WGAN-GP with our spectral loss did not vastly improve the performance of our network and we therefore deemed the significantly higher computational cost unwarranted. Nevertheless, this is an interesting avenue of research that we plan to explore in future work. 

\section{Conclusion}
This paper is a proof-of-concept of using a CNN to correct for ring artifacts, while preserving fine details, in the material basis images in photon-counting spectral CT. This was achieved by training a basic UNet with a spectral loss that combines a perceptual loss operating on 70 $\mathrm{keV}$ virtual monoenergetic images with a L1-loss operating on the material basis images. We have demonstrated that the network is capable of producing ring corrected virtual monoenergetic images at a range of energy levels.  

\section*{Acknowledgment}
We thank the PDC Center for High Performance Computing, KTH Royal Institute of Technology, Sweden, for providing access to computing resources used in this research. Model training was enabled by the Berzelius resource provided by the Knut and Alice Wallenberg Foundation at the National Supercomputer Centre.

%Bibliography
\bibliographystyle{unsrt}  
\bibliography{references}

\end{document}